\documentclass[manuscript]{aastex}
\usepackage{txfonts}
\usepackage{graphicx}

\slugcomment{Accepted to ApJS}

\shorttitle{Molecular environment of 51 Planck cold clumps in Orion complex}

\shortauthors{Liu et al.}

\begin{document}

\title{Molecular environments of 51 Planck cold clumps in Orion complex}

\author{Tie Liu\altaffilmark{1}, Yuefang Wu\altaffilmark{1}, Huawei Zhang\altaffilmark{1} }

\altaffiltext{1}{Department of Astronomy, Peking University, 100871,
Beijing China; liutiepku@gmail.com, ywu@pku.edu.cn}

\begin{abstract}
A mapping survey towards 51 Planck cold clumps projected on Orion complex was performed with J=1-0 lines of $^{12}$CO and $^{13}$CO at the 13.7 m telescope of Purple Mountain Observatory. The mean column densities of the Planck gas clumps range from 0.5 to 9.5$\times10^{21}$ cm$^{-2}$, with an average value of (2.9$\pm$1.9)$\times10^{21}$ cm$^{-2}$. While the mean excitation temperatures of these clumps range from 7.4 to 21.1 K, with an average value of 12.1$\pm$3.0 K. The averaged three-dimensional velocity dispersion $\sigma_{3D}$ in these molecular clumps is 0.66$\pm$0.24 km~s$^{-1}$. Most of the clumps have $\sigma_{NT}$ larger than or comparable with $\sigma_{Therm}$. The H$_{2}$ column density of the molecular clumps calculated from molecular lines correlates with the aperture flux at 857 GHz of the dust emission. Through analyzing the distributions of the physical parameters, we suggest turbulent flows can shape the clump structure and dominate their density distribution in large scale, but not affect in small scale due to the local fluctuations. Eighty two dense cores are identified in the molecular clumps. The dense cores have an averaged radius and LTE mass of 0.34$\pm$0.14 pc and 38$_{-30}^{+5}$ M$_{\sun}$, respectively. And structures of low column density cores are more affected by turbulence, while those of high column density cores are more concerned by other factors, especially by gravity. The correlation of the velocity dispersion versus core size is very weak for the dense cores. The dense cores are found most likely gravitationally bounded rather than pressure confined. The relationship between $M_{vir}$ and $M_{LTE}$ can be well fitted with a power law. The core mass function here is much more flatten than the stellar initial mass function. The lognormal behavior of the core mass distribution is most likely determined by the internal turbulence.
\end{abstract}

\keywords{Massive core:pre-main sequence-ISM: molecular-ISM:
kinematics and dynamics-stars: formation}

\clearpage

\section{Introduction}
Orion complex is the most excellent laboratory for studies of star formation. Thousands of low-mass stars as well as a number of high mass stars formed in this region
within the last few million years \citep{ba05,Hil97}. Massive stars interact with the molecular clouds in this region through the powerful ionizing radiation, strong stellar wind and supernova explosion \citep{cow79,ba87}, reshape and compress the clouds into filamentary structures which contains three 10$^{5}$ M$_{\sun}$ giant molecular clouds (Orion A, B and Mon R2) \citep{ba05,wil05}. Triggered star formation has been suggested to explain the large spatial scale age gradients of the distinguished star clusters in this region \citep{el77}. Thus it is important to study the properties of the molecular clouds and the star forming activities in this region. In previous works, the molecular clouds in this region have been widely studied through mapping surveys in transitions of CO and its isotopes \citep{ma86,ba87,cas90,kra96,sak96,na98,wil01,wil05,sh11}. However, most of these studies focus on particular regions with active star formation (e.g. Orion A and Orion B) or have poor spatial resolution, which seldom pay attention to the clouds comprising pre-stellar cores. The studies towards pre-stellar cores can help understand the formation and evolution of dense cores, as well as the cause of the initial mass function (IMF) \citep{ade11a}. Studies towards pre-stellar cores in this region are urgently needed. Recently, millieter/sub-millimeter continuum surveys have revealed some quiescent cores in Orion \citep{Li07,Sad10}, and the core mass function for these quiescent cores has a power index of $\sim-0.85$ \citep{Li07}. But continuum observations can not provide us the velocity information of these cores, and limit our understandings towards the core properties such as velocity dispersions, core stabilities and so on.

The investigations of the density and temperature distributions, and the pressure supports in the pre-stellar cores are critically needed. Unfortunately, the properties of pre-stellar cores were not known too much due to lacking of samples before Planck satellite. Working at submillimetre/millimetre bands, Planck satellite is unique and suited for systemic surveys of cold clumps, which already have provided a preliminary catalogue
of 10,783 cold clumps (the cold core Catalogue of Planck Objects, C3PO) \citep{ade11a}. The planck cold clumps in the C3PO sample were found with low column densities N$_{H_{2}}\sim0.1$ to 1.6$\times10^{22}$ cm$^{-2}$ and dust temperatures between 10 and 15 K \citep{ade11a}. However, the physical properties of those cold clumps are still poorly known, especially for their molecular environments. Recently a single-point survey toward 674 Planck cold clumps of the Early Cold clump Catalogue (ECC)
in the J=1-0 transitions of $^{12}$CO, $^{13}$CO and C$^{18}$O has been carried out using the Purple Mountain Observatory (PMO) 13.7 m telescope \citep{wu12}. However, mapping observations are needed to reveal the structures and properties of these clumps in detail.

In this paper, we report the results of a mapping survey in $^{12}$CO (1-0) and it's isotopes toward 51 Planck cold clumps selected from the survey of \cite{wu12}. The selected cold clumps are associated with the Orion giant molecular cloud (GMC) \citep{da87} in projection and their coordinates and systemic velocities are listed in Table 1. The clumps having two velocity components are distinguished with "a" and "b" at the end of the core names. The 9th column of Table 1 presents the aperture flux density at 857 GHz (apflux857) observed by Planck satellite \citep{ade11b}. The last column presents the associations with these cold clumps identified with SIMBAD\footnote{SIMBAD database is operated at CDS, Strasbourg, France}. It can be seen that most of those cold clumps are associated with dark clouds or very weak IRAS point sources (the flux at 100 $\micron$ ranging from 2.3 to 35 Jy), indicating they have low column densities and less star forming activities. In this paper, the distances of these cold clumps are assumed to be 450 pc except for G180.81-19.66 and G185.80-09.12, which are associated with $\lambda$ Orion region and have distances of 400 pc \citep{ade11b}.

The observations will be introduced in the next section. The basic analysis and results of the molecular line observations will be presented in the third section. We discussed the properties of the molecular environments of these cold clumps in section 4. Section 5 summarizes this paper.

\section{Observations}
The observations toward 51 Planck cold clumps in Orion complex in
$^{12}$CO (1-0), $^{13}$CO (1-0) and C$^{18}$O (1-0) were carried out
with the PMO 13.7 m radio telescope
from April to June in 2011. The new 9-beam array receiver system in single-sideband
(SSB) mode was used as front end. FFTS spectrometers were used as back ends, which have a
total bandwidth of 1 GHz and 16384 channels, corresponding to a
velocity resolution of 0.16 km~s$^{-1}$ for $^{12}$CO (1-0) and 0.17
km~s$^{-1}$ for $^{13}$CO (1-0) and C$^{18}$O (1-0). $^{12}$CO (1-0) was observed
at upper sideband, while $^{13}$CO (1-0) and C$^{18}$O (1-0) were observed simultaneously at lower
sideband.  The half-power beam width (HPBW) is 56$\arcsec$ and the main beam efficiency is
$\sim0.5$. The pointing accuracy of the telescope was better than
4$\arcsec$. The typical system temperature (T$_{sys}$) in SSB mode
is around 110 K and varies about 10\% for each beam. The
On-The-Fly (OTF) observing mode was applied. The antenna
continuously scanned a region of 22$\arcmin\times22\arcmin$ centered
on the Planck cold clumps with a scan speed of 20$\arcsec$~s$^{-1}$.
However, the edges of the OTF maps are very noisy and thus only the central 14$\arcmin\times14\arcmin$
regions are selected to be further analyzed. The typical rms noise level was
0.2 K in T$_{A}^{*}$ for $^{12}$CO (1-0), and 0.1 K for $^{13}$CO (1-0) and C$^{18}$O (1-0).
Using the GILDAS software package including CLASS and GREG \citep{gui00}, the OTF data were
converted to 3-D cube data with a grid spacing of 30$\arcsec$ and the baselines
were corrected by fitting with linear or sinusoidal functions.
Then the data were exported to MIRIAD \citep{sau95} for further analysis.

\section{Results}
\subsection{Distribution of the cold clumps in Orion GMC}
There are 82 cold clumps in the ECC catalog projected on the Orion complex \citep{wu12}. We showed the distributions of the aperture flux density at 857 GHz (apflux857) for the mapped cold cores and for all the Planck cold clumps projected in Orion complex in the left panel of Figure 1. One can see our mapped sub-sample can well represent the whole sample in Orion complex. As shown in the right panel of Figure 1, the locations of the Planck cold clumps in the mapping survey are plotted as green "crosses". The background image represents the H$_{\alpha}$ emission \citep{fin03}. The red and blue contours present the CO (1-0) \citep{dame01} and IRAS 100 $\micron$ emission, respectively. The CO emission roughly coincides with the IRAS 100 $\micron$ emission in space, but is not associated with H$_{\alpha}$ emission. Nearly all the cold clumps are associated with the CO emission and distribute at the boundary or far from the H$_{\alpha}$ emission. The cold clumps form two large loops as denoted by the two dashed ellipses.

\subsection{Overall pictures of the molecular clumps}
\subsubsection{LTE analysis}
With the theory of radiation transfer and molecular excitation \citep{win79,gar91}, the analysis of the parameters of each clump was performed under local thermal equilibrium (LTE) assumption. Assuming $^{12}$CO (1-0) emission to be optically thick and the beam-filling factor to be unit, the excitation temperature T$_{ex}$ can be calculated directly. Then the column densities of $^{13}$CO (1-0) were straightforward calculated using the first equation in \cite{gar91}. The column densities of H$_{2}$ were obtained by adopting typical abundance ratios [H$_{2}$]/[$^{12}$CO]=$10^{4}$ and [$^{12}$CO]/[$^{13}$CO]=60 in the interstellar medium.

In Figure 2 the excitation temperatures are presented in color scale and the column densities of H$_{2}$ in contours. One can see most of the clumps are very diffuse and temperature gradients are seen in many of them. The mean values of the column density and excitation temperature of each clump were obtained through analyzing the pixies within the 30\% contours in the column density maps and presented in column 2 and 5 in Table 2. The mean column densities of these clumps range from 0.5 to 9.5$\times10^{21}$ cm$^{-2}$, with an average value of (2.9$\pm$1.9)$\times10^{21}$ cm$^{-2}$. While the mean excitation temperatures of these clumps range from 7.4 to 21.1 K, with an average value of 12.1$\pm$3.0 K. The column densities revealed by dust emission were found to range from 0.1 to 1.6$\times10^{22}$ cm$^{-2}$ and dust temperature from 10 to 15 K in the C3PO samples \citep{ade11a}, which are consistent with the excitations and column densities obtained here from CO emission. In 135 clumps associated with IRAS sources, \cite{wang09} found an averaged excitation temperature of 9.7 K and an averaged H$_{2}$ column density of 8.9$\times10^{21}$ cm$^{-2}$. The infrared dark clumps (IRDCs) were found to have a typical excitation temperature of 10 K and typical column density of several $\times10^{22}$ cm$^{-2}$ \citep{du08}. Comparing with the IRAS sources and IRDCs, the Planck cold clumps have slightly larger excitation temperatures but much smaller column densities, indicating these Planck cold clumps represent an earlier evolutionary phase in star formation.

\subsubsection{First moment maps}
The intensity weighted velocity maps (First moment maps) of the clumps are shown in Figure 3 in color scale overlayed with the contours of the column densities of H$_{2}$. Velocity gradients are found in nearly all the clumps. Taking G185.80-09.12 for example, two compact cores are revealed in the map. The northern one has an averaged velocity of -2.3$\pm$0.1 km~s$^{-1}$, while the central one has an averaged velocity of -2.8$\pm$0.1 km~s$^{-1}$.

\subsubsection{Velocity dispersion}
The maps of one dimensional velocity dispersion of $^{13}$CO (1-0) are shown in Figure 4 in color scale overlayed with the contours of the column densities of H$_{2}$. One can found the velocity dispersion is usually larger in the dense regions than in the less dense regions.  The non-thermal ($\sigma_{NT}$ ) and thermal ($\sigma_{Therm}$) one dimensional velocity dispersions in each clump are calculated as following:
\begin{equation}
\sigma_{NT} = \left[\sigma_{^{13}CO}^{2}-\frac{kT_{ex}}{m_{^{13}CO}}\right]^{1/2}
\end{equation}

\begin{equation}
\sigma_{Therm} = \sqrt{\frac{kT_{ex}}{m_{H}\mu}}
\end{equation}
where $\sigma_{^{13}CO}$ and $T_{ex}$ are the one dimensional velocity dispersion of $^{13}$CO (1-0) and excitation temperature of each pixel in each clump, respectively. k is Boltzmann's constant, $m_{^{13}CO}$ is the mass of $^{13}CO$, $m_{H}$ is the mass of atomic hydrogen, and $\mu$=2.72 is the mean molecular weight of the gas. Then the pixel value of three-dimensional velocity dispersion $\sigma_{3D}$ can be estimated as:
\begin{equation}
\sigma_{3D} = \sqrt{3(\sigma_{Therm}^{2}+\sigma_{NT}^{2})}
\end{equation}
The mean values of $\sigma_{Therm}$, $\sigma_{NT}$ and $\sigma_{3D}$ in each clump are presented in Table 2. The mean thermal one dimensional velocity dispersion of in these clumps range from 0.15 to 0.25 km~s$^{-1}$, with an average value of 0.19$\pm$0.02 km~s$^{-1}$. The mean non-thermal one dimensional velocity dispersion of these clumps range from 0.1 to 0.79 km~s$^{-1}$, with an average value of 0.32$\pm$0.16 km~s$^{-1}$. The three-dimensional velocity dispersion $\sigma_{3D}$ ranges from 0.35 to 1.41 km~s$^{-1}$, with an averaged value of 0.66$\pm$0.24 km~s$^{-1}$. There are 44 clumps with $\sigma_{NT}$ larger than $\sigma_{Therm}$, and in the remaining clumps $\sigma_{NT}$ and $\sigma_{Therm}$ are comparable. The mean ratio of $\sigma_{NT}$ to $\sigma_{Therm}$ in these clumps is 1.65$\pm$0.76.

Star forming activities such as infall, outflow, and rotation could increase the non-thermal velocity dispersion. However, no significant star forming activities were found in those cold and low density Planck clumps. These clumps are more quiescent than the other typical star forming regions. Thus, the non-thermal motions in the
Planck cold clumps are mainly determined by turbulence and the non-thermal velocity dispersion can be used as a measurement of the turbulent strength.

\subsection{The properties of dense cores}
In spite of the diffuse aspects, there also exist dense parts in the clumps. The individual dense cores are identified within the 50\% contours of the column density distribution. Thirteen clumps are too diffuse to identify any dense core and fifteen clumps contain only one dense core. In the other clumps, more than one dense cores are identified. In total of 82 dense cores are identified. The dense cores are fitted with a two-dimensional gaussian function. The positions of each dense core are shown in column 2 of Table 3. The radii of the cores are defined as $R=\sqrt{(a\cdot b)}/2$, where a and b are the sizes of the minor and major axes, respectively. The systemic velocities of each dense core are obtained by averaging the pixel values within their radii in the first moment images. The deconvolved sizes, the radii and the systemic velocities are shown in column 2 to 4 in Table 4, respectively. The radii of the cores range from 0.07 to 0.55 pc, with an averaged value of 0.27$\pm$0.12 pc. The statistical results of column densities of H$_{2}$, excitation temperatures and velocity dispersions within the radii of cores are summarized in Table 3. The averaged column density of H$_{2}$ and excitation temperature of the dense cores are (3.3$\pm$2.1)$\times10^{21}$ cm$^{-2}$ and 12.5$\pm$3.5 K, respectively, which are slightly larger than the averaged values over the whole clumps. The average values of $\sigma_{Therm}$, $\sigma_{NT}$ and $\sigma_{3D}$ are 0.19$\pm$0.03, 0.33$\pm$0.14 and 0.67$\pm$0.22 km~s$^{-1}$, respectively, which are the same as the values averaged over the whole clumps.

The volume densities of each core are inferred as $n=N_{H_{2}}^{peak}/2R$, where $N_{H_{2}}^{peak}$ is the peak H$_{2}$ column density. The volume densities range from 0.9 to 5.6$\times10^{3}$ cm$^{3}$, with an average value of $(2.4\pm1.1)\times10^{3}$ cm$^{3}$. The LTE masses of the cores are estimated as $M_{LTE}=\frac{4}{3}\pi R^{3}\cdot n\cdot m_{H_{2}}\cdot\mu_{g}$, where $m_{H_{2}}$ is the mass of a hydrogen molecule and
$\mu_{g}$=1.36 is the mean atomic weight of the gas.. The LTE masses range from 0.3 to 270 M$_{\sun}$, with an average value of 38$_{-30}^{+5}$ M$_{\sun}$.

\section{Discussion}

\subsection{The probability distributions of the derived parameters in the GMC scale}
The lognormal behaviors of volume or column density in molecular clouds were frequently reported in recent observations \citep{rid06,fro07,goo09}, which are often interpreted as a consequence of supersonic turbulence in the observed clouds \citep{va94}. However clouds that have already
formed stars also exhibit power law like tails at large column densities besides of the lognormal like shape at low column densities \citep{ka09,fro10}. In simulations, the power law like tails are often attributed to the formation of local collapsing sites in turbulent flows \citep{kr11,ba11}. The developing of power-law tails at high densities is thus a consequence of the transition from more diffuse, turbulence-dominated clouds to denser, star-forming, collapsing clouds \citep{ba11}. Thus, the distributions of volume or column densities in clumps can be used as a indicator of their evolutionary states. Additionally, to investigate the distributions of the other parameters such as velocity dispersion and excitation temperature can help understand the formation of the column or volume density distributions. For example, if the non-thermal velocity dispersion rather than the thermal velocity dispersion (or excitation temperature) has the similar distribution as the column density, we can argue that the non-thermal motions (turbulence) play a more important role in the formation of density distribution as well as the cloud structure.

We investigate the probability distributions of the derived parameters in the GMC scale. The mean parameters used to generate the cumulative distributions in Figure 5 are derived by averaging all the pixel values within the 30\% contours of the column density image in each clump. The Kolmogorov-Smirnov (K-S) test is applied to check whether the distributions of the parameters follow a normal or lognormal distribution. The motivation of modeling the distributions with a normal shape is to test whether the variables are randomly changed. Moreover a variable having a log-normal distribution can be thought of as the multiplicative product of many independent random variables. Thus it is worth modeling the distributions of the parameters with a normal distribution for comparison. The null hypothesis is the distribution of the derived parameter can be described with normal or lognormal distribution. The decision to reject or accept the null hypothesis is based on comparing the P-value with the desired significance level, which is 0.05 in this paper. Thus if the P-values from K-S test is larger than 0.05, the derived parameter should follow the reference distribution.

As shown in Figure 5, the distributions of the derived parameters (N$_{H_{2}}$, apflux857, T$_{ex}$, $\sigma_{Therm}$, $\sigma_{NT}$, $\sigma_{3D}$) of the whole region including all the mapped clumps can all be fitted with a lognormal distribution. Besides $\sigma_{Therm}$, the other five parameters also follow a normal distribution. However, the P-values of K-S test for normal distribution hypothesis are much smaller than that for lognormal distribution hypothesis, indicating the underlaying distributions of these parameters more likely have a lognormal shape. The P-values for lognormal distribution hypothesis of N$_{H_{2}}$ and apflux857 are as high as $\sim0.9$, indicating perfect lognormal distributions. In previous works, the lognormal behaviors of column density distribution in clumps without star formation was interpreted to be determined by turbulent motions. We also noticed that the P-value for lognormal distribution hypothesis of $\sigma_{NT}$ is much larger than that of  $\sigma_{Therm}$, also suggesting turbulence dominates the density distribution in a sense.

We also investigate the parameter distributions of the dense cores in this region in GMC scale. As shown in Figure 6, the the distributions of the derived parameters (N$_{H_{2}}$, n, T$_{ex}$, $\sigma_{Therm}$, $\sigma_{NT}$, $\sigma_{3D}$) of the cold clumps are also fitted with normal and lognormal distributions. Besides $\sigma_{Therm}$, the distributions of the other five parameters can be well described by a lognormal distribution with P-values larger than 0.3. Volume density n, $\sigma_{NT}$, and $\sigma_{3D}$ also can be fitted with a normal distribution with much lower P-values, indicating their distributions prefer lognormal shape rather than normal shape. The volume density has a remarkable lognormal distribution with P-value as large as 0.83. The column densities of these dense cores also can be fitted with a lognormal distribution (see panel (a) of Figure 6), but the P-value (0.44) is much lower than that of the whole clumps (0.88) (see panel (a) of Figure 5). The significant variance reflects that the dense cores are more evolved regions in the clumps, which begin to decouple from the general turbulent field and are more affected by gravity. $\sigma_{NT}$ distribution has a lognormal behaviors with a larger P-value of 0.89 than that of $\sigma_{Therm}$ (0.03), indicating turbulent motions dominate shaping the clump structures and inducing density fluctuations in GMC scale.

\subsection{The probability distributions of the derived parameters in the clump and core scale}
The distributions of the parameters in each clump are studied separately. For each clump, the distributions of the pixel values of the derived parameters (N$_{H_{2}}$, T$_{ex}$, $\sigma_{Therm}$, $\sigma_{NT}$, $\sigma_{3D}$) were investigated. The distributions are tested for normal and lognormal distribution hypothesis with K-S test. The P-values from K-S test are summarized in Table 2. The cumulative distribution of the P-values for normal and lognormal distribution hypothesis are shown in panel (a) and panel (b) of Figure 7, respectively. It can be seen that the distributions of the five parameters in more than a half of the clumps can be fitted with a normal distribution with P-value larger than 0.05. And 31 clumps have normal-like column density distributions with P-value larger than 0.05. In contrast, as shown in panel (b) of Figure 7, only 19 clumps show lognormal-like column density distributions. In panel (c) of Figure 7, we plot the distributions of the ratios P(lognormal)/P(normal) of P-values for lognormal hypothesis to P-values for normal hypothesis. We can see the column density distributions in more than 65\% clumps have P-values for normal distribution larger than that for lognormal distribution. And more than 85\% clumps have $\sigma_{NT}$ distribution with P-values for normal distribution larger than that for lognormal distribution. Except for $\sigma_{3D}$, the other four parameters in these clumps are more likely to be fitted with a normal distribution rather than lognormal distribution. Those parameters seem to prefer a normal distribution in small scale but favor a lognormal distribution in large scale (see section 4.1). It seems the local values of these parameters (N$_{H_{2}}$, T$_{ex}$, $\sigma_{Therm}$, $\sigma_{NT}$) are essentially random, leading to a normal distribution. If the fluctuations of the parameters in small scale are independent of one another, the distributions of the average parameters of each clump are thus lognormally distributed \citep{va94}. In panel (d) of Figure 7, the bin-averaged p-values are plotted versus bin-averaged column densities. The width of the bins is varied to guarantee that the numbers of clumps in each bin are similar. Except $\sigma_{3D}$, the bin-averaged P-values for normal hypothesis are larger than that for lognormal hypothesis in each N$_{H_{2}}$ bin, which is revealed by that the solid lines in panel (d) are always above the dashed lines. We also noticed the bin-averaged P-values of N$_{H_{2}}$, T$_{ex}$, $\sigma_{Therm}$, and $\sigma_{NT}$ for normal distribution hypothesis roughly decrease with column density.

We also investigate the distributions of the pixel values of the derived parameters (N$_{H_{2}}$, T$_{ex}$, $\sigma_{Therm}$, $\sigma_{NT}$, $\sigma_{3D}$) in each dense cores. The distributions are also tested for normal and lognormal distribution hypothesis with K-S test and the P-values from K-S test are summarized in Table 3. The cumulative distribution of the P-values for normal and lognormal distribution hypothesis are shown in panel (a) and panel (b) of Figure 8, respectively. In panel (a), we can see the cumulative density distributions almost have a linear shape. And more than 90\% dense cores show P-values larger than 0.05. In panel (b), we can see also in a large number ($>$60\%) of these cores the five parameters are lognormally distributed. However, as shown in panel (c), in about a half of these dense cores, the distributions of T$_{ex}$ and $\sigma_{Therm}$ prefer to follow a normal behavior. In more than 75\% of the dense cores, the $\sigma_{3D}$ favors to a normal distribution rather than lognormal distribution, that is P(lognormal)/P(normal)$<$1. And in more than 90\% of the dense cores, the $\sigma_{NT}$ and N$_{H_{2}}$ can be fitted better by normal distribution rather than by lognormal distribution. All the above indicates in small scale, the local values of these parameters are more likely normal distributed. In panel (d), we also investigated the bin-averaged P-values versus column density. It can be seen all the solid lines (normal distribution) locate above the dashed lines (lognormal distribution), which also reveals that these parameters favor normal distribution in small scale. All the P-values of these parameters seem roughly decrease with column density with the maximum value peaking at (2-2.5)$\times$10$^{21}$ cm$^{-1}$. This evolutionary behavior reflects that the structures of low column density cores are more affected by turbulence, but the structures of high column density cores are more affected by other factors, especially by gravity.

\subsection{Turbulence dominated clumps}
As discussed in section 4.1 and 4.2, the distributions of the parameters especially the density distribution have lognormal behaviors in large scale (GMC) and normal behaviors in small scale (clump \& dense cores), which indicate these clumps are turbulence dominated, in other word, the clump and core structures are more likely shaped by turbulent flows \citep{va94}.

The left panel of Figure 9 shows the relationship between H$_{2}$ column density and three dimensional velocity dispersion averaged over the whole clumps. It can be seen the velocity dispersion increases with the H$_{2}$ column density. The relationship can be better described as power law rather than linear relations. The linear fitting is $\frac{\sigma_{3D}}{(km~s^{-1})}=(0.11\pm0.01)\frac{N_{H_{2}}}{(10^{21} cm^{-2})}+(0.36\pm0.04)$, with R$^{2}$=0.64. While the power law fitting has the form as
$\frac{\sigma_{3D}}{(km~s^{-1})}=(0.41\pm0.02)\left[\frac{N_{H_{2}}}{(10^{21} cm^{-2})}\right]^{0.47\pm0.04}$, with R$^{2}$=0.74.

We also noticed the ratio of non-thermal to thermal velocity dispersion increases with the H$_{2}$ column density. As shown in the right panel of Figure 9, the relationship can be better fitted with power law rather than linear relation. The linear fitting is $\frac{\sigma_{NT}}{\sigma_{Therm}}=(0.25\pm0.05)\frac{N_{H_{2}}}{(10^{21} cm^{-2})}+(0.93\pm0.15)$, with R$^{2}$=0.37. While the power law fitting is as
$\frac{\sigma_{NT}}{\sigma_{Therm}}=(0.98\pm0.07)\left[\frac{N_{H_{2}}}{(10^{21} cm^{-2})}\right]^{0.49\pm0.07}$, with R$^{2}$=0.49.

The ratio of non-thermal to thermal pressure in one clump can be estimated as $R_{p}=\frac{\sigma_{NT}^{2}}{\sigma_{Therm}^{2}}$. As discussed in section 3.2.3, most of the clumps has $\sigma_{NT}$ larger than or comparable with $\sigma_{Therm}$. Thus $R_{p}$ is larger than unit in most of the clumps, suggesting the clumps associated with Planck cold clumps are mostly non-thermally dominated. Since the ratio of non-thermal to thermal velocity dispersion increases with the H$_{2}$ column density, the non-thermal pressure becomes more dominated in dense clumps, i.e., the denser clumps are more turbulent. However, once gravity dominates pressure, the gas can collapse until the densest parts become optically thick, allowing the gas to heat up adiabatically and to increase the gas pressure dramatically while the supersonic turbulence may decay quickly on a dynamical timescale \citep{shu87,zin07}. The low column densities and excitation temperatures as well as the large ratio of non-thermal to thermal pressure indicate those clumps are still turbulence dominated and not greatly affected by gravity. In other words, most parts of these clumps are still quiescent and have not suffered gravitational contracting.

\subsection{Correlations between dust and gas emission}
Dust and molecular gas are considered to coexist in molecular clumps. In sub-mm band, dust emission is always optically thin and can be treated as a good tracer of column density. In Figure 10, we plot the H$_{2}$ column densities of the clumps obtained from $^{13}$CO data as function of the aperture flux at 857 GHz. The column density increases with the flux at 857 GHz. The linear fitting is as $\frac{N_{H_{2}}}{(10^{21} cm^{-2})}=0.01\frac{apflux857}{(Jy)}+(1.29\pm0.30)$, with R$^{2}$=0.47. While the power law fitting has the form of
$\frac{N_{H_{2}}}{(10^{21} cm^{-2})}=(0.13\pm0.06)\left[\frac{apflux857}{(Jy)}\right]^{0.61\pm0.09}$, with R$^{2}$=0.47. As discussed in section 4.1, the column density and the flux at 857 GHz of the clumps are both following a lognormal distribution with a large p-values. All above indicate the dust and gas are well mixed in these clumps.

\subsection{Larson relationship}
The correlation of the velocity dispersion versus region size has a power law form and the so called Larson relationship was found to be held for size from 0.1 to 100 pc \citep{lar81}. However, in several other surveys Larson relationship is found to be not valid or very weak \citep{bu09,kra96,on02}. The plot of the three dimensional velocity dispersion against radius of the dense cores is shown in Figure 11. The correlation can be represented as $\frac{\sigma_{3D}}{(km~s^{-1})}=(0.92\pm0.08)\left[\frac{R}{(pc)}\right]^{0.31\pm0.07}$, with R$^{2}$=0.21. The correlation is very weak, and the power index found here is also smaller than that in \cite{lar81}, which is 0.38. The weak correlation could be due to the small range in R (0.08-0.65 pc) and $\sigma_{3D}$ (0.34-1.51 km~s$^{-1}$) and the large scatter in the values as well as the uncertainties of the distance. However, Larson relationship may be not valid in small scale. As discussed in section 4.1 and 4.2, the turbulence dominates the clump structure and density distribution in large scale but not in small scale. Small scale structures are more easily to be affected by the fluctuations of density and temperature.

\subsection{Gravitational stabilities of the dense cores}
Assuming the dense cores are gravitationally bound isothermal spheres
with a density profile of $\rho\propto R^{a}$, the
virial mass M$_{vir}$ can be calculated following \cite{ma88,wil94}:
\begin{equation}
M_{vir}=\frac{5R\sigma_{3D}}{3\gamma G}
\end{equation}
where G is the gravitational constant. Assuming the density profile is of  $\rho\propto R^{-2}$, $\gamma=5/3$.
The virial masses are listed in column 7 of Table 4.

In molecular clumps, many factors including thermal pressure, turbulence, and magnetic
field support the gas against gravity collapse. The Jeans mass, which takes into account of thermal and turbulent support, can be expressed as \citep{hen08}:
\begin{equation}
M_{J}\approx1.0a_{J}(\frac{T_{eff}}{10~K})^{3/2}(\frac{\mu}{2.33})^{-1/2}(\frac{n}{10^4~cm^{-3}})^{-1/2}M_{\sun}
\end{equation}
where $a_{J}$ is a dimensionless parameter of order unity which
takes into account the geometrical factor, $\mu=2.72$ is the mean molecular weight, n is the volume density of H$_{2}$
and $T_{eff}=\frac{C_{s,eff}^{2} \mu m_{H}}{k}$ is the effective kinematic temperature. The effective sound speed $C_{s,eff}$ including turbulent support
can be calculated as:
\begin{equation}
C_{s,eff}=\left[(\sigma_{NT})^{2}+(\sigma_{Therm})^{2}\right]^{1/2}
\end{equation}
The calculated Jeans masses are listed in the 8th column of Table 4.

About 64 (78\%) dense cores have virial masses larger than LTE masses. And only 10 (12\%) dense cores have virial masses larger than three times of the LTE masses. The others (88\%) have virial masses consistent with LTE masses within a factor of three. The average ratio of virial mass to LTE mass is 1.36. About 45\% of the dense cores have Jeans masses larger than LTE masses. More than 84\% cores have Jeans masses consistent with LTE mass within a factor a three. The average ratio of Jeans mass to LTE mass is 1.89. There are 83\% dense cores have both virial and Jeans masses consistent with LTE masses within a factor of three. Considering the uncertainties in estimating the masses, it seems that most of the cores are gravitationally bounded, and the internal thermal and turbulent pressure can support them against gravitational collapse.

The left panel of Figure 12 shows the relationship between $M_{vir}$ and $M_{LTE}$. The relationship can be well fitted with a power law, $\frac{M_{vir}}{(M_{\sun})}=(4.96\pm0.49)\left[\frac{M_{LTE}}{(M_{\sun})}\right]^{0.61\pm0.03}$, with R$^{2}$=0.83. The power index obtained here is very close to the value obtained in Orion B (0.67) \citep{Ike09} and NGC 2071 (0.6) \citep{bu09}. The pressure-confined cores were thought to have $M_{vir}\propto M_{LTE}^{1/3}$ \citep{be92}. The power law index for the pressure-confined cores is significant smaller than that for the dense cores obtained here, indicating the dense cores are most likely not pressure confined, but gravitationally bounded \citep{Ike09}. In the right panel of Figure 12, we plot $M_{J}$ versus $M_{LTE}$, whose relationship can be also well fitted with a power law. The power law has a form of $\frac{M_{J}}{(M_{\sun})}=(5.40\pm0.80)\left[\frac{M_{LTE}}{(M_{\sun})}\right]^{0.43\pm0.05}$, with R$^{2}$=0.51.

As shown in the last column of Table 4, only 11 cores are found associated with IRAS point sources. However, these IRAS point sources are very weak with 100 $\micron$ flux ranging from 2 to 30 Jy. It seems star forming activities haven't taken in these cores, and most of them may be star-less.

\subsection{Core mass function}
The differential core mass function (CMF) is usually
found to follow a power-law spectrum like $\frac{dN}{dM}\sim
M^{-\alpha}$. However, the shape of the mass spectrum is far from
fixed and broken power law appearance is also suggested in some surveys \citep{bu09}.
In the left panel of Figure 13, we plot the core mass function of the dense cores. The core mass function can be well fitted with a power law
for $15<M_{LTE}/M_{\sun}<200$. The power law gives $\alpha$=1.32$\pm$0.08.  Compared with $\alpha$ (2.35) of the stellar initial mass function (IMF) \citep{sal55}, the power law shape of the CMF here is much more flatten. As shown in the right panel of Figure 13, the masses of the cores are also found lognormally distributed with a P-value from K-S test as large as 0.87. The mean ($\mu$) and standard deviation ($\sigma$) of the lognormal distribution are 2.9 and 1.3, corresponding to a mass of 18 and 4 M$_{\sun}$, respectively. As discussed in section 4.1, the non-thermal velocity distribution also prefers to follow a lognormal shape, while the thermal dispersion doesn't, indicating non-thermal motions (turbulence) dominate the evolution of clumps. We argue that the supersonic turbulence can create dense cores with a lognoraml distribution of densities and masses in the clumps, which is consistent with the case in simulations \citep{ve11}.

\section{Summary}
We performed a mapping survey in molecular lines of CO, $^{13}$CO and C$^{18}$O J=1-0 towards 51 Planck cold clumps projected on Orion complex. The main findings in this work are as below:

(1).The mean H$_{2}$ column densities of the Planck cold clumps range from 0.5 to 9.5$\times10^{21}$ cm$^{-2}$, with an average value of (2.9$\pm$1.9)$\times10^{21}$ cm$^{-2}$. While the mean excitation temperatures of these clumps range from 7.4 to 21.1 K, with an average value of 12.1$\pm$3.0 K. The H$_{2}$ column densities and excitation temperatures obtained here are comparable to those values obtained from dust emission in the C3PO samples. However, the Planck cold clumps have slightly larger excitation temperatures but much smaller column densities when comparing with the IRAS sources and IRDCs, indicating these Planck cold clumps represent an earlier evolutionary phase in star formation. The three-dimensional velocity dispersion $\sigma_{3D}$ in these molecular clumps ranges from 0.35 to 1.41 km~s$^{-1}$, with an averaged value of 0.66$\pm$0.24 km~s$^{-1}$. Most of the clumps have $\sigma_{NT}$ larger than or comparable with $\sigma_{Therm}$. The three dimensional velocity dispersion and the ratio of $\sigma_{NT}$ to $\sigma_{Therm}$ increase with the H$_{2}$ column density, indicating the clumps are mostly non-thermally dominated and the denser clumps are more turbulent.

(2).We identified 82 dense cores in the molecular clumps. The averaged column density of H$_{2}$ and excitation temperature of the dense cores are larger than that averaged over the whole clumps. The radii of the cores range from 0.08 to 0.65 pc, with an averaged value of 0.24$\pm$0.14 pc. Their LTE masses range from 0.3 to 270 M$_{\sun}$, with an average value of 38$_{-30}^{+5}$ M$_{\sun}$.

(4).We found that N$_{H_{2}}$ , apflux857, T$_{ex}$, $\sigma_{Therm}$, $\sigma_{NT}$, $\sigma_{3D}$ and volume density n have a lognormal distribution in large scale (in the whole GMC), but N$_{H_{2}}$, T$_{ex}$, $\sigma_{Therm}$, $\sigma_{NT}$ more likely favor normal distribution in small scale (in each clump). It seems turbulent flows can shape the clump structure and dominate their density distribution in large scale, but not function in small scale due to the local fluctuations. In each clump or dense cores, we also noticed that the distributions of N$_{H_{2}}$, T$_{ex}$, $\sigma_{Therm}$, $\sigma_{NT}$ deviate from normal or lognormal distributions at high column densities, indicating the structures of low column density cores are more affected by turbulence, but the structures of high column density cores are more affected by other factors, especially by gravity.

(5).The H$_{2}$ column density of the molecular clumps calculated from molecular lines correlate with the aperture flux at 857 GHz of the dust emission.

(6).The correlation of the velocity dispersion versus core size, .i.e., the Larson relationship is very weak for the dense cores. The power index of the relation is found to be $0.31\pm0.07$, smaller than the value obtained by \citep{lar81}. Larson relationship seems not valid in small scale, which also indicate that the turbulence dominates the clump structure and density distribution in large scale but not in small scale.

(7).The dense cores are found to be mostly gravitationally bounded through analyzing their virial masses and Jeans masses. The relationship between $M_{vir}$ and $M_{LTE}$ can be well fitted with a power law with a power index of $0.61\pm0.03$, similar to that in Orion B (0.67) and NGC 2071 (0.6), also indicating the dense cores are most likely gravitationally bounded rather than pressure confined. The relationship between $M_{J}$ and $M_{LTE}$ also can be well fitted with a power law with a power index of $0.43\pm0.05$. Only 11 cores are found associated with weak IRAS point sources. Most of the cores may be star-less and have not affected by star forming activities.

(8). The core mass function can be well fitted with a power law
for $15<M_{LTE}/M_{\sun}<200$, whose slope $\alpha$=1.32$\pm$0.08. The masses of the dense cores are also lognormally distributed. The lognormal behavior of the core mass distribution is determined by the internal turbulence, whose distribution also shows lognormal shape.

\section*{Acknowledgment}
\begin{acknowledgements}
We are grateful to the staff at the Qinghai Station of PMO for their assistance
during the observations. Thanks for the Key Laboratory for Radio Astronomy, CAS to partly support the telescope operating. This work was supported by China Ministry of Science and Technology under State
Key Development Program for Basic Research (2012CB821800) and grant No. 11073003.
\end{acknowledgements}

\begin{figure}
\begin{minipage}[c]{0.5\textwidth}
  \centering
  \includegraphics[width=80mm,height=60mm,angle=0]{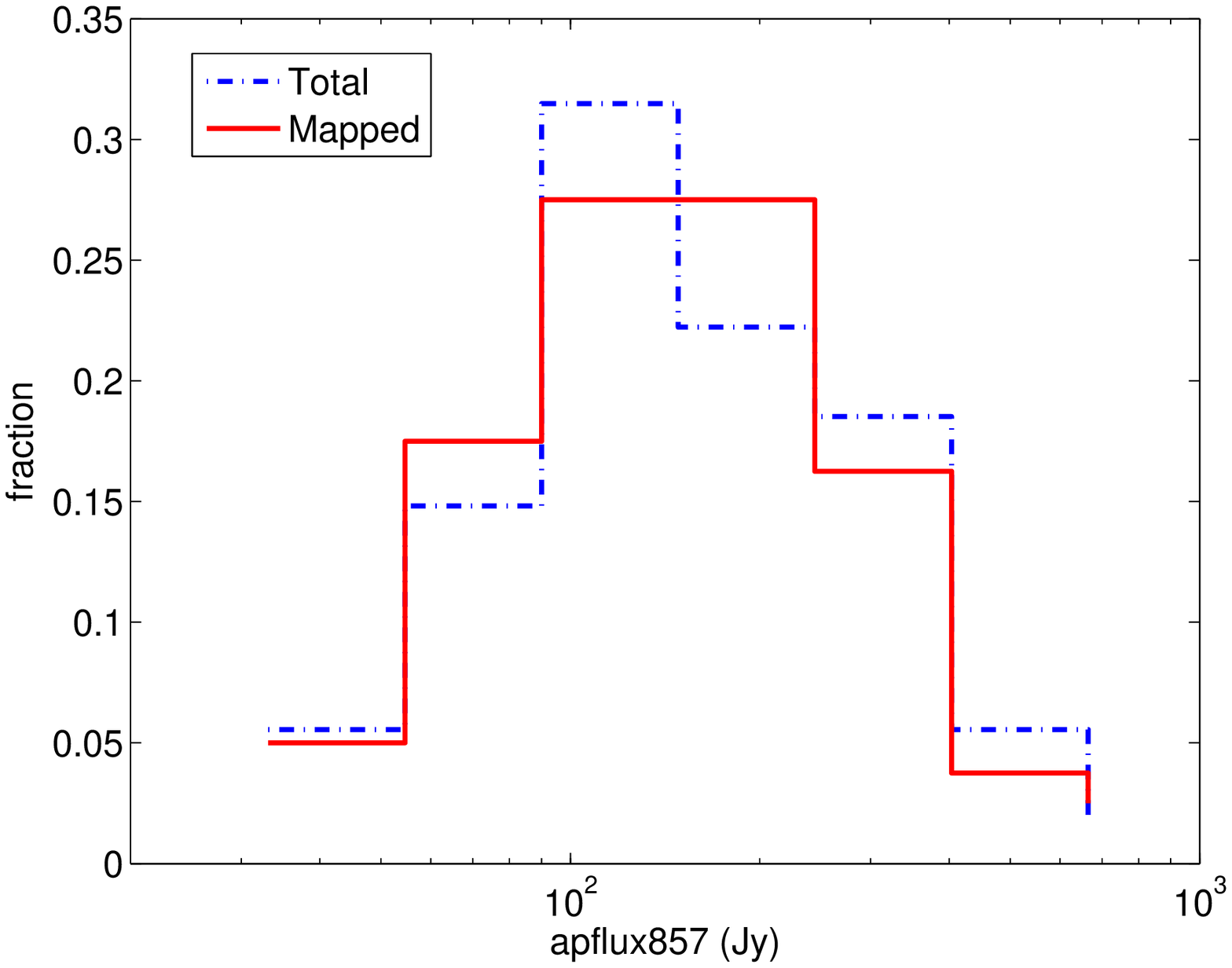}
\end{minipage}
\begin{minipage}[c]{0.5\textwidth}
  \centering
  \includegraphics[width=80mm,height=120mm,angle=0]{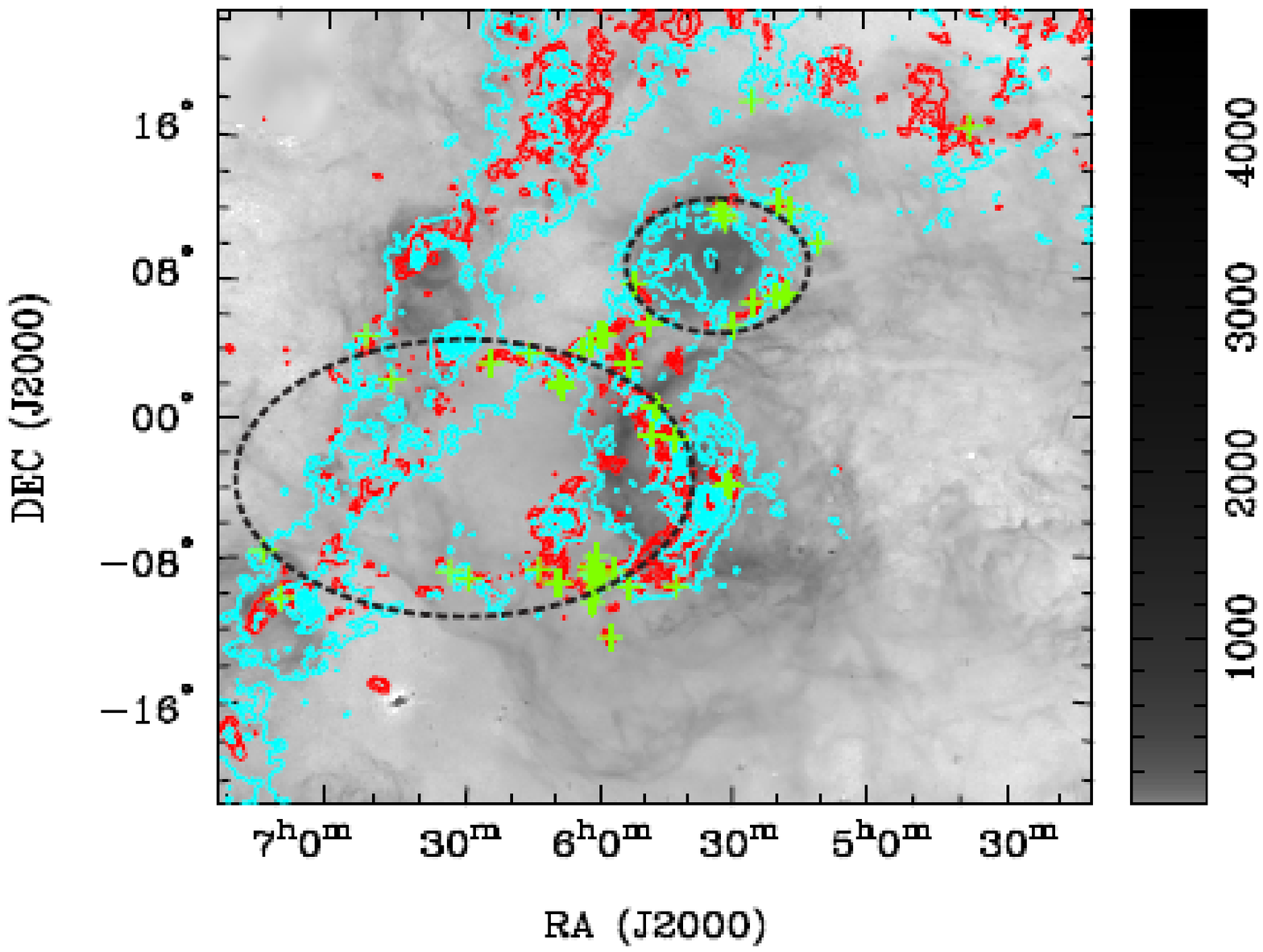}
\end{minipage}
\caption{Left: The distributions of the aperture flux density at 857 GHz (apflux857) for the mapped sources and for all the Planck cold clumps projected in Orion complex. Right: Distribution of the Planck cold clumps in Orion complex. Their locations are marked with green "crosses". The background image represents the H$_{\alpha}$ emission \citep{fin03}. The red contours present the CO (1-0) emission \citep{dame01}. The contour levels are (0.5,1,2,3,4,5,6,7,8,9)$\times$10 K~km~s$^{-1}$. The blue contours show the IRAS 100 $\micron$ emission. The contour levels are (0.5,1,2,3,4,5,6,7,8,9)$\times$50 mJy/sr.}
\end{figure}

\clearpage

\begin{figure}
\includegraphics[angle=0,scale=0.6]{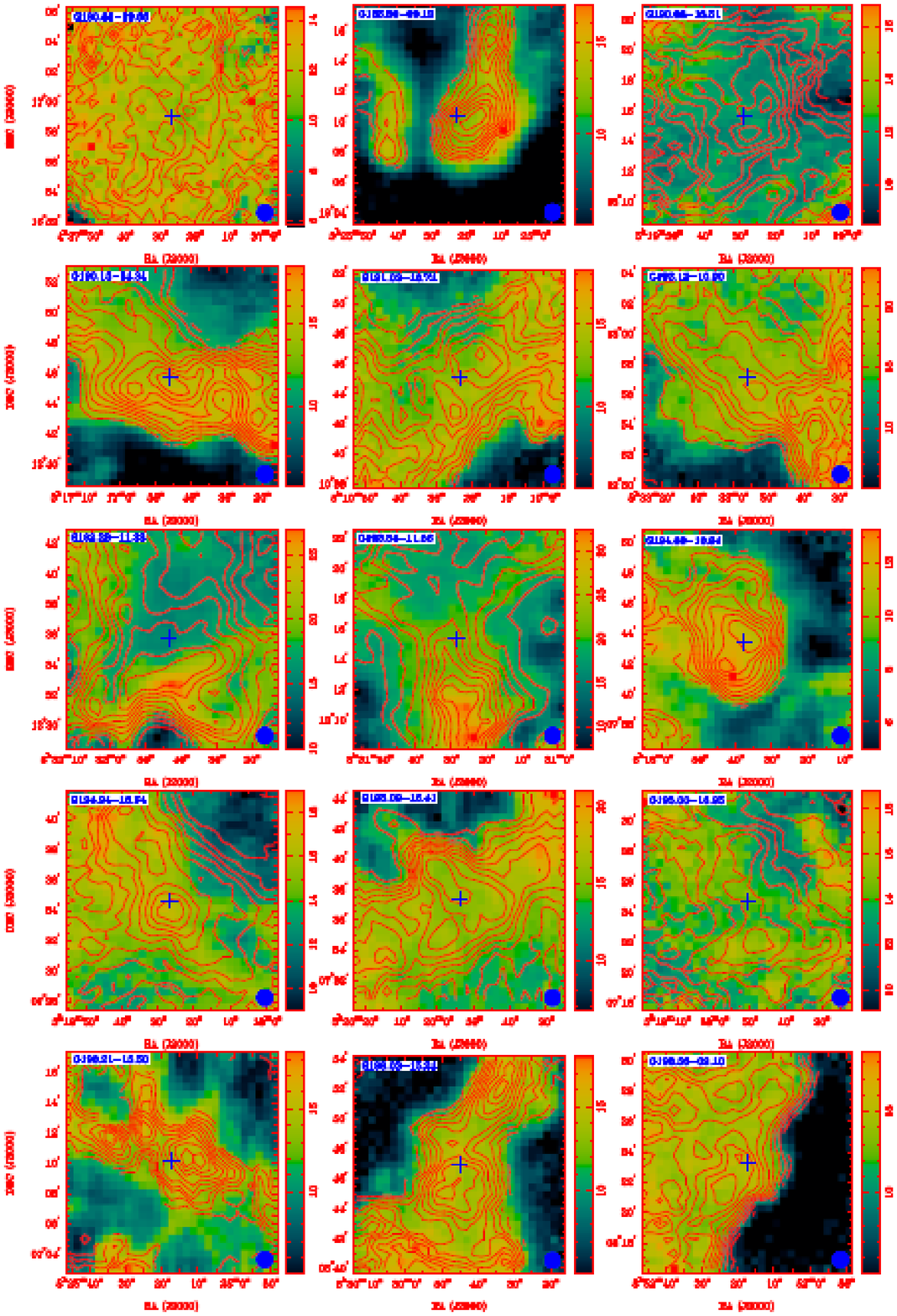}
\caption{The contours represent the column density distribution. The contour levels are from 10\% to 90\% in steps of 10\% of the peak value. The image in color scale shows the distribution of excitation temperature. The cloud names are labeled in the upper-left corner in each panel. (First page of Figure 2, the others are only available on line.)}
\end{figure}

\clearpage
\begin{figure}
\includegraphics[angle=0,scale=0.6]{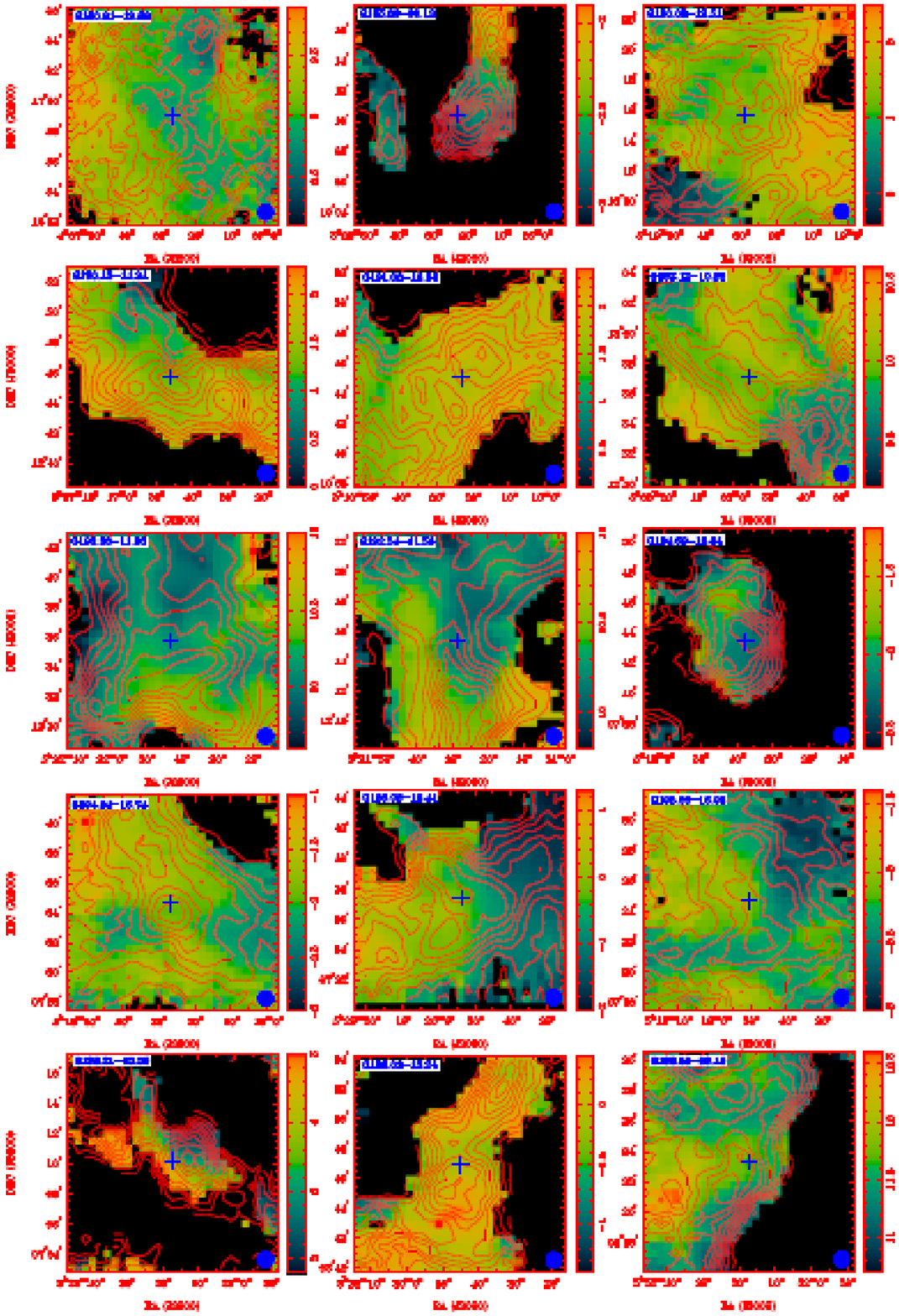}
\caption{The contours represent the column density distribution. The contour levels are from 10\% to 90\% in steps of 10\% of the peak value. The first momentum maps of $^{13}$CO (1-0) emission are shown in color scale. The cloud names are labeled in the upper-left corner in each panel. (First page of Figure 3, the others are only available on line.) }
\end{figure}

\clearpage
\begin{figure}
\includegraphics[angle=0,scale=0.6]{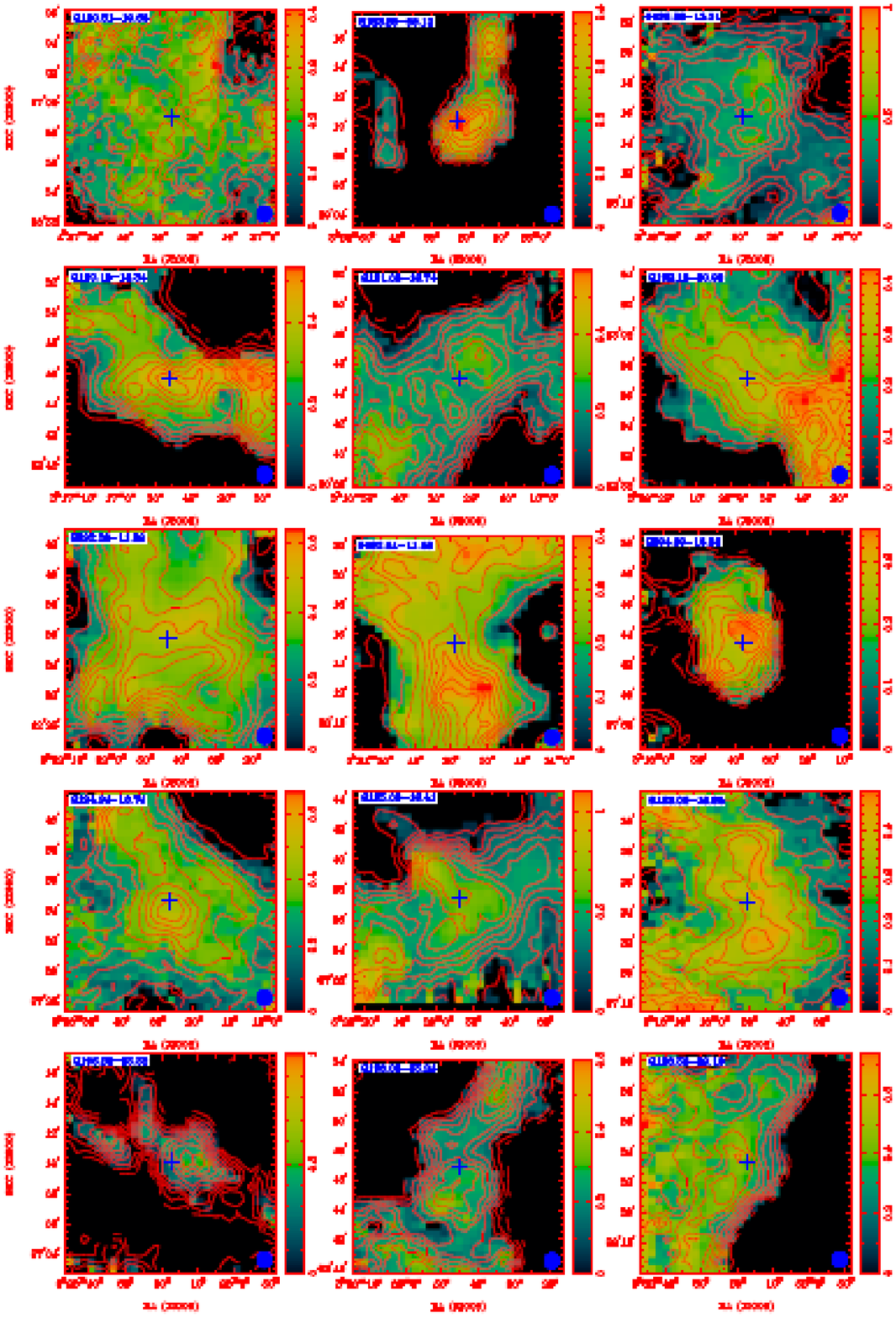}
\caption{The contours represent the column density distribution. The contour levels are from 10\% to 90\% in steps of 10\% of the peak value. The second momentum (velocity dispersion) maps of $^{13}$CO (1-0) emission are shown in color scale. The cloud names are labeled in the upper-left corner in each panel. (First page of Figure 4, the others are only available on line.)}
\end{figure}

\clearpage
\begin{figure}
\begin{minipage}[c]{0.5\textwidth}
  \centering
  \includegraphics[width=80mm,height=60mm,angle=0]{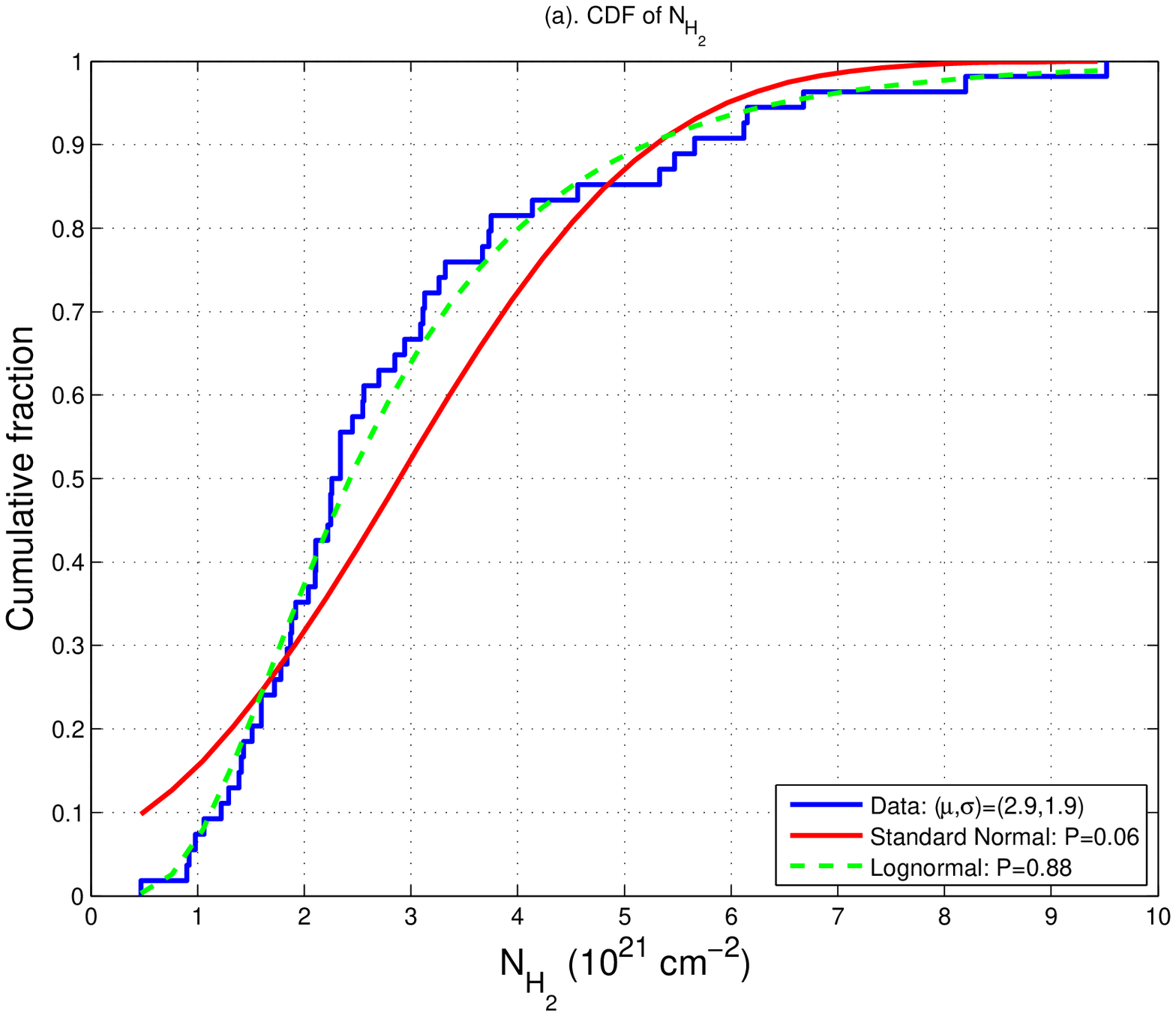}
\end{minipage}
\begin{minipage}[c]{0.5\textwidth}
  \centering
  \includegraphics[width=80mm,height=60mm,angle=0]{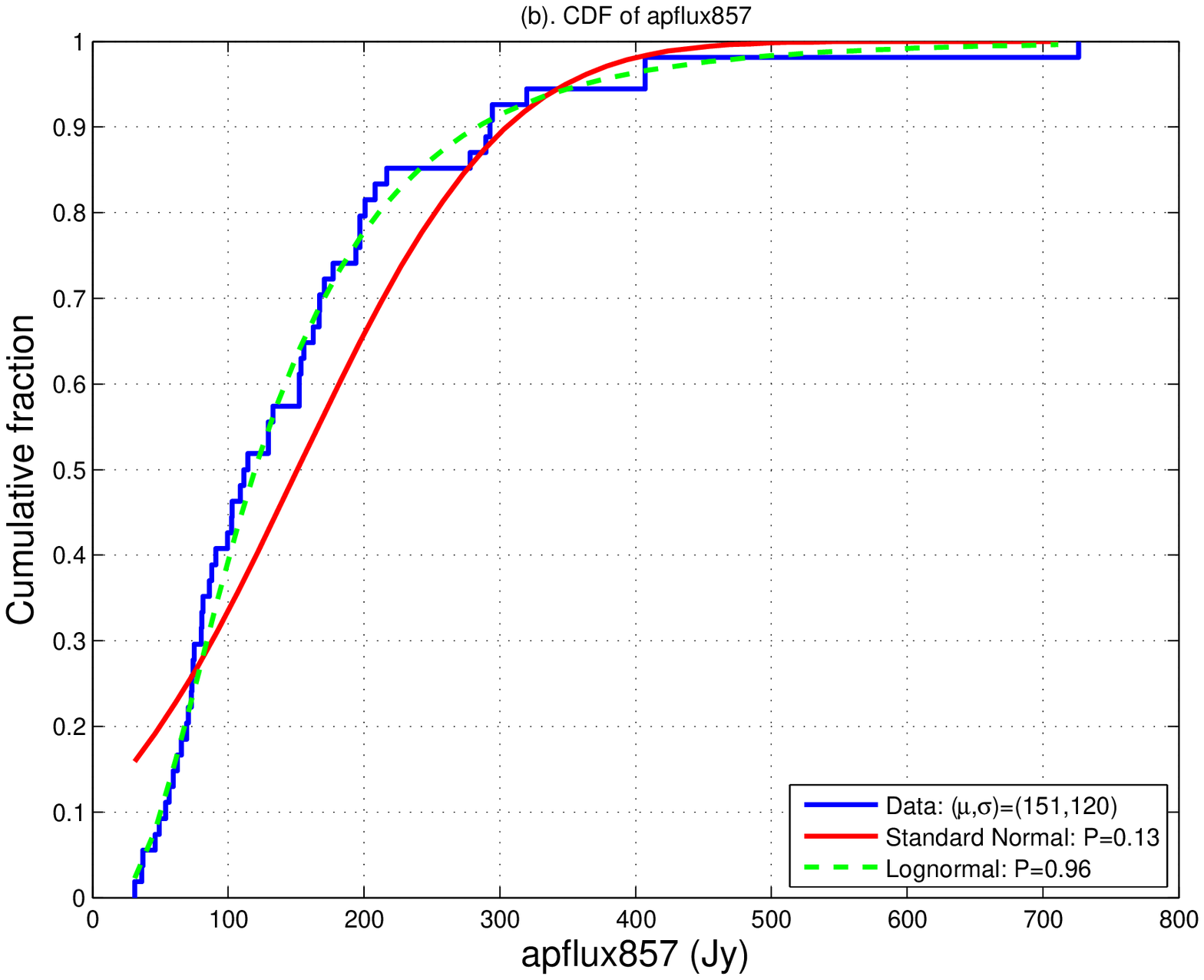}
\end{minipage}
\begin{minipage}[c]{0.5\textwidth}
  \centering
  \includegraphics[width=80mm,height=60mm,angle=0]{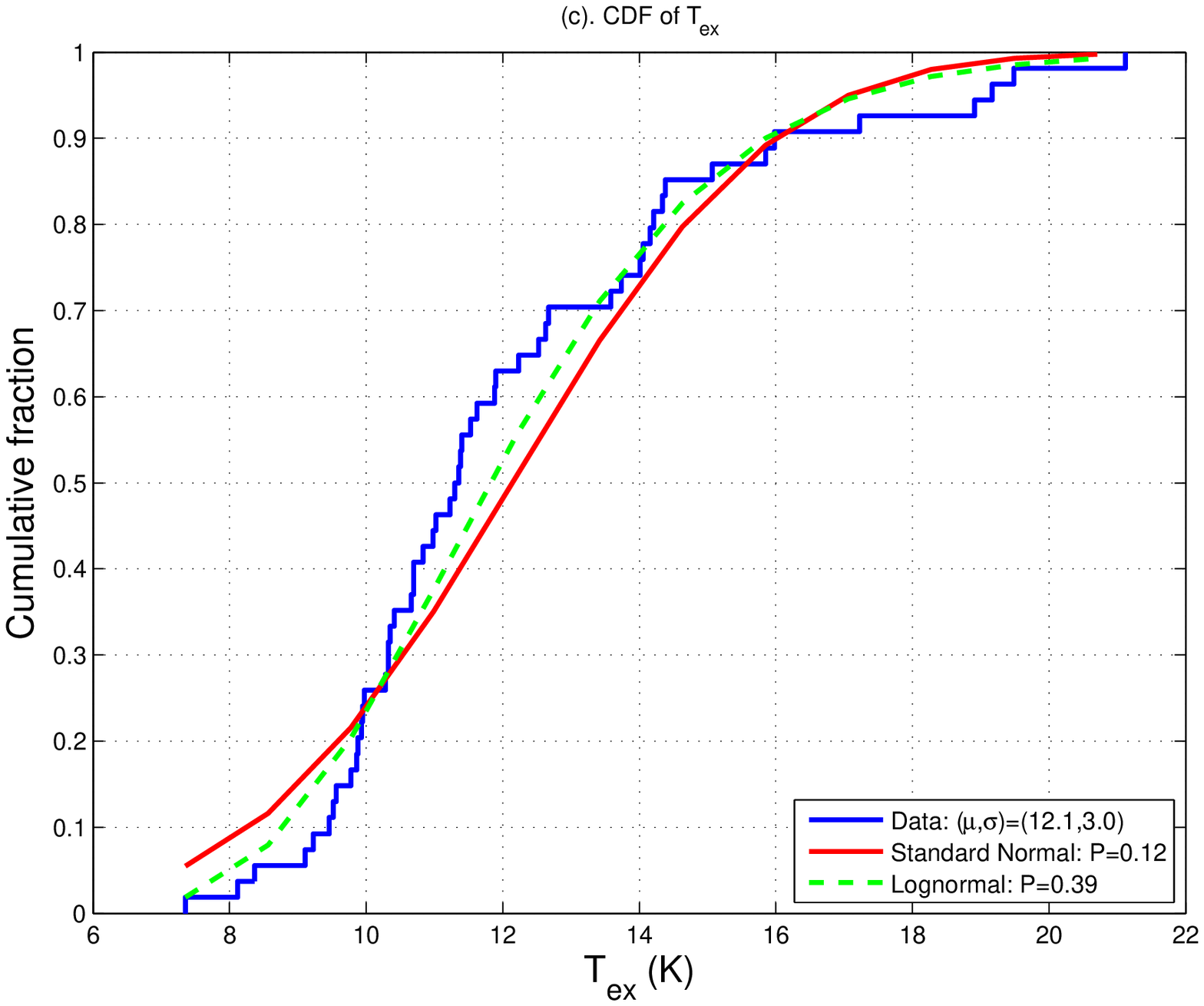}
\end{minipage}
\begin{minipage}[c]{0.5\textwidth}
  \centering
  \includegraphics[width=80mm,height=60mm,angle=0]{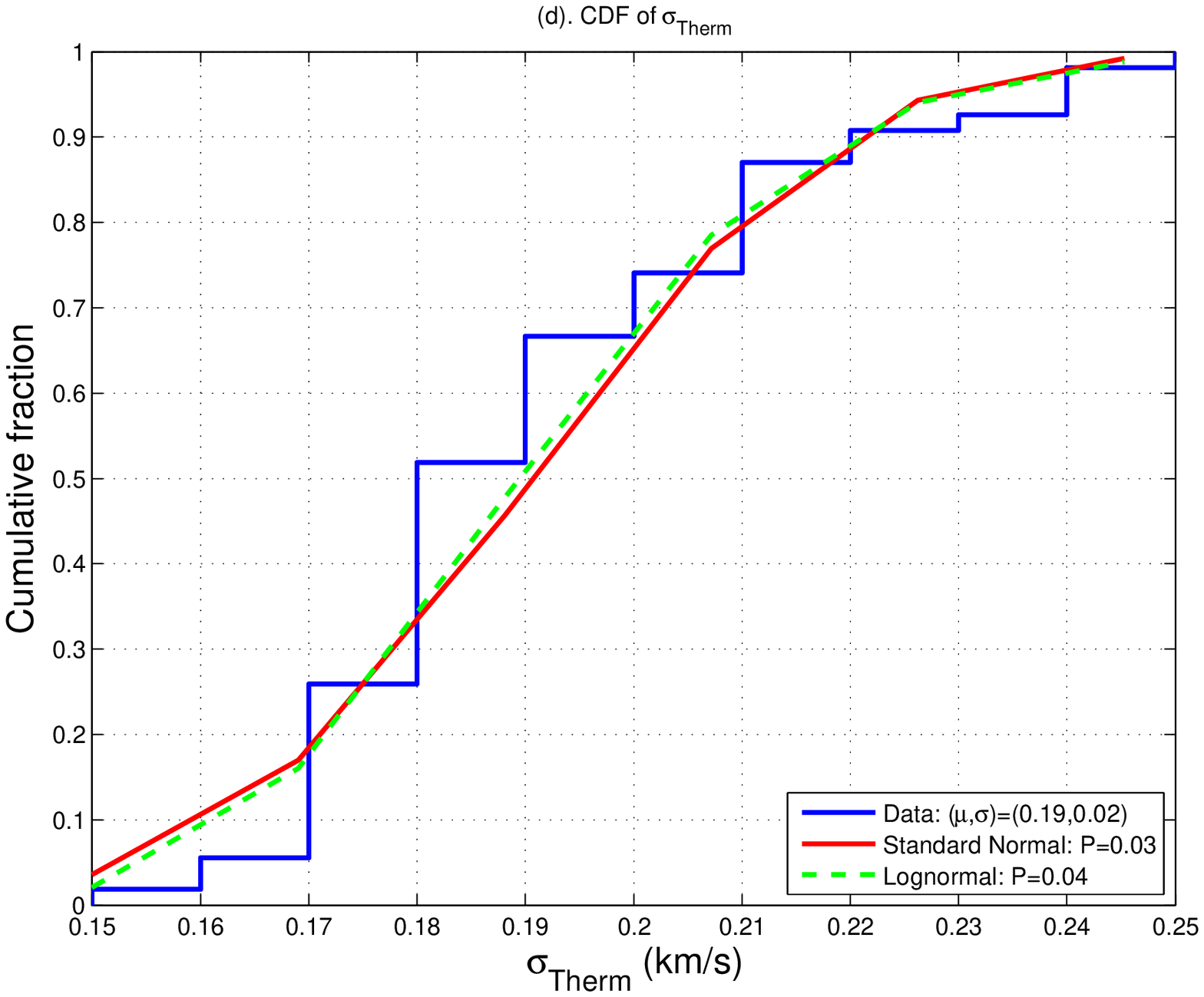}
\end{minipage}
\begin{minipage}[c]{0.5\textwidth}
  \centering
  \includegraphics[width=80mm,height=60mm,angle=0]{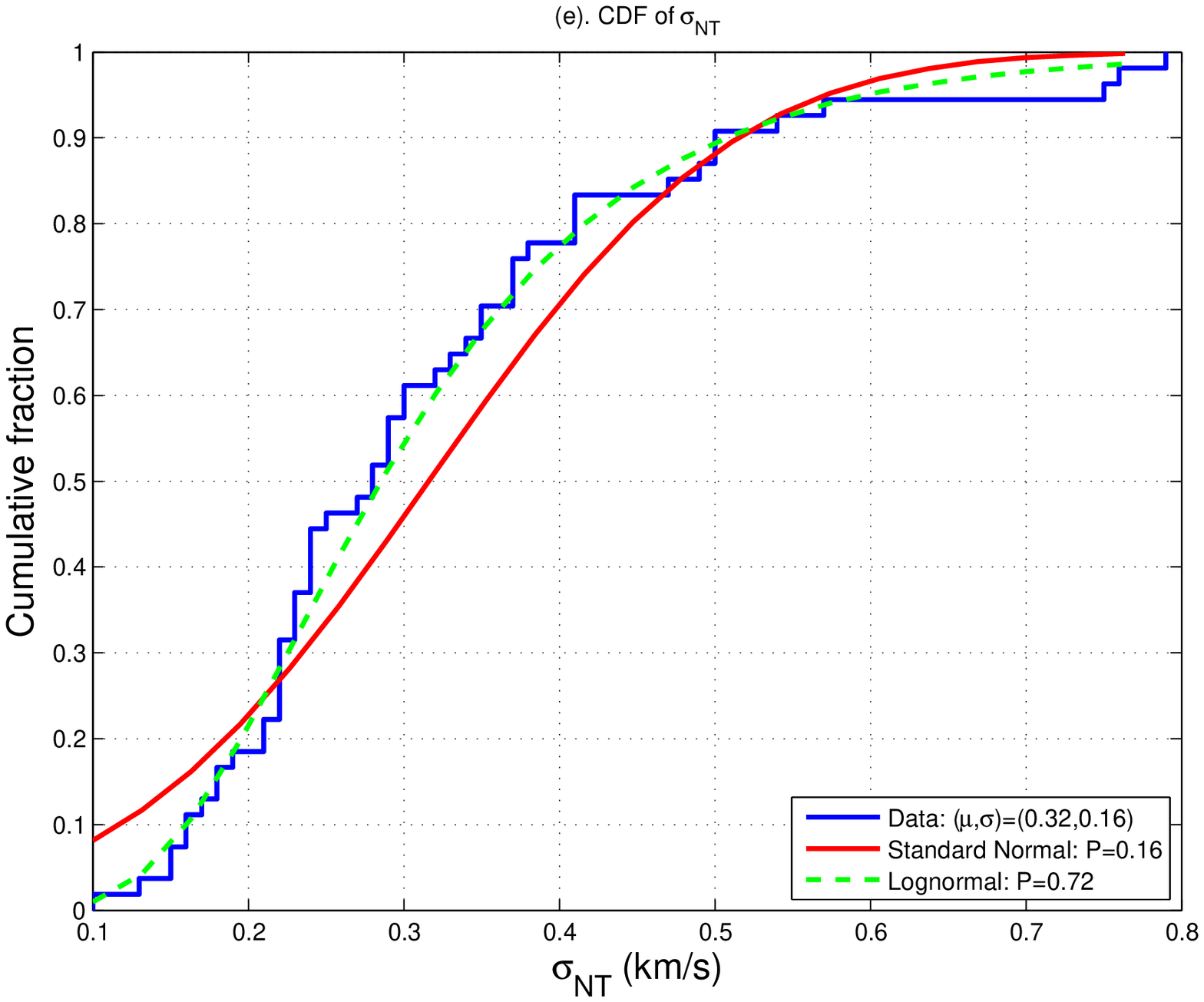}
\end{minipage}
\begin{minipage}[c]{0.5\textwidth}
  \centering
  \includegraphics[width=80mm,height=60mm,angle=0]{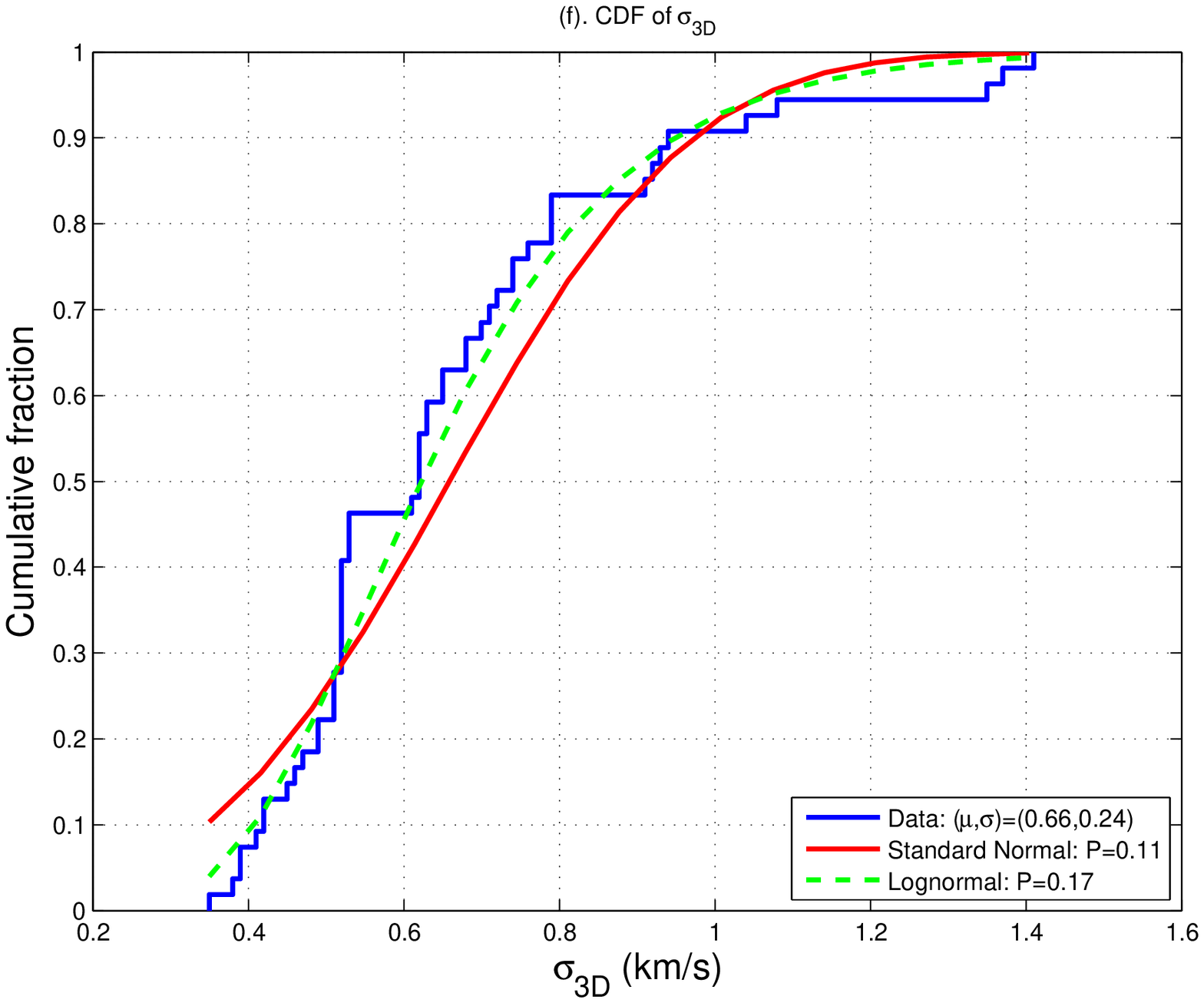}
\end{minipage}
\caption{Cumulative distributions of the derived parameters averaged over the whole clouds. The names of the parameters are labeled on the top of each panel. The blue curve is the data distribution. The red solid and green dashed lines represent the best normal and lognormal distribution fits, respectively. }
\end{figure}

\clearpage
\begin{figure}
\begin{minipage}[c]{0.5\textwidth}
  \centering
  \includegraphics[width=80mm,height=60mm,angle=0]{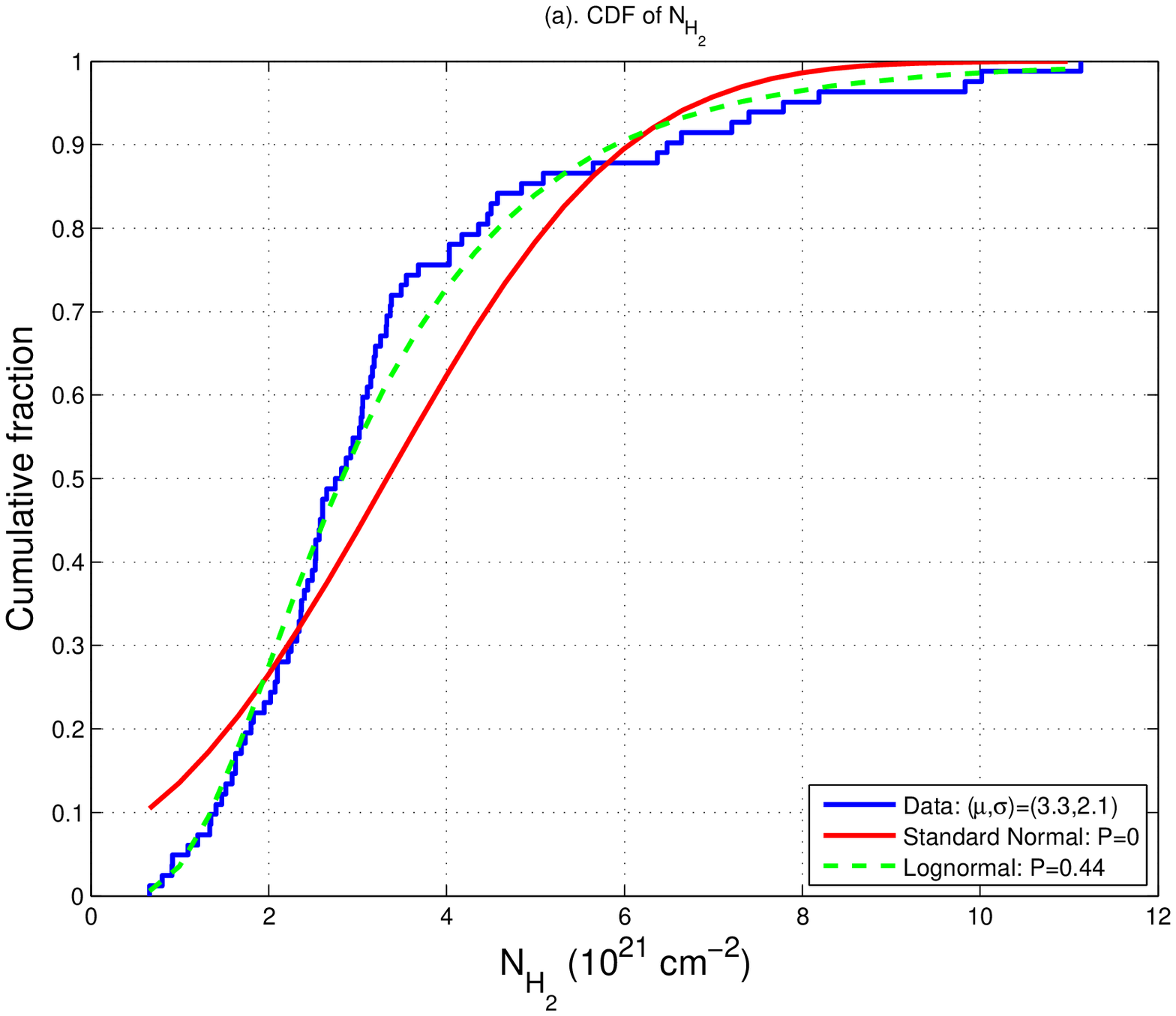}
\end{minipage}
\begin{minipage}[c]{0.5\textwidth}
  \centering
  \includegraphics[width=80mm,height=60mm,angle=0]{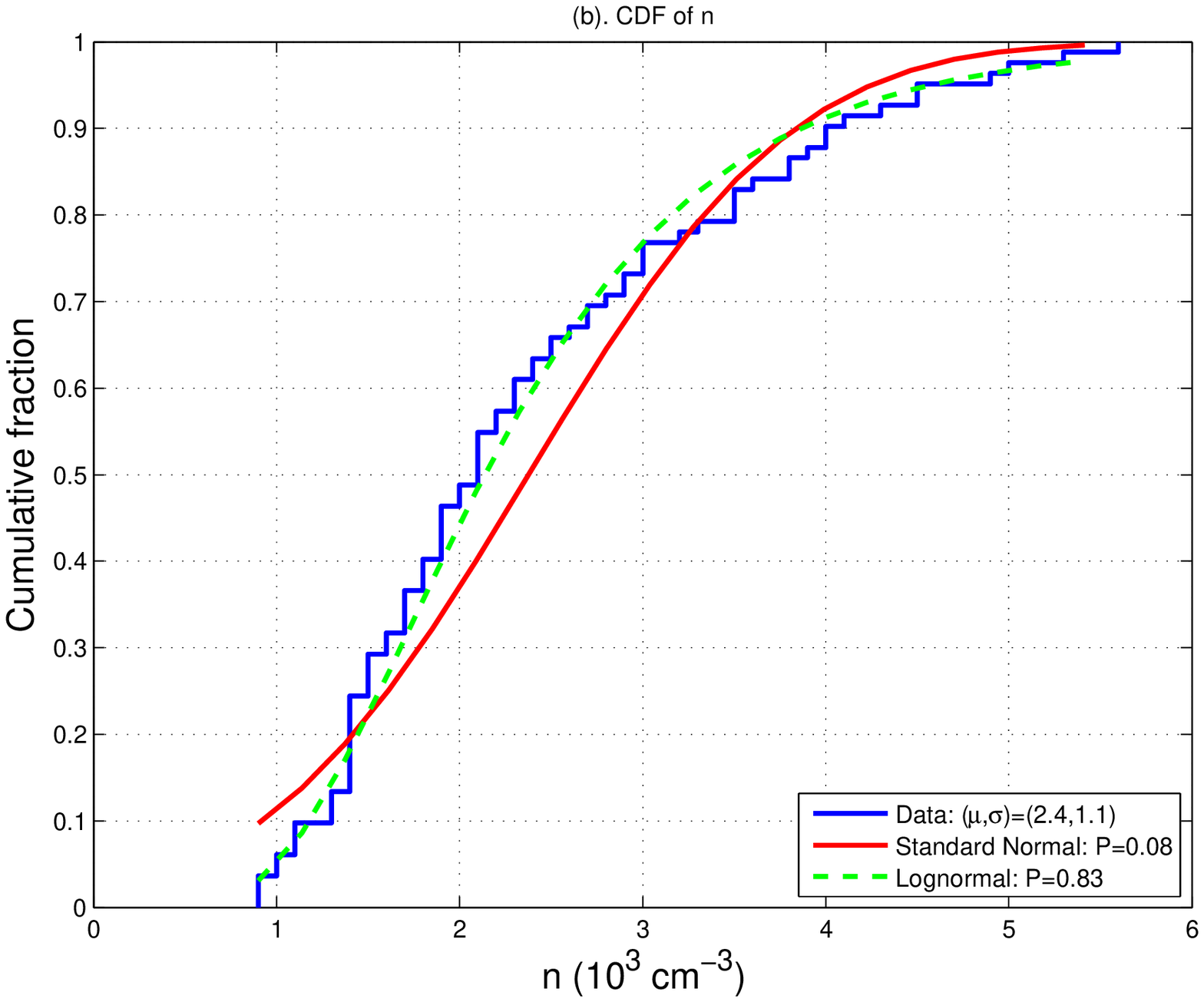}
\end{minipage}
\begin{minipage}[c]{0.5\textwidth}
  \centering
  \includegraphics[width=80mm,height=60mm,angle=0]{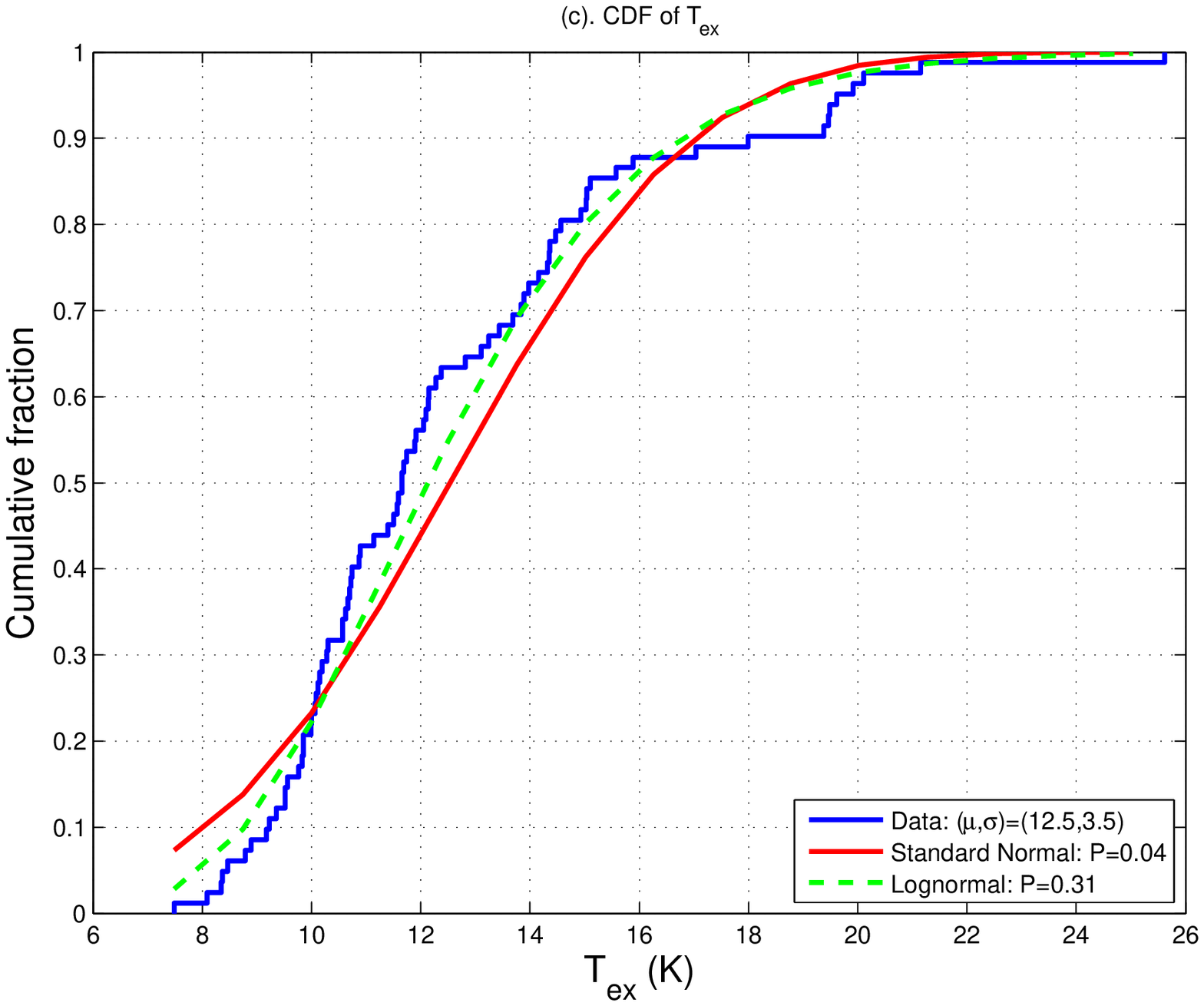}
\end{minipage}
\begin{minipage}[c]{0.5\textwidth}
  \centering
  \includegraphics[width=80mm,height=60mm,angle=0]{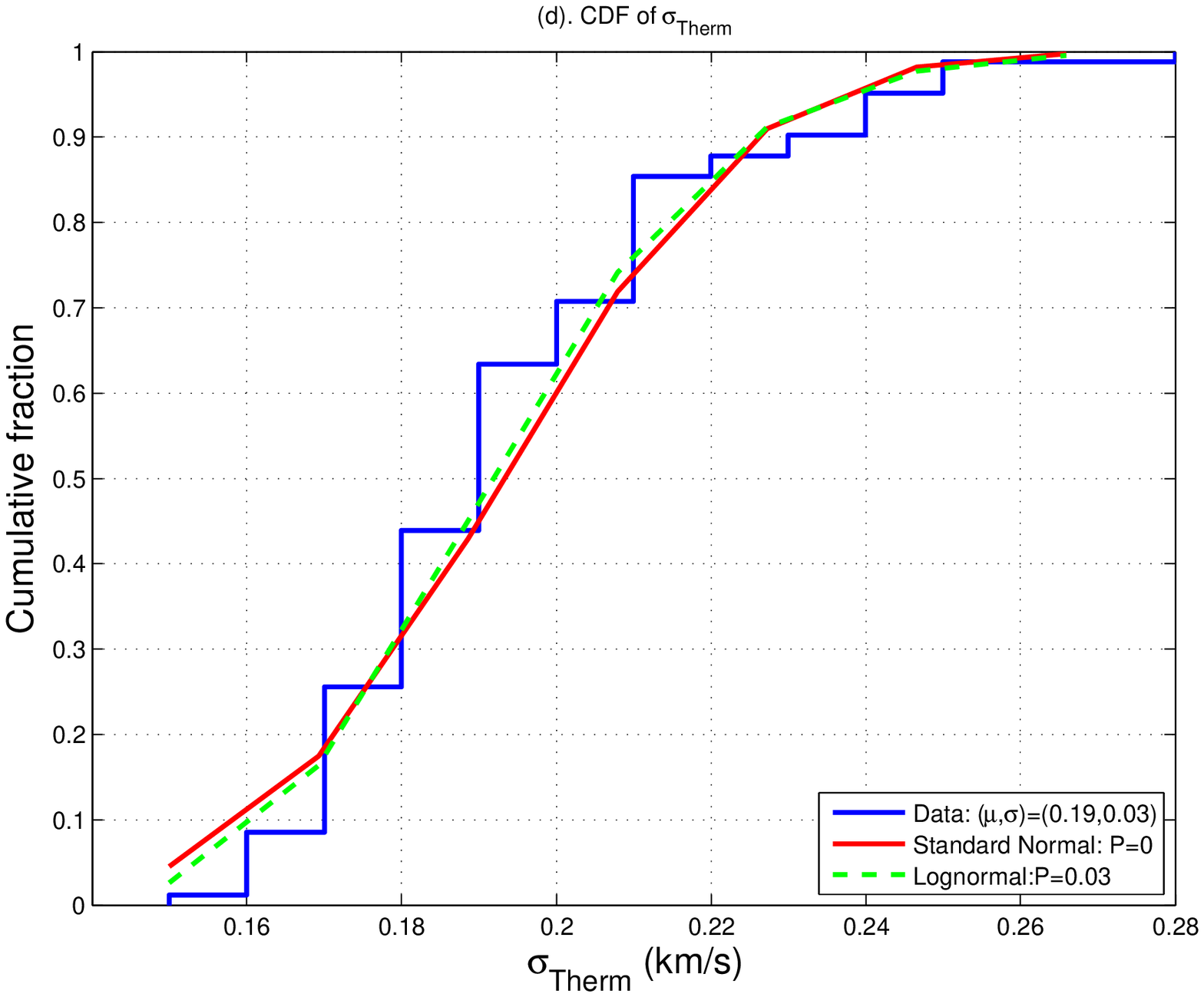}
\end{minipage}
\begin{minipage}[c]{0.5\textwidth}
  \centering
  \includegraphics[width=80mm,height=60mm,angle=0]{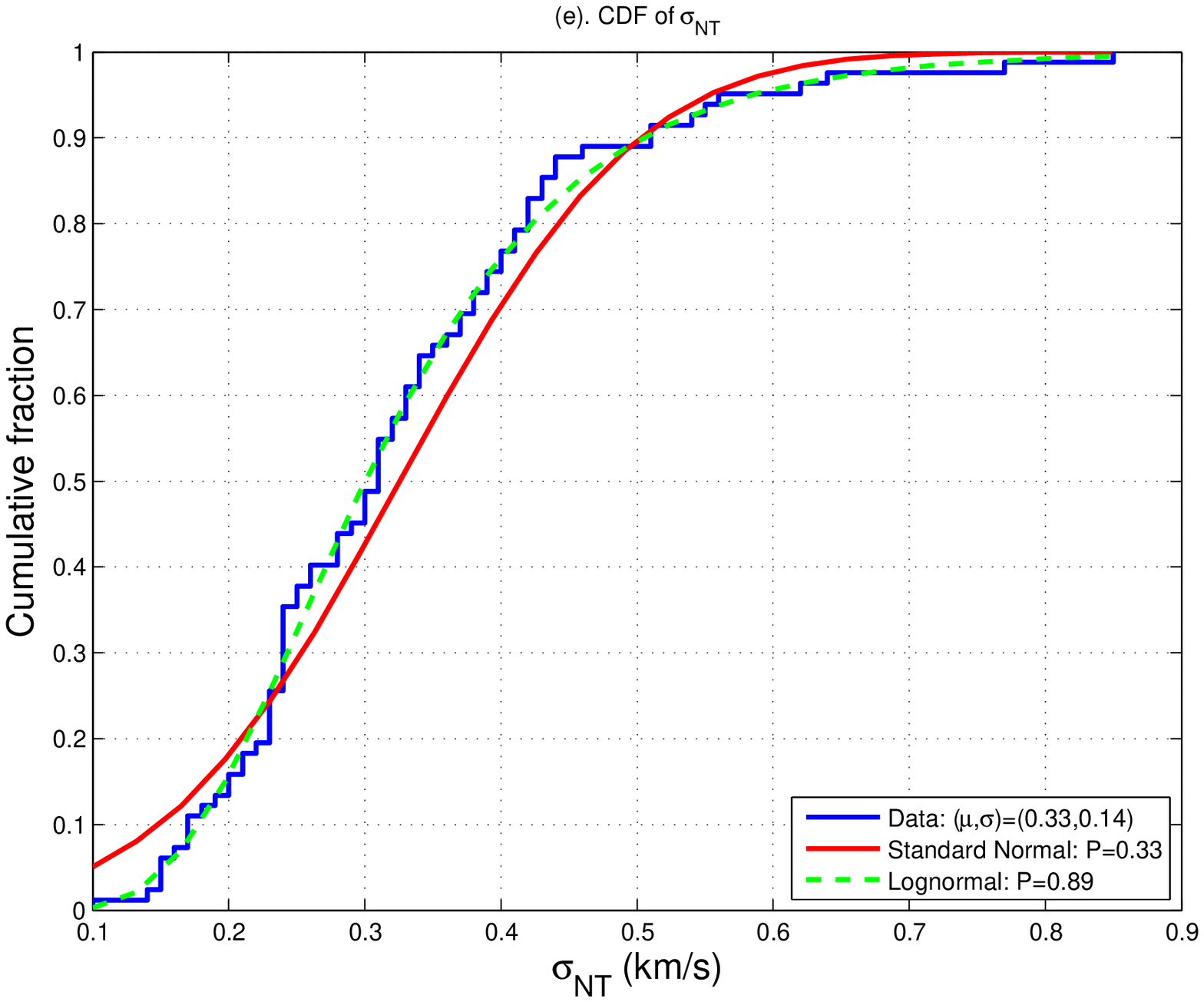}
\end{minipage}
\begin{minipage}[c]{0.5\textwidth}
  \centering
  \includegraphics[width=80mm,height=60mm,angle=0]{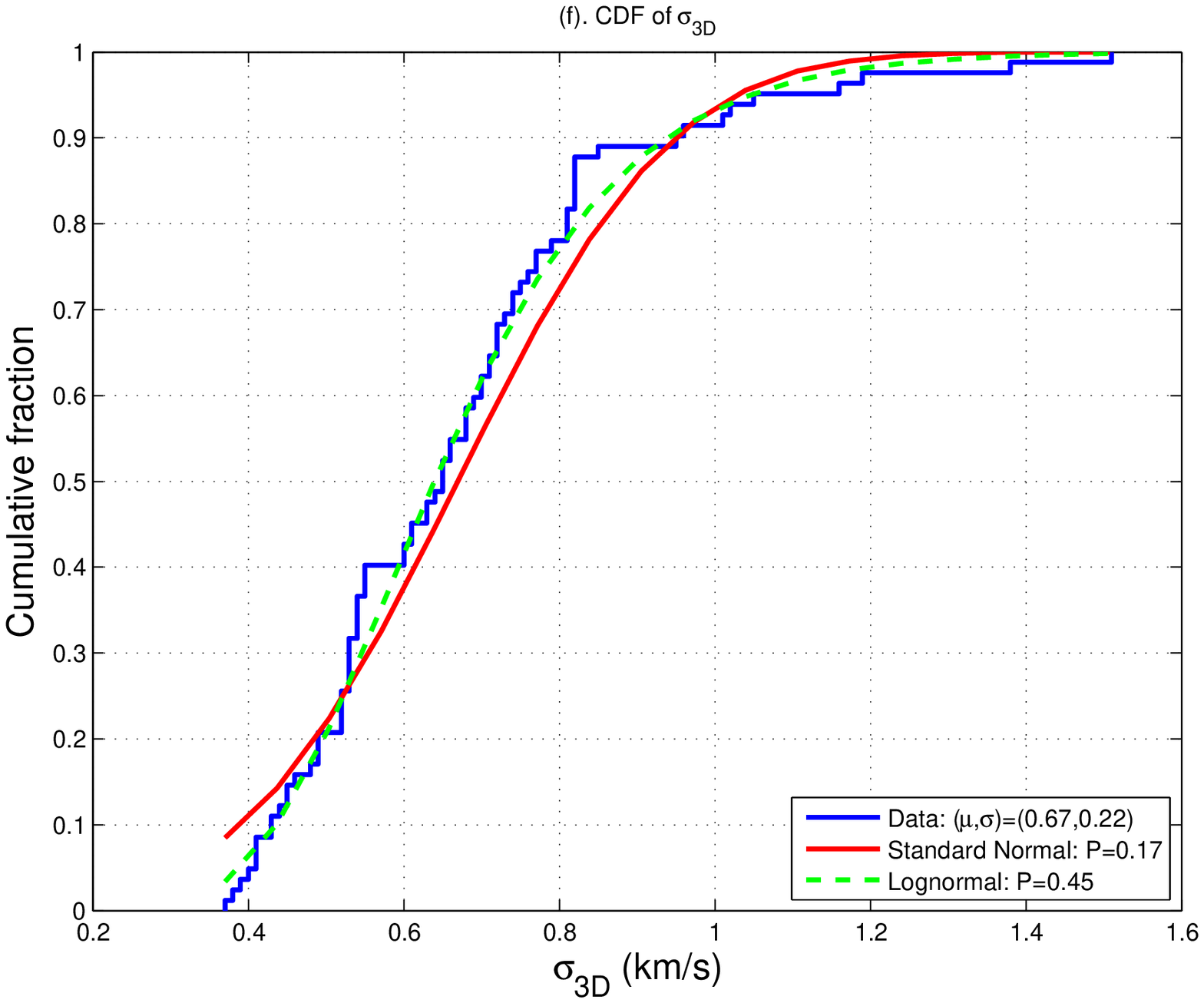}
\end{minipage}
\caption{Cumulative distributions of the derived parameters averaged over the dense cores. The names of the parameters are labeled on the top of each panel. The blue curve is the data distribution. The red solid and green dashed lines represent the best normal and lognormal distribution fits, respectively.}
\end{figure}

\clearpage

\begin{figure}
\begin{minipage}[c]{0.5\textwidth}
  \centering
  \includegraphics[width=90mm,height=70mm,angle=0]{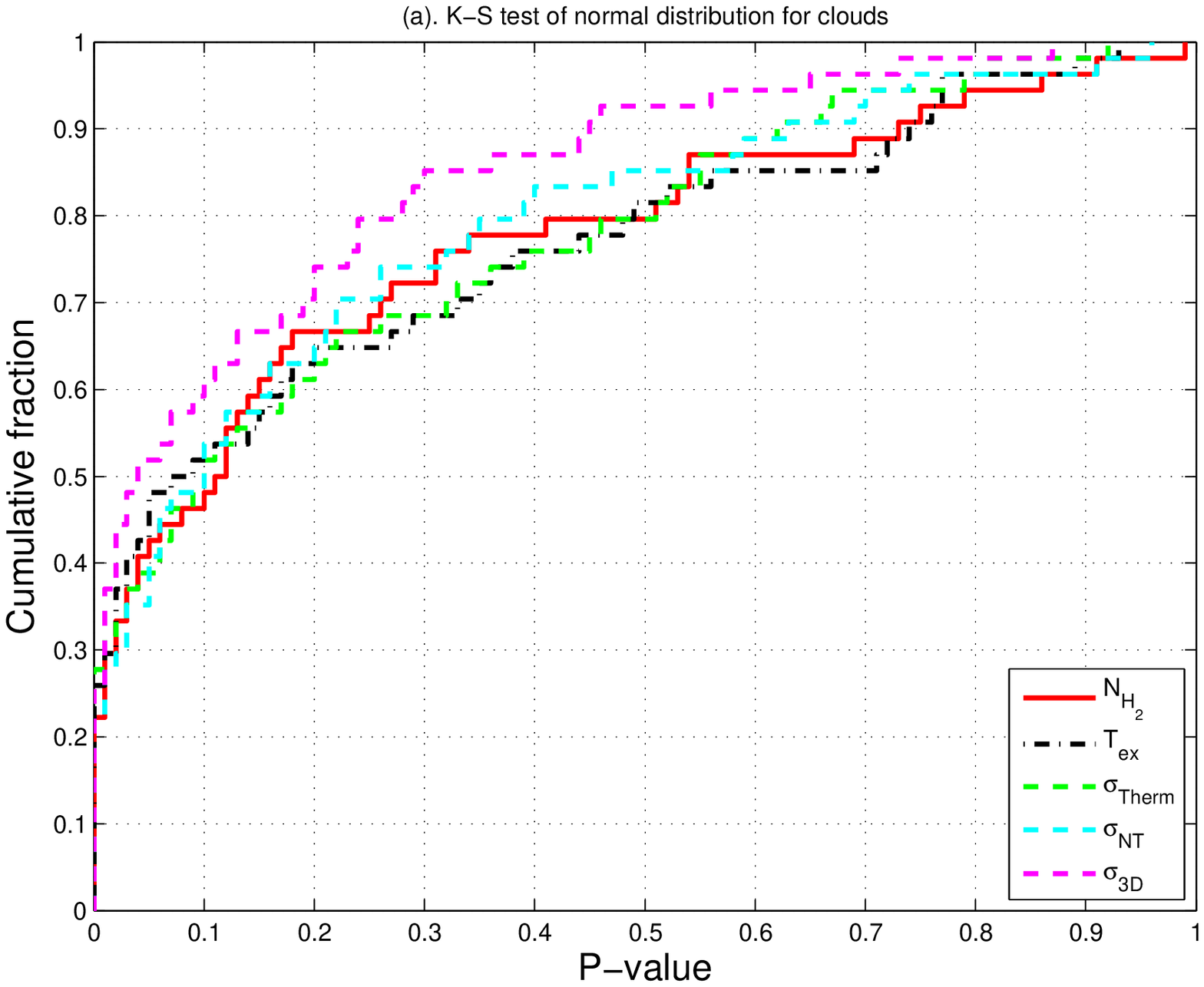}
\end{minipage}
\begin{minipage}[c]{0.5\textwidth}
  \centering
  \includegraphics[width=90mm,height=70mm,angle=0]{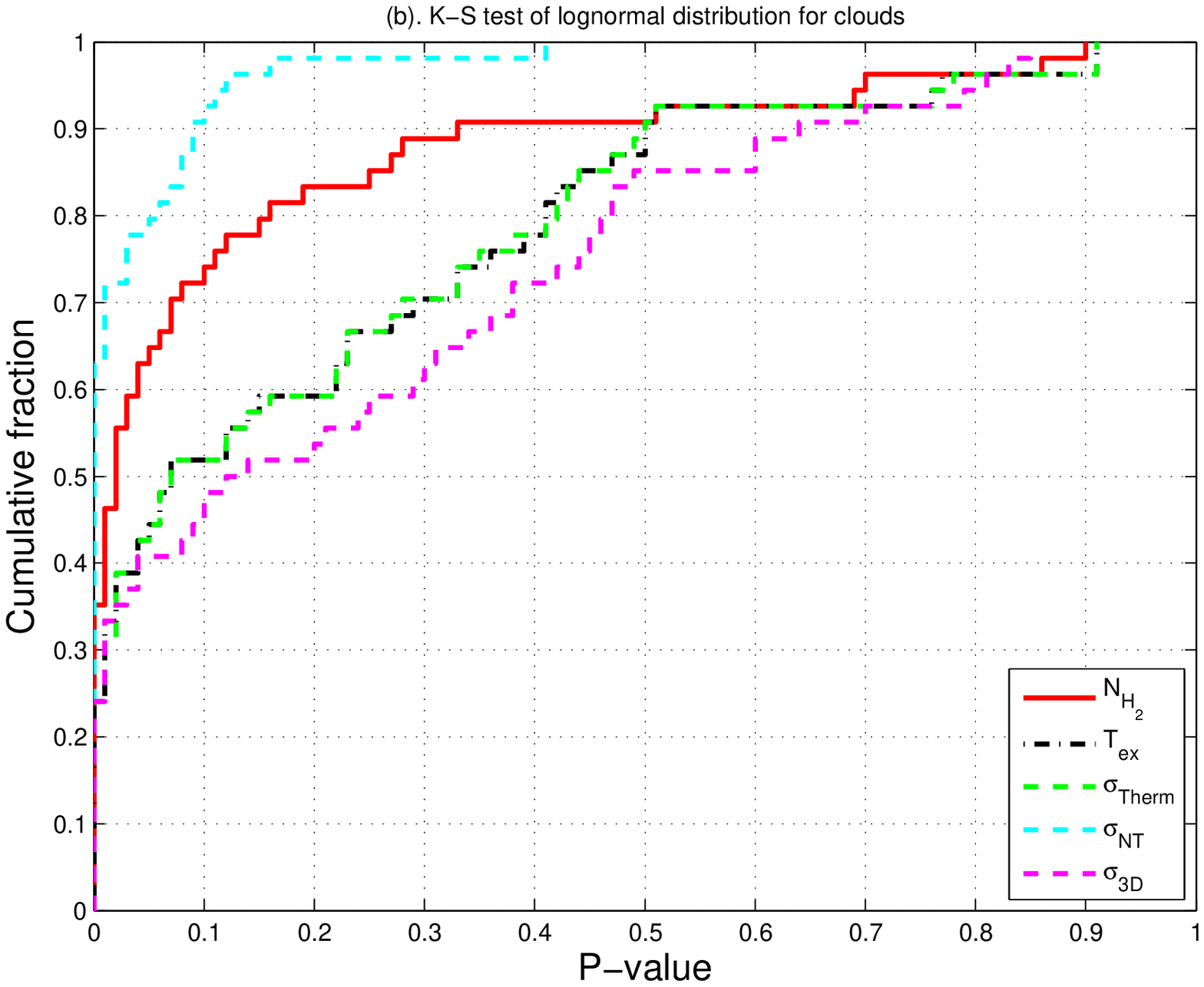}
\end{minipage}
\begin{minipage}[c]{0.5\textwidth}
  \centering
  \includegraphics[width=90mm,height=70mm,angle=0]{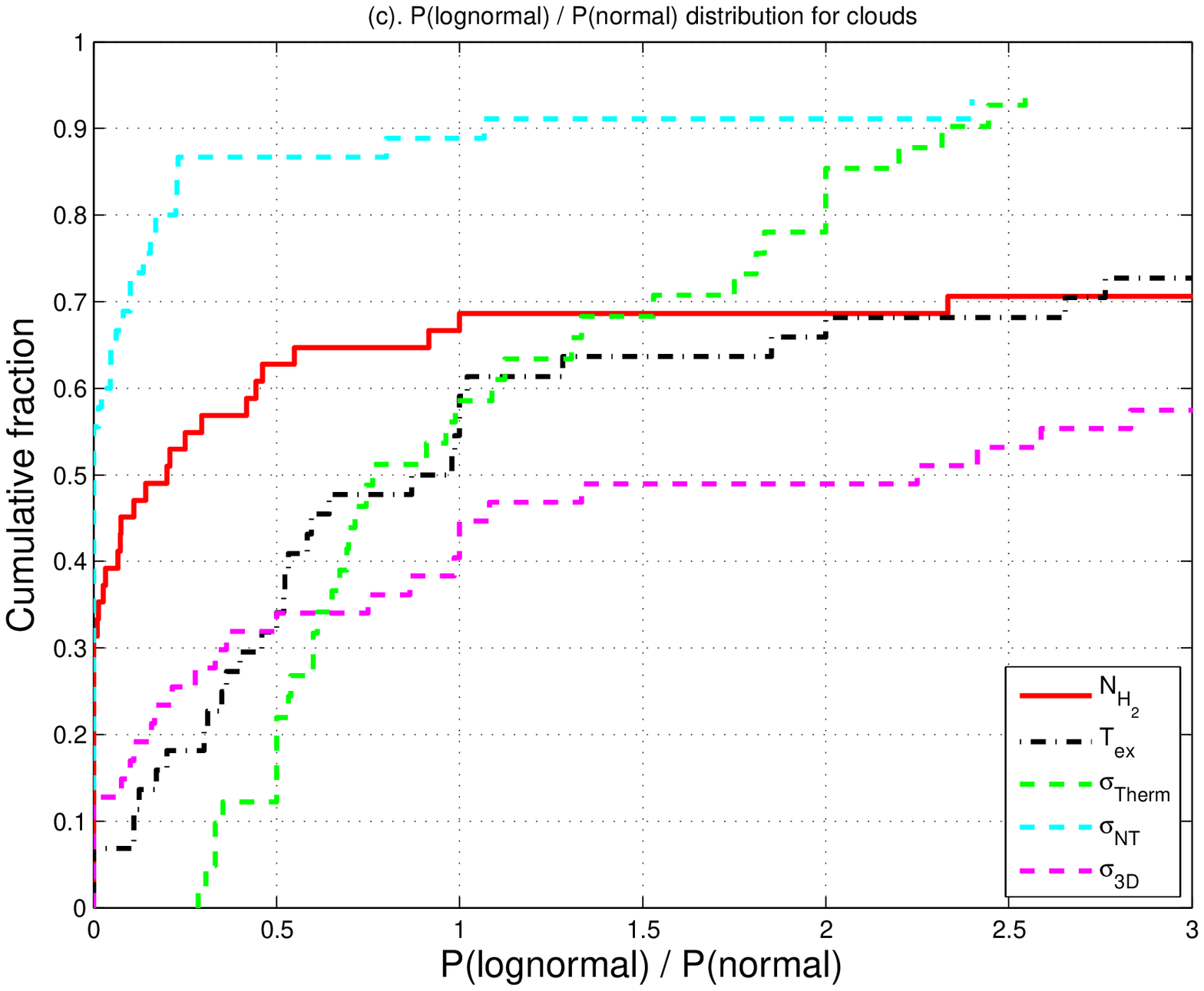}
\end{minipage}
\begin{minipage}[c]{0.5\textwidth}
  \centering
  \includegraphics[width=90mm,height=70mm,angle=0]{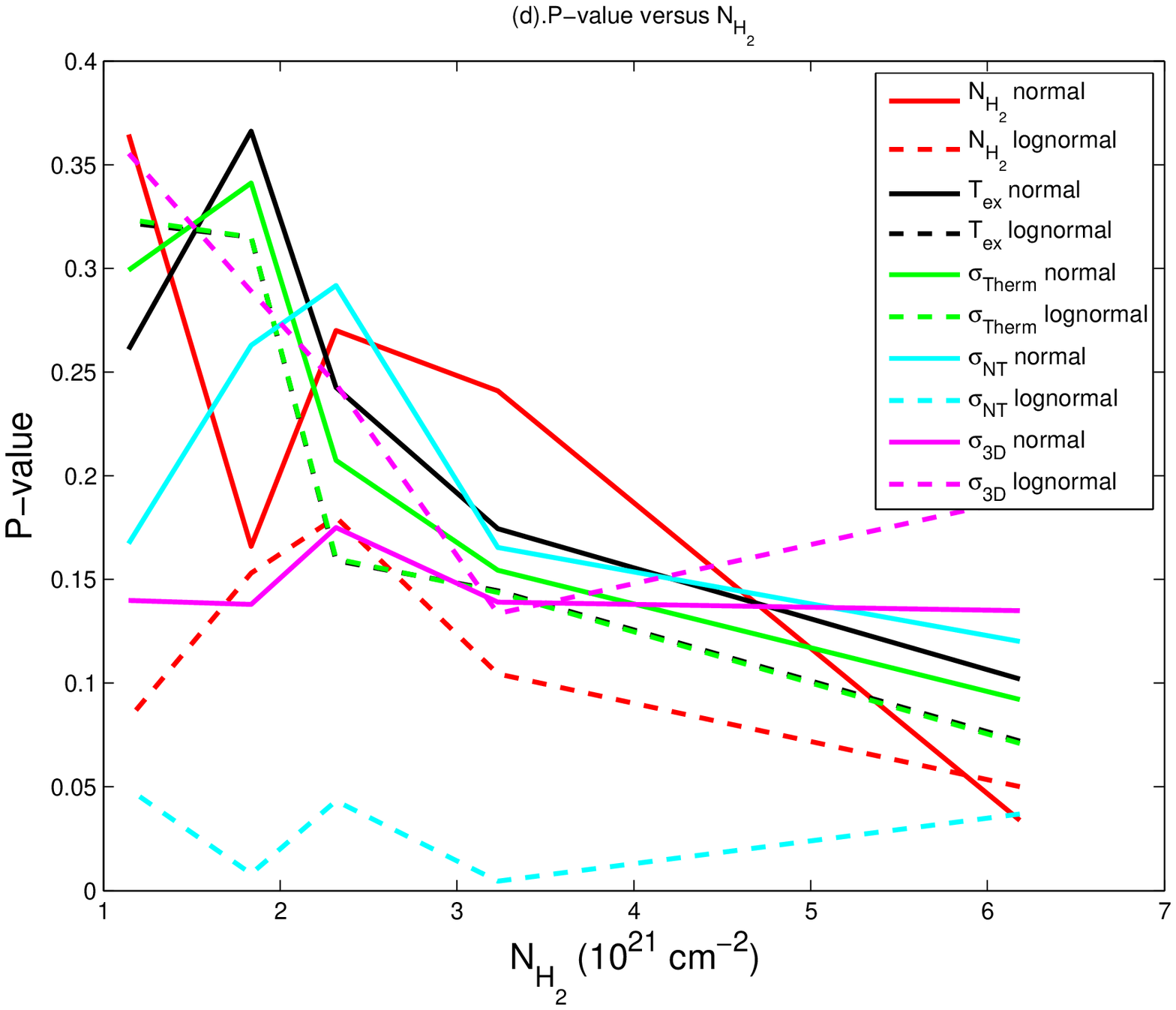}
\end{minipage}
\caption{Statistics of P-values of K-S test for parameter distributions in each cloud. (a). P-value distribution from K-S test for normal distribution hypothesis. (b). P-value distribution from K-S test for lognormal distribution hypothesis. (c). Distribution of the ratio P(lognormal)/P(normal). (d). Bin averaged P-values versus Bin averaged column densities. }
\end{figure}

\begin{figure}
\begin{minipage}[c]{0.5\textwidth}
  \centering
  \includegraphics[width=90mm,height=70mm,angle=0]{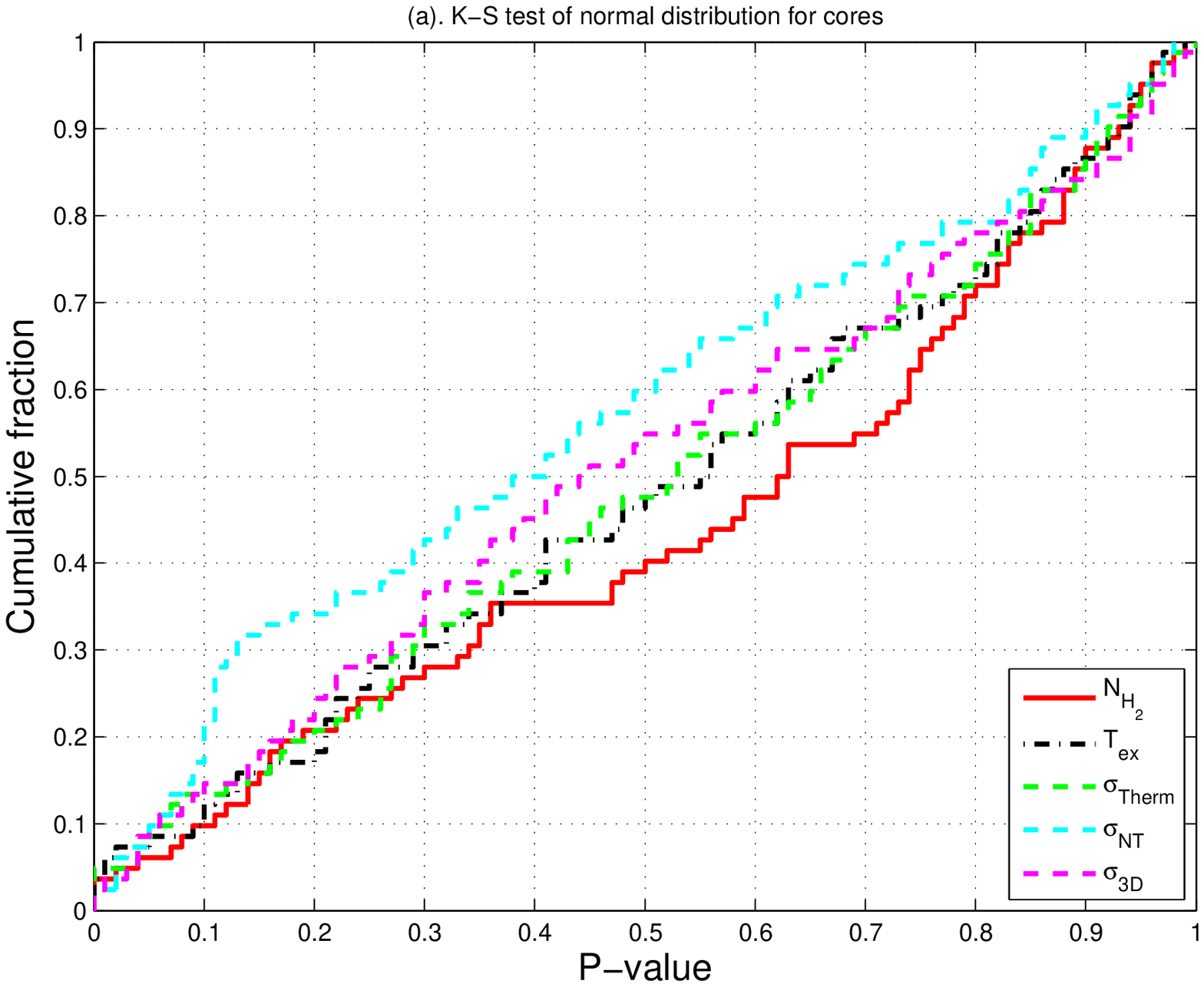}
\end{minipage}
\begin{minipage}[c]{0.5\textwidth}
  \centering
  \includegraphics[width=90mm,height=70mm,angle=0]{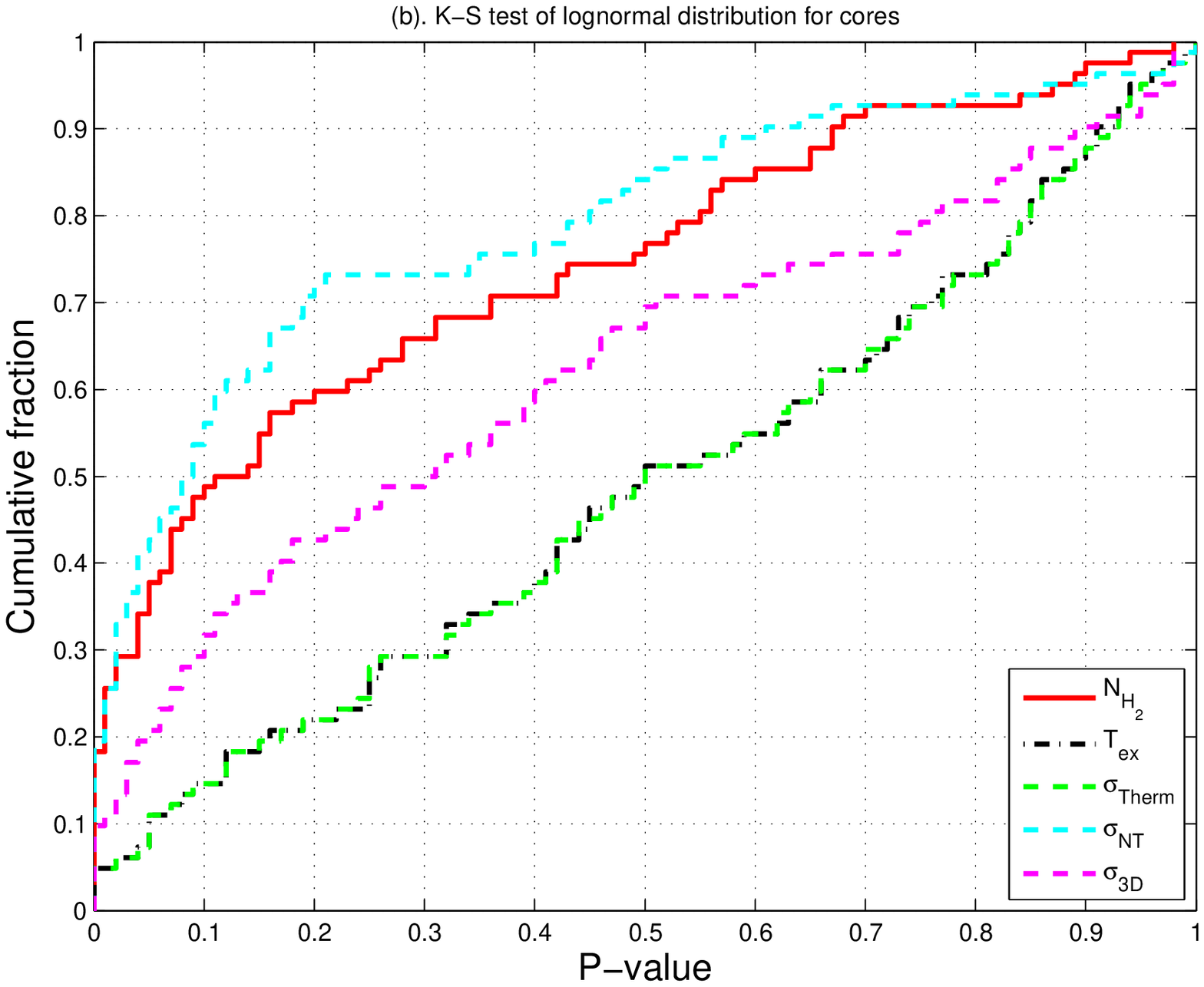}
\end{minipage}
\begin{minipage}[c]{0.5\textwidth}
  \centering
  \includegraphics[width=90mm,height=70mm,angle=0]{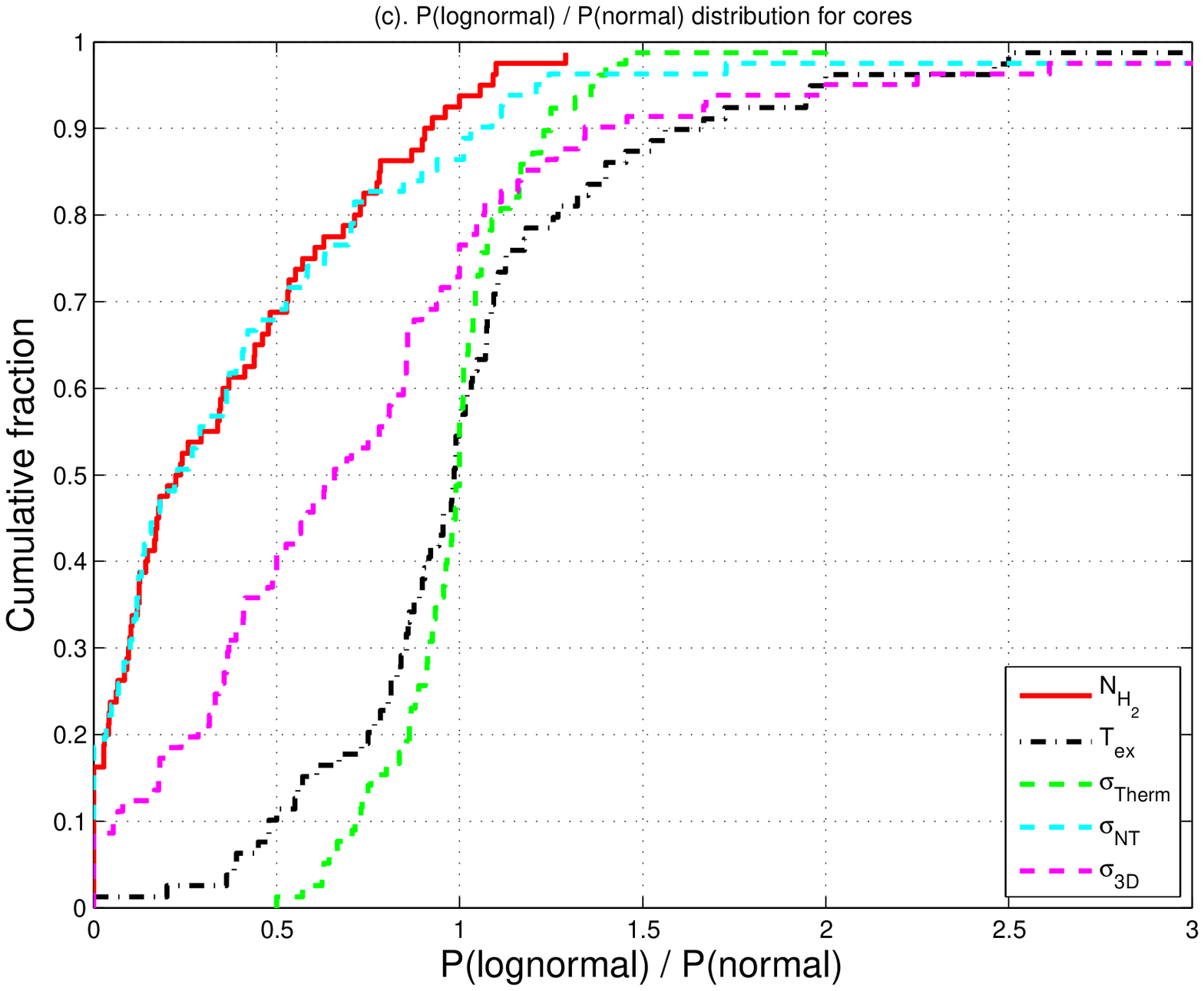}
\end{minipage}
\begin{minipage}[c]{0.5\textwidth}
  \centering
  \includegraphics[width=90mm,height=70mm,angle=0]{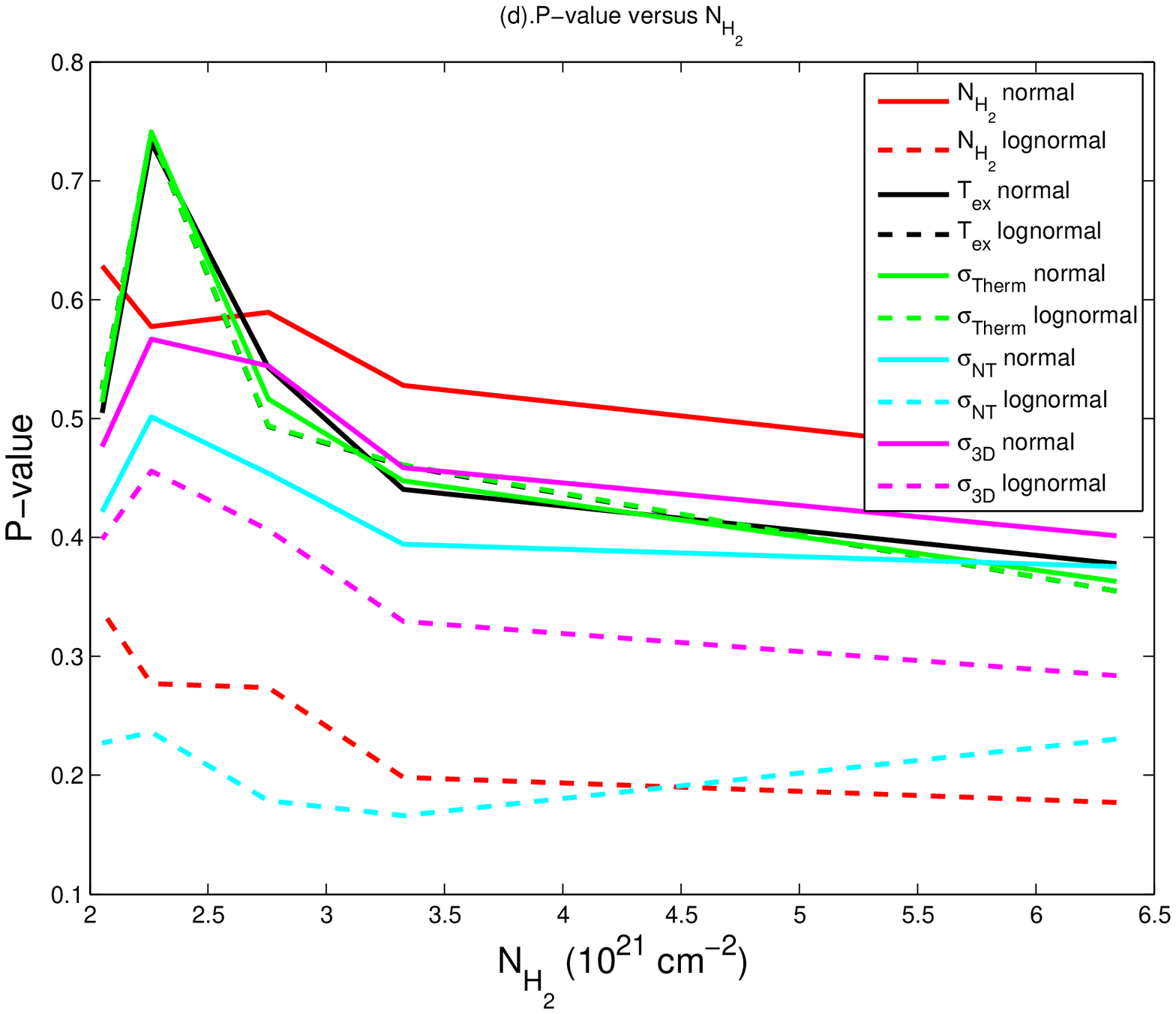}
\end{minipage}
\caption{Statistics of P-values of K-S test for parameter distributions in each dense core. (a). P-value distribution from K-S test for normal distribution hypothesis. (b). P-value distribution from K-S test for lognormal distribution hypothesis. (c). Distribution of the ratio P(lognormal)/P(normal). (d). Bin averaged P-values versus Bin averaged column densities.}
\end{figure}

\begin{figure}
\begin{minipage}[c]{0.5\textwidth}
  \centering
  \includegraphics[width=90mm,height=70mm,angle=0]{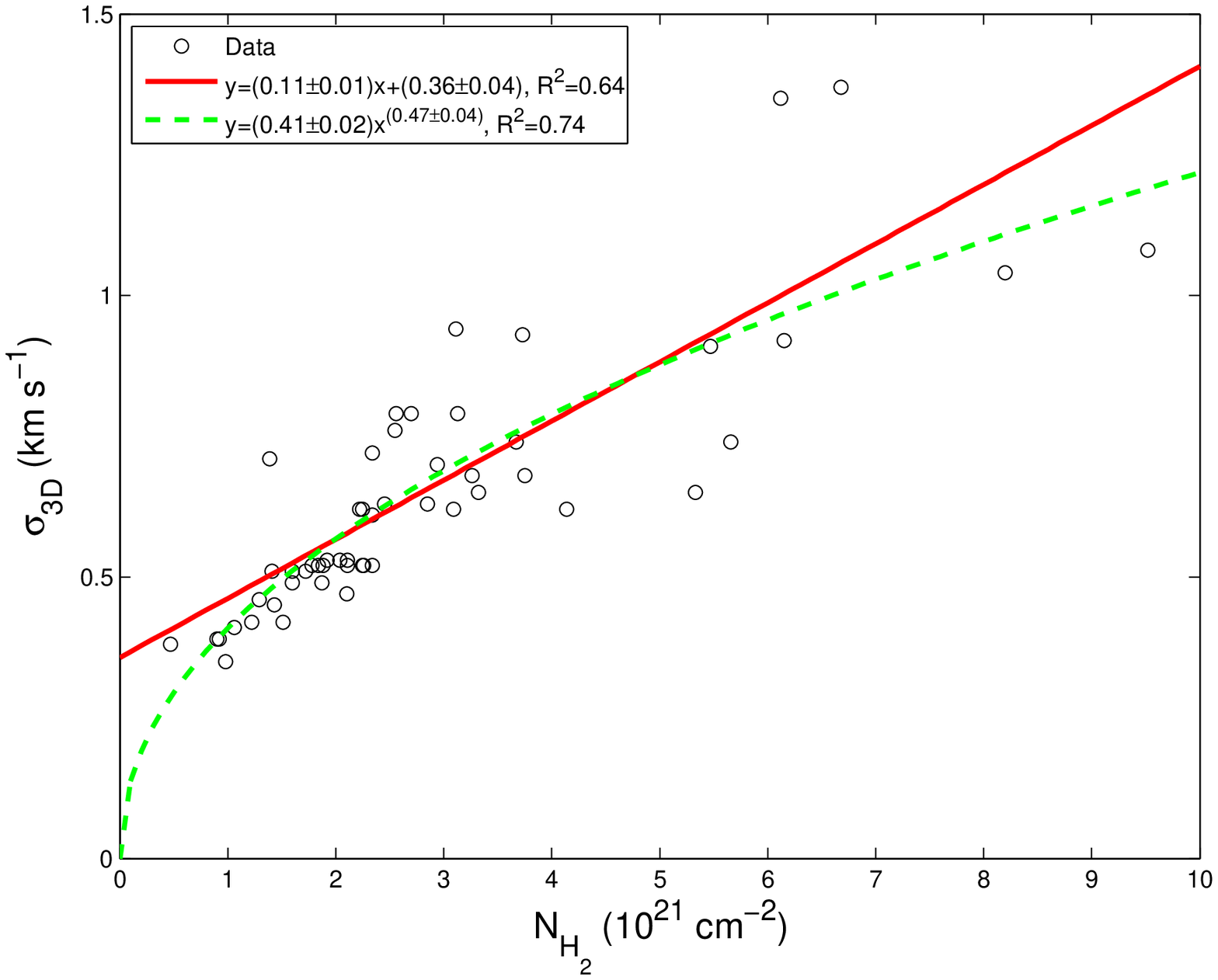}
\end{minipage}
\begin{minipage}[c]{0.5\textwidth}
  \centering
  \includegraphics[width=90mm,height=70mm,angle=0]{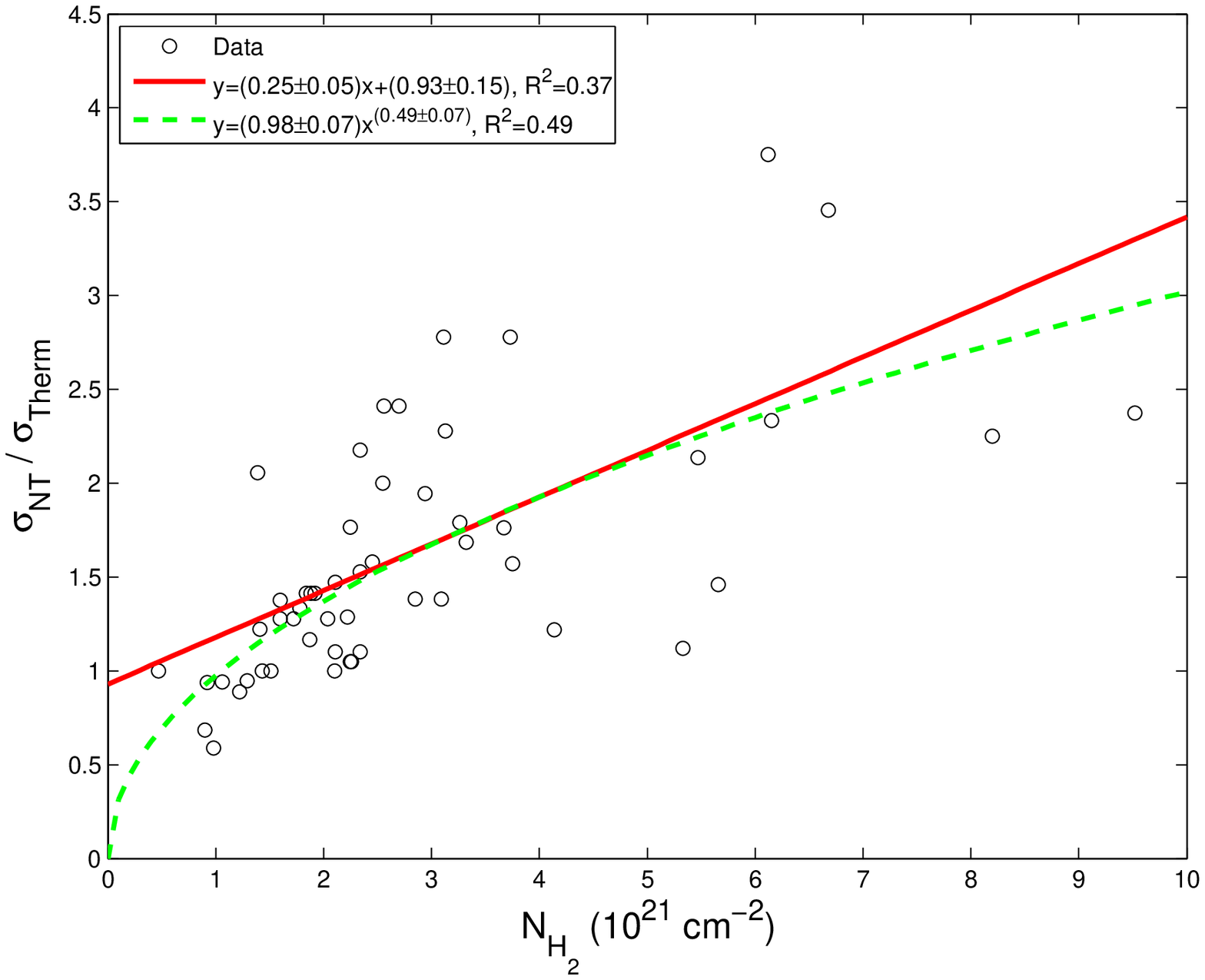}
\end{minipage}
\caption{N$_{H_{2}}$-$\sigma_{3D}$ (left) and N$_{H_{2}}$-$\sigma_{NT}/\sigma_{Therm}$ (right) relations for the clouds. Linear fittings are shown as dashed lines, and power law fittings shown in solid lines. The equations of the fittings are shown in the upper-left corners in each plot.}
\end{figure}

\begin{figure}
\includegraphics[angle=0,scale=.50]{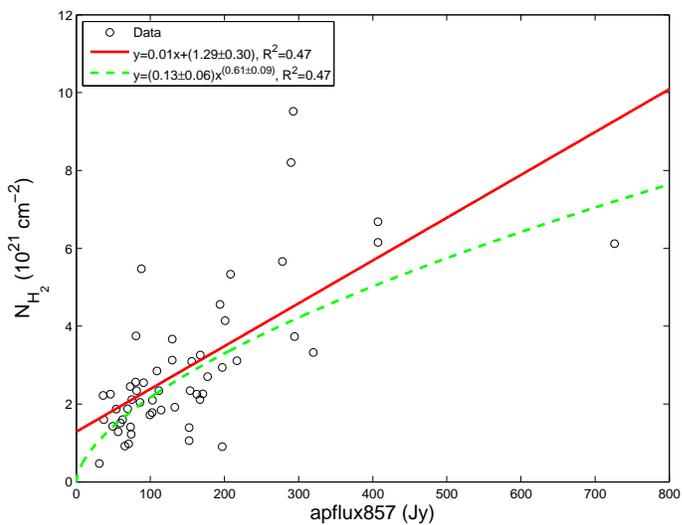}
\caption{Plots of aperture fluxes at 857 GHz versus H$_{2}$ column densities for the clouds.}
\end{figure}

\begin{figure}
\includegraphics[angle=0,scale=.50]{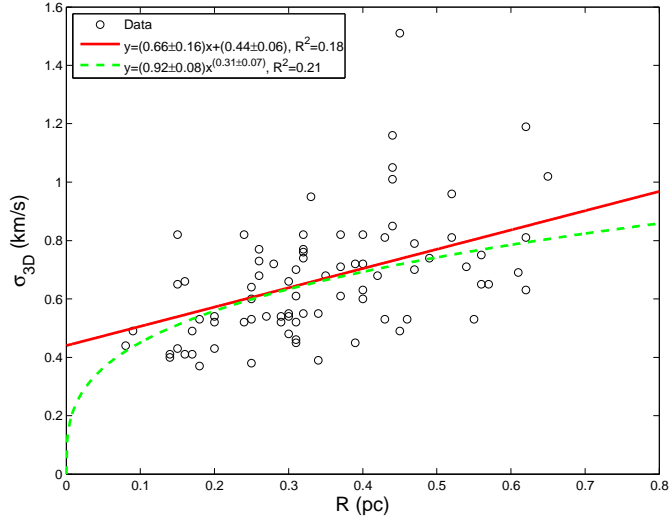}
\caption{$\sigma_{3D}$ versus R relation, i.e., the Larson relationship for the dense cores.}
\end{figure}

\begin{figure}
\begin{minipage}[c]{0.5\textwidth}
  \centering
  \includegraphics[width=70mm,height=90mm,angle=90]{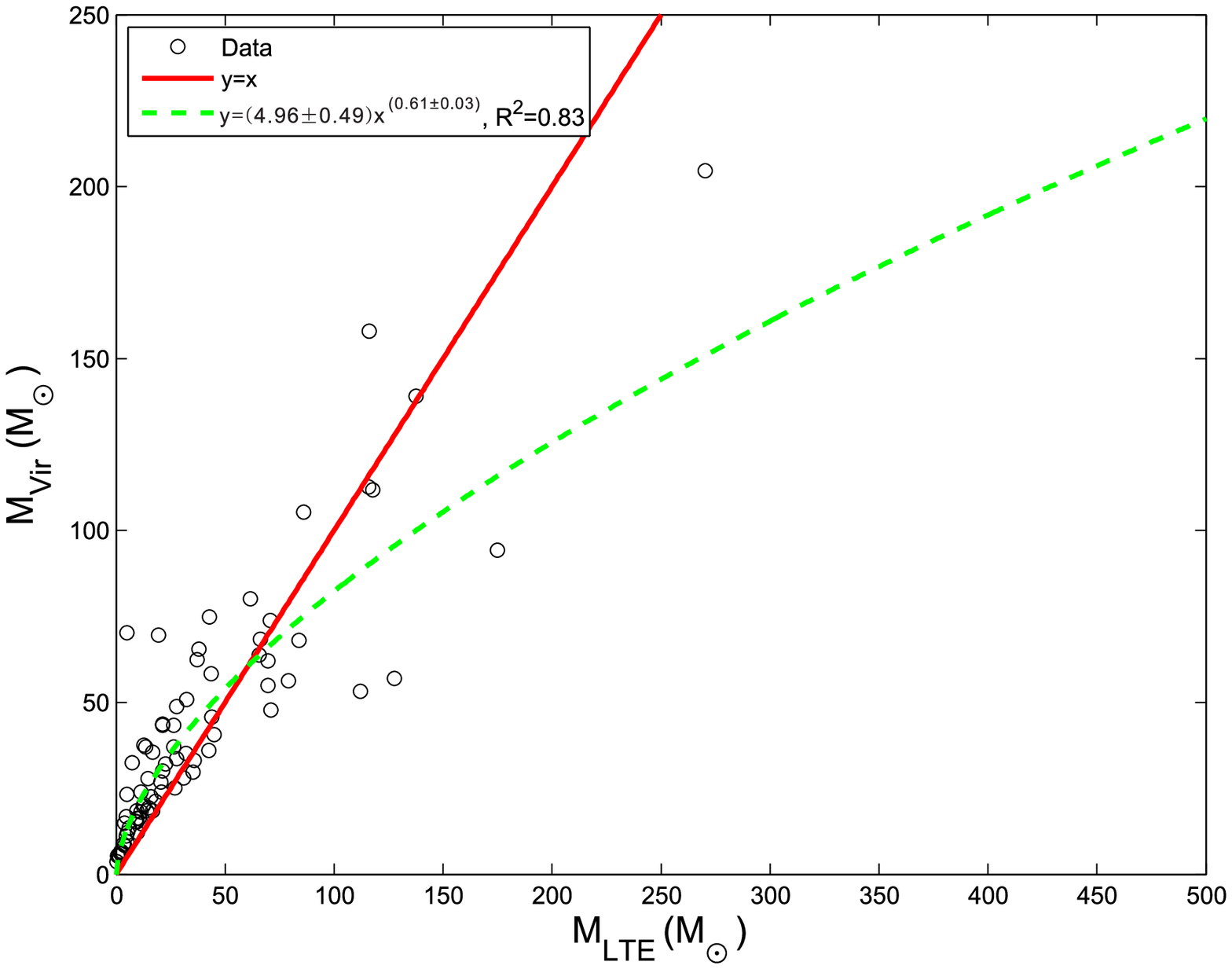}
\end{minipage}
\begin{minipage}[c]{0.5\textwidth}
  \centering
  \includegraphics[width=70mm,height=90mm,angle=90]{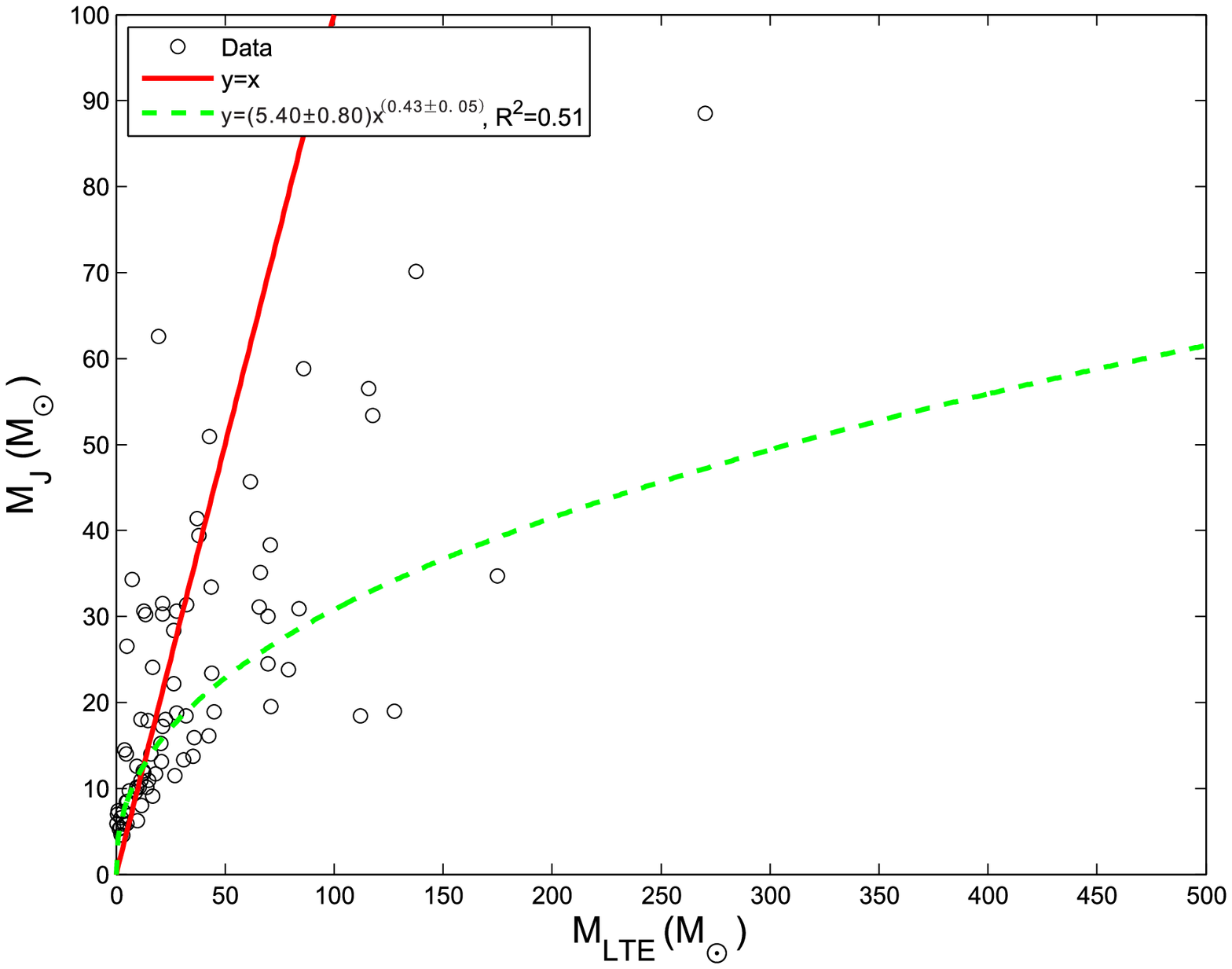}
\end{minipage}
\caption{Left: Plots of virial mass versus LTE mass for the dense cores; Right: Plots of Jeans mass versus LTE mass for the dense cores}
\end{figure}

\begin{figure}
\begin{minipage}[c]{0.5\textwidth}
  \centering
  \includegraphics[width=65mm,height=85mm,angle=90]{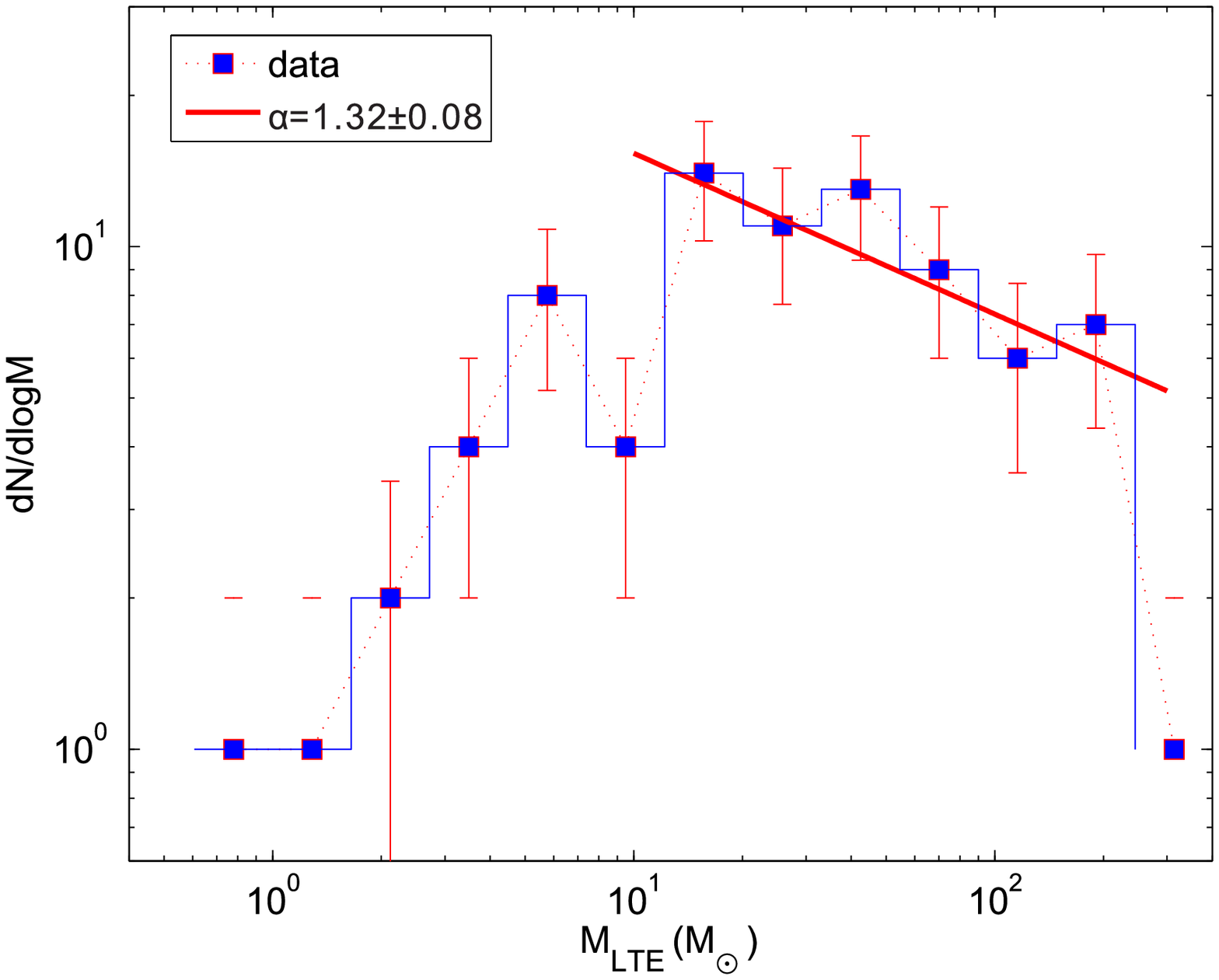}
\end{minipage}
\begin{minipage}[c]{0.5\textwidth}
  \centering
  \includegraphics[width=65mm,height=85mm,angle=90]{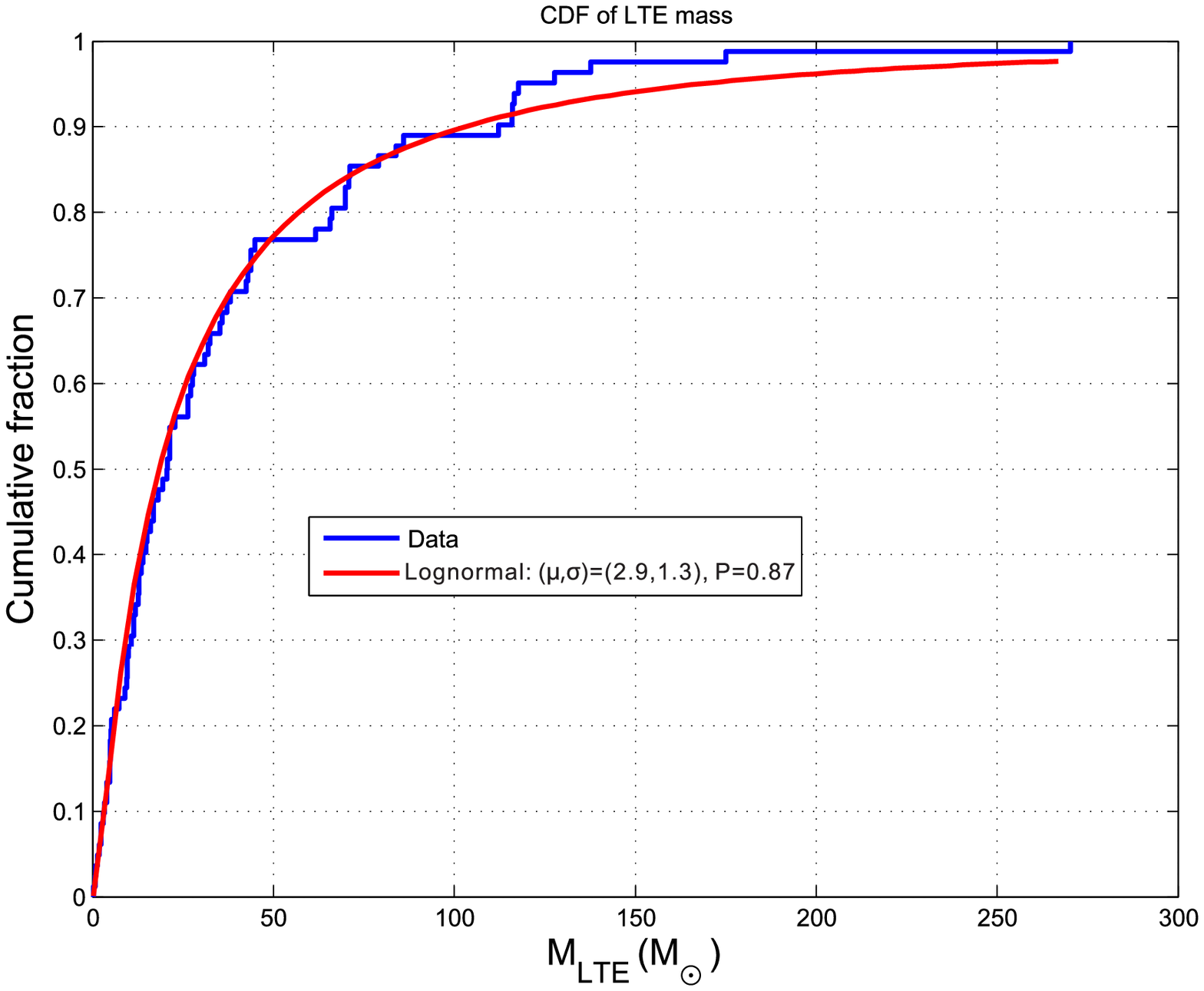}
\end{minipage}
\caption{Left: core mass function (CMF) of the dense cores. The red solid line is the power-law fit. Right: cumulative distribution of the core masses. The red line represents the best lognormal distribution fit.}
\end{figure}

\clearpage
\begin{deluxetable}{ccrrrrrrrrrrrrrrrrrrcrl}
\setlength{\tabcolsep}{0.05in} \rotate
\tabletypesize{\scriptsize}  \tablecaption{The parameters of the Planck cold clumps surveyed (First page of Table 1, the full table is only available on line)}
 \tablewidth{0pt} \tablehead{
  Name& Glon & Glat &Ra(J2000)  &Dec(J2000) &Ra(B1950)  &Dec(B1950) &V$_{lsr}$ & apflux857 &Remarks\tablenotemark{a}\\
      &($\arcdeg$) &($\arcdeg$) &(h m s) &(d m s) &(h m s) &(d m s) &(km~s$^{-1}$) & (Jy) }
\startdata
G180.81-19.66     & 180.81297    & -19.669159       & 04 37 26.49  & +16 59 01.22  & 04 34 33.78  &  +16 53 02.05   & 8.91  & 102.8 &DNe \\
G185.80-09.12     & 185.80077    & -9.1214981       & 05 25 22.61  & +19 10 27.12  & 05 22 25.97  &  +19 07 49.92   & -2.75 & 74.9  &smm \\
G190.08-13.51     & 190.08543    & -13.516411       & 05 19 28.71  & +13 15 43.32  & 05 16 39.53  &  +13 12 40.97   & 1.28  & 72.9  &DNe,Cld \\
G190.15-14.34     & 190.15135    & -14.342709       & 05 16 45.83  & +12 45 44.57  & 05 13 57.29  &  +12 42 30.58   & 1.32  & 108.8 & \\
G191.03-16.74     & 191.03026    & -16.743393       & 05 10 23.30  & +10 45 03.71  & 05 07 37.24  &  +10 41 22.53   & 1.75  & 46.0  &PoC \\
G192.12-10.90     & 192.12889    & -10.901871       & 05 32 55.83  & +12 57 08.60  & 05 30 06.88  &  +12 55 04.38   & 10.02 & 200.7 &IRAS 05300+1253, smm\\
G192.28-11.33     & 192.2827     & -11.339053       & 05 31 43.36  & +12 35 44.79  & 05 28 54.85  &  +12 33 35.35   & 10.13 & 278.1 &Cld,DNe,PoC \\
G192.54-11.56     & 192.54637    & -11.567412       & 05 31 28.55  & +12 15 22.39  & 05 28 40.45  &  +12 13 11.90   & 10.21 & 208.2 &DNe \\
G194.69-16.84     & 194.69969    & -16.840801       & 05 17 37.50  & +07 43 26.85  & 05 14 54.91  &  +07 40 16.77   & -2.41 & 53.8  &IRAS 05150+0739 \\
G194.94-16.74     & 194.94139    & -16.743393       & 05 18 26.83  & +07 34 38.62  & 05 15 44.40  &  +07 31 32.08   & -1.81 & 80.7  &DNe,PoC \\
G195.09-16.41     & 195.0952     & -16.412575       & 05 19 53.82  & +07 37 24.15  & 05 17 11.32  &  +07 34 23.84   & -0.1  & 87.8  &DNe \\
G195.00-16.95     & 195.00731    & -16.957758       & 05 17 50.47  & +07 24 43.08  & 05 15 08.24  &  +07 21 33.94   & -1.81 & 155.7 &PoC \\
G196.21-15.50     & 196.21581    & -15.50084        & 05 25 17.29  & +07 10 11.87  & 05 22 35.29  &  +07 07 34.82   & 3.76  & 36.3  &PoC,MoC \\
G198.03-15.24     & 198.03954    & -15.249377       & 05 29 45.07  & +05 46 55.24  & 05 27 04.67  &  +05 44 37.55   & 0.01  & 170.9 &DNe,MoC \\
G198.56-09.10     & 198.56688    & -9.1026087       & 05 52 17.25  & +08 23 18.98  & 05 49 33.69  &  +08 22 39.30   & 11.42 & 129.5 &DNe \\
G200.34-10.97     & 200.34666    & -10.977856       & 05 49 08.80  & +05 56 35.00  & 05 46 28.14  &  +05 55 41.71   & 13.56 & 153.8 &IRAS 05464+0554, DNe, PoC \\
\enddata
\tablenotetext{a}{Simbad identifiers within 5$\arcmin$. DNe: Dark Cloud (nebula); smm: sub-millimetric source; Cld: Cloud; PoC: Part of Cloud; Rad: Radio-source; RNe: Reflection Nebula; MoC: Molecular Cloud; IRAS: IRAS point source.}
\end{deluxetable}

\clearpage

\begin{deluxetable}{ccccccccccccccccccccccc}
\tabletypesize{\scriptsize} \tablecolumns{16}
\tablewidth{0pc} \rotate \setlength{\tabcolsep}{0.04in}
\tablecaption{Derived parameters of gas emission over the whole clouds (First page of Table 2, the full table is only available on line)} \tablehead{
 Name & \multicolumn{3}{c}{N$_{H_{2}}$}  &\multicolumn{3}{c}{T$_{ex}$} &\multicolumn{3}{c}{$\sigma_{Therm}$} &\multicolumn{3}{c}{$\sigma_{NT}$} &\multicolumn{3}{c}{$\sigma_{3D}$}\\
\cline{2-16}
& mean & \multicolumn{2}{c}{P-value} & mean & \multicolumn{2}{c}{P-value} & mean & \multicolumn{2}{c}{P-value} & mean & \multicolumn{2}{c}{P-value}& mean & \multicolumn{2}{c}{P-value}\\
\cline{2-16} \\
 & (10$^{21}$ cm$^{-2}$) &normal &lognormal & (K) &normal &lognormal & (km~s$^{-1}$) &normal &lognormal &(km~s$^{-1}$) &normal &lognormal &(km~s$^{-1}$) &normal &lognormal
}\startdata
G180.81-19.66  & 2.10$_{-0.32}^{+0.39}$   & 0.16 &0.00  &  12.23$_{-0.55}^{+0.53}$ & 0.77 &0.76  & 0.19$_{-0.00}^{+0.00}$  &0.79 &0.76   & 0.19$_{-0.03}^{+0.02}$   & 0.00  & 0.00  & 0.47$_{-0.04}^{+0.02}$  &0.00  &0.00      \\
G185.80-09.12  & 2.11$_{-0.79}^{+0.66}$   & 0.51 &0.28  &  12.67$_{-1.06}^{+0.93}$ & 0.09 &0.33  & 0.20$_{-0.01}^{+0.01}$  &0.18 &0.33   & 0.22$_{-0.05}^{+0.07}$   & 0.47  & 0.01  & 0.52$_{-0.08}^{+0.09}$  &0.07  &0.31      \\
G190.08-13.51  & 2.45$_{-0.64}^{+0.68}$   & 0.00 &0.10  &  11.88$_{-0.54}^{+0.44}$ & 0.00 &0.01  & 0.19$_{-0.00}^{+0.00}$  &0.00 &0.01   & 0.30$_{-0.09}^{+0.09}$   & 0.03  & 0.00  & 0.63$_{-0.14}^{+0.12}$  &0.00  &0.09      \\
G190.15-14.34  & 2.85$_{-0.79}^{+0.82}$   & 0.15 &0.01  &  14.34$_{-1.09}^{+1.13}$ & 0.20 &0.07  & 0.21$_{-0.01}^{+0.01}$  &0.14 &0.07   & 0.29$_{-0.07}^{+0.05}$   & 0.16  & 0.01  & 0.63$_{-0.10}^{+0.08}$  &0.01  &0.47        \\
G191.03-16.74  & 2.25$_{-0.57}^{+0.52}$   & 0.12 &0.00  &  13.74$_{-1.11}^{+0.92}$ & 0.36 &0.04  & 0.20$_{-0.01}^{+0.01}$  &0.13 &0.04   & 0.21$_{-0.04}^{+0.04}$   & 0.01  & 0.00  & 0.52$_{-0.05}^{+0.05}$  &0.56  &0.12        \\
G192.12-10.90  & 4.14$_{-1.23}^{+1.13}$   & 0.02 &0.02  &  17.23$_{-1.38}^{+1.23}$ & 0.04 &0.29  & 0.23$_{-0.01}^{+0.01}$  &0.11 &0.28   & 0.28$_{-0.05}^{+0.05}$   & 0.00  & 0.11  & 0.62$_{-0.08}^{+0.08}$  &0.00  &0.01        \\
G192.28-11.33  & 5.66$_{-1.41}^{+1.35}$   & 0.06 &0.00  &  18.90$_{-2.04}^{+1.37}$ & 0.02 &0.01  & 0.24$_{-0.01}^{+0.01}$  &0.02 &0.01   & 0.35$_{-0.05}^{+0.06}$   & 0.05  & 0.00  & 0.74$_{-0.07}^{+0.08}$  &0.04  &0.00        \\
G192.54-11.56  & 5.33$_{-1.34}^{+0.58}$   & 0.00 &0.01  &  21.12$_{-2.25}^{+0.79}$ & 0.00 &0.00  & 0.25$_{-0.01}^{+0.01}$  &0.00 &0.00   & 0.28$_{-0.03}^{+0.03}$   & 0.59  & 0.10  & 0.65$_{-0.04}^{+0.04}$  &0.73  &0.79        \\
G194.69-16.84  & 1.87$_{-0.65}^{+0.58}$   & 0.27 &0.12  &  10.98$_{-0.73}^{+0.73}$ & 0.52 &0.27  & 0.18$_{-0.01}^{+0.01}$  &0.39 &0.27   & 0.21$_{-0.04}^{+0.04}$   & 0.07  & 0.00  & 0.49$_{-0.07}^{+0.06}$  &0.36  &0.10        \\
G194.94-16.74  & 3.75$_{-1.05}^{+0.84}$   & 0.02 &0.15  &  15.07$_{-0.91}^{+0.95}$ & 0.16 &0.02  & 0.21$_{-0.01}^{+0.01}$  &0.06 &0.02   & 0.33$_{-0.05}^{+0.05}$   & 0.34  & 0.00  & 0.68$_{-0.08}^{+0.07}$  &0.87  &0.29        \\
G195.09-16.41  & 5.47$_{-1.56}^{+1.49}$   & 0.05 &0.00  &  15.85$_{-0.95}^{+0.76}$ & 0.14 &0.14  & 0.22$_{-0.01}^{+0.01}$  &0.26 &0.14   & 0.47$_{-0.09}^{+0.08}$   & 0.05  & 0.00  & 0.91$_{-0.15}^{+0.12}$  &0.02  &0.25        \\
G195.00-16.95  & 3.09$_{-0.71}^{+0.66}$   & 0.10 &0.02  &  14.21$_{-0.66}^{+0.71}$ & 0.56 &0.36  & 0.21$_{-0.00}^{+0.01}$  &0.52 &0.35   & 0.29$_{-0.05}^{+0.06}$   & 0.21  & 0.00  & 0.62$_{-0.07}^{+0.08}$  &0.11  &0.04        \\
G196.21-15.50  & 2.22$_{-0.79}^{+0.74}$   & 0.41 &0.03  &  14.38$_{-0.74}^{+0.75}$ & 0.74 &0.44  & 0.21$_{-0.01}^{+0.01}$  &0.59 &0.44   & 0.27$_{-0.09}^{+0.10}$   & 0.74  & 0.06  & 0.62$_{-0.13}^{+0.13}$  &0.20  &0.45        \\
G198.03-15.24  & 2.26$_{-0.65}^{+0.68}$   & 0.31 &0.01  &  14.06$_{-0.73}^{+0.80}$ & 0.48 &0.15  & 0.21$_{-0.01}^{+0.01}$  &0.32 &0.16   & 0.22$_{-0.05}^{+0.05}$   & 0.69  & 0.01  & 0.52$_{-0.07}^{+0.06}$  &0.44  &0.38        \\
\enddata
\tablecomments{The errors throughout all the tables are calculated from the first and third quartiles. }
\end{deluxetable}

\clearpage

\begin{deluxetable}{ccccccccccccccccccccccccccccccccccccccccc}
\tabletypesize{\scriptsize} \tablecolumns{35} \rotate
\tablewidth{0pc} \setlength{\tabcolsep}{0.02in}
\tablecaption{Derived parameters of the dense cores (First page of Table 3, the full table is only available on line)} \tablehead{
 Name& \multicolumn{2}{c}{Offset}  & \multicolumn{4}{c}{N$_{H_{2}}$}  &\multicolumn{3}{c}{T$_{ex}$} &\multicolumn{3}{c}{$\sigma_{Therm}$} &\multicolumn{3}{c}{$\sigma_{NT}$} &\multicolumn{3}{c}{$\sigma_{3D}$} \\
\cline{3-19}
& \multicolumn{2}{c}{}&max & mean & \multicolumn{2}{c}{P-value} & mean & \multicolumn{2}{c}{P-value} & mean & \multicolumn{2}{c}{P-value} & mean & \multicolumn{2}{c}{P-value}& mean & \multicolumn{2}{c}{P-value}\\
\cline{3-19} \\
 & \multicolumn{2}{c}{($\arcsec,\arcsec$)} & (10$^{21}$ cm$^{-2}$)& (10$^{21}$ cm$^{-2}$) &normal &lognormal & (K) &normal &lognormal & (km~s$^{-1}$) &normal &lognormal &(km~s$^{-1}$) &normal &lognormal &(km~s$^{-1}$) &normal &lognormal
}\startdata
G185.80-09.12  & \multicolumn{2}{c}{(-69,-19)}   &4.37  & 2.95$_{-0.31}^{+0.72}$   &0.47  &0.02  & 13.10$_{-0.92}^{+0.91}$  &0.21  &0.32  & 0.20$_{-0.01}^{+0.01}$  & 0.26 &0.32  & 0.28$_{-0.04}^{+0.05}$    & 0.22 &0.03  & 0.60$_{-0.05}^{+0.06}$   & 0.30 & 0.11   \\
               & \multicolumn{2}{c}{(-134,290)}  &3.36  & 2.25$_{-0.61}^{+0.62}$   &0.90  &0.90  & 11.89$_{-0.35}^{+0.51}$  &0.48  &0.42  & 0.19$_{-0.00}^{+0.00}$  & 0.45 &0.42  & 0.23$_{-0.04}^{+0.06}$    & 0.86 &0.61  & 0.52$_{-0.06}^{+0.08}$   & 0.87 & 0.91    \\
G190.08-13.51  & \multicolumn{2}{c}{(110,106)}   &4.86  & 3.11$_{-0.67}^{+0.64}$   &0.12  &0.00  & 11.91$_{-0.40}^{+0.35}$  &0.22  &0.32  & 0.19$_{-0.00}^{+0.00}$  & 0.27 &0.33  & 0.36$_{-0.06}^{+0.07}$    & 0.38 &0.00  & 0.71$_{-0.09}^{+0.10}$   & 0.56 & 0.03    \\
               & \multicolumn{2}{c}{(25,-62)}    &4.86  & 3.49$_{-0.40}^{+0.48}$   &0.72  &0.25  & 11.69$_{-0.21}^{+0.20}$  &0.62  &0.73  & 0.19$_{-0.00}^{+0.00}$  & 0.68 &0.74  & 0.39$_{-0.04}^{+0.05}$    & 0.29 &0.02  & 0.75$_{-0.06}^{+0.08}$   & 0.36 & 0.06    \\
               & \multicolumn{2}{c}{(-17,-89)}   &4.86  & 4.03$_{-0.26}^{+0.40}$   &0.99  &0.89  & 11.66$_{-0.18}^{+0.14}$  &0.96  &0.96  & 0.19$_{-0.00}^{+0.00}$  & 0.96 &0.96  & 0.44$_{-0.03}^{+0.04}$    & 0.97 &0.98  & 0.82$_{-0.05}^{+0.06}$   & 0.98 & 0.98    \\
               & \multicolumn{2}{c}{(-22,62)}    &4.86  & 3.04$_{-0.60}^{+0.60}$   &0.14  &0.00  & 11.59$_{-0.31}^{+0.38}$  &0.25  &0.12  & 0.19$_{-0.00}^{+0.00}$  & 0.18 &0.12  & 0.35$_{-0.09}^{+0.08}$    & 0.11 &0.00  & 0.69$_{-0.13}^{+0.12}$   & 0.17 & 0.00    \\
G190.15-14.34  & \multicolumn{2}{c}{(122,-27)}   &5.27  & 3.32$_{-0.48}^{+0.66}$   &0.74  &0.05  & 15.03$_{-0.83}^{+0.85}$  &0.37  &0.50  & 0.21$_{-0.01}^{+0.01}$  & 0.43 &0.50  & 0.31$_{-0.04}^{+0.04}$    & 0.97 &1.00  & 0.66$_{-0.06}^{+0.05}$   & 0.74 & 0.98    \\
               & \multicolumn{2}{c}{(-18,-31)}   &5.29  & 3.38$_{-0.61}^{+0.71}$   &0.88  &0.09  & 14.93$_{-0.87}^{+0.96}$  &0.67  &0.45  & 0.21$_{-0.01}^{+0.01}$  & 0.55 &0.44  & 0.33$_{-0.06}^{+0.07}$    & 0.41 &0.51  & 0.68$_{-0.08}^{+0.10}$   & 0.27 & 0.45    \\
               & \multicolumn{2}{c}{(-340,-5)}   &5.29  & 4.03$_{-0.42}^{+0.53}$   &0.84  &0.65  & 15.04$_{-0.73}^{+0.80}$  &0.29  &0.25  & 0.21$_{-0.01}^{+0.01}$  & 0.27 &0.25  & 0.42$_{-0.05}^{+0.07}$    & 0.30 &0.16  & 0.82$_{-0.08}^{+0.10}$   & 0.32 & 0.26    \\
G191.03-16.74  & \multicolumn{2}{c}{(-81,72)}    &3.69  & 2.49$_{-0.44}^{+0.42}$   &0.55  &0.07  & 13.44$_{-1.09}^{+0.92}$  &0.00  &0.00  & 0.20$_{-0.01}^{+0.01}$  & 0.00 &0.00  & 0.23$_{-0.02}^{+0.03}$    & 0.11 &0.02  & 0.53$_{-0.04}^{+0.04}$   & 0.56 & 0.23    \\
               & \multicolumn{2}{c}{(74,-80)}    &3.69  & 2.44$_{-0.28}^{+0.42}$   &0.04  &0.00  & 13.25$_{-0.94}^{+0.90}$  &0.05  &0.07  & 0.20$_{-0.01}^{+0.01}$  & 0.06 &0.07  & 0.23$_{-0.03}^{+0.03}$    & 0.07 &0.00  & 0.53$_{-0.04}^{+0.04}$   & 0.25 & 0.06    \\
G192.12-10.90  & \multicolumn{2}{c}{(-39,-33)}   &7.33  & 4.36$_{-0.71}^{+0.74}$   &0.92  &0.68  & 17.04$_{-0.73}^{+0.73}$  &0.82  &0.63  & 0.23$_{-0.00}^{+0.00}$  & 0.73 &0.63  & 0.28$_{-0.02}^{+0.02}$    & 0.02 &0.00  & 0.63$_{-0.03}^{+0.03}$   & 0.01 & 0.02    \\
               & \multicolumn{2}{c}{(-212,-113)} &7.33  & 5.09$_{-1.00}^{+1.12}$   &0.75  &0.09  & 17.99$_{-0.55}^{+0.85}$  &0.34  &0.19  & 0.23$_{-0.00}^{+0.01}$  & 0.26 &0.19  & 0.33$_{-0.05}^{+0.05}$    & 0.29 &0.05  & 0.70$_{-0.07}^{+0.08}$   & 0.38 & 0.13    \\
               & \multicolumn{2}{c}{(-326,-338)} &7.81  & 5.65$_{-0.57}^{+0.64}$   &0.75  &0.31  & 19.61$_{-1.64}^{+1.47}$  &0.48  &0.39  & 0.24$_{-0.01}^{+0.01}$  & 0.43 &0.39  & 0.34$_{-0.02}^{+0.02}$    & 0.43 &0.52  & 0.72$_{-0.04}^{+0.03}$   & 0.36 & 0.42    \\
G192.28-11.33  & \multicolumn{2}{c}{(240,-235)}  &9.60  & 6.37$_{-1.21}^{+1.66}$   &0.08  &0.01  & 20.11$_{-1.47}^{+1.49}$  &0.24  &0.47  & 0.25$_{-0.01}^{+0.01}$  & 0.34 &0.47  & 0.34$_{-0.07}^{+0.07}$    & 0.09 &0.02  & 0.74$_{-0.10}^{+0.10}$   & 0.14 & 0.07    \\
               & \multicolumn{2}{c}{(-164,-264)} &9.96  & 6.48$_{-1.02}^{+1.32}$   &0.22  &0.00  & 21.15$_{-1.83}^{+1.91}$  &0.25  &0.12  & 0.25$_{-0.01}^{+0.01}$  & 0.17 &0.12  & 0.33$_{-0.02}^{+0.03}$    & 0.46 &0.12  & 0.72$_{-0.04}^{+0.05}$   & 0.62 & 0.39    \\
               & \multicolumn{2}{c}{(-7,-105)}   &10.04 & 7.21$_{-0.83}^{+0.94}$   &0.28  &0.00  & 19.38$_{-2.48}^{+2.47}$  &0.01  &0.08  & 0.24$_{-0.02}^{+0.02}$  & 0.04 &0.08  & 0.40$_{-0.04}^{+0.04}$    & 0.49 &0.04  & 0.81$_{-0.05}^{+0.06}$   & 0.45 & 0.08    \\
\enddata
\end{deluxetable}

\begin{deluxetable}{ccccccccccccccccccccccccccccccccccc}
\tabletypesize{\scriptsize} \tablecolumns{9}
\tablewidth{0pc} \setlength{\tabcolsep}{0.05in}
\tablecaption{Derived parameters of the dense cores (First page of Table 4, the full table is only available on line)} \tablehead{
 Name & \multicolumn{3}{c}{Deconvolved Size} & R & V$_{lsr}$ & n  &\colhead{M$_{LTE}$} &\colhead{M$_{vir}$} &\colhead{M$_{J}$} &Remarks\\
 & \multicolumn{3}{c}{($\arcsec\times\arcsec$($\arcdeg$))}& (pc)  &(km~s$^{-1}$) &(10$^{3}$ cm$^{-3}$) &\colhead{(M$_{\sun}$)} &\colhead{(M$_{\sun}$)} &\colhead{(M$_{\sun}$)}
}\startdata
G185.80-09.12  &\multicolumn{3}{c}{291$\times$176(-42.8)} & 0.22   & -2.8(0.1) & 3.2  &9.6    &18.4   &12.6   &   \\
               &\multicolumn{3}{c}{313$\times$105(-7.6)}  & 0.18   & -2.3(0.1) & 3.1  &4.8    &11.1   &8.4    &   \\
G190.08-13.51  &\multicolumn{3}{c}{577$\times$429(75.8)}  & 0.54   & 1.2(0.2)  & 1.4  &65.6   &63.8   &31.1   &   \\
               &\multicolumn{3}{c}{829$\times$322(-25.5)} & 0.56   & 1.3(0.2)  & 1.4  &70.8   &73.9   &38.3   &   \\
               &\multicolumn{3}{c}{153$\times$121(-39.4)} & 0.15   & 1.3(0.1)  & 5.3  &4.9    &23.3   &26.5   &   \\
               &\multicolumn{3}{c}{627$\times$504(4.1)}   & 0.61   & 1.4(0.3)  & 1.3  &83.8   &68.0   &30.9   &   \\
G190.15-14.34  &\multicolumn{3}{c}{332$\times$223(85.3)}  & 0.30   & 1.5(0.2)  & 2.9  &21.3   &30.1   &17.2   &   \\
               &\multicolumn{3}{c}{518$\times$293(80.0)}  & 0.42   & 1.4(0.1)  & 2.0  &43.8   &45.8   &23.4   &   \\
               &\multicolumn{3}{c}{293$\times$160(78.6)}  & 0.24   & 1.5(0.2)  & 3.6  &13.5   &37.0   &30.2   &   \\
G191.03-16.74  &\multicolumn{3}{c}{447$\times$345(-65.9)} & 0.43   & 1.8(0.1)  & 1.4  &31.0   &28.0   &13.3   &   \\
               &\multicolumn{3}{c}{440$\times$397(33.6)}  & 0.46   & 1.7(0.2)  & 1.3  &35.2   &29.8   &13.7   &   \\
G192.12-10.90  &\multicolumn{3}{c}{948$\times$337(51.4)}  & 0.62   & 10.0(0.1) & 1.9  &127.7  &57.0   &19.0   &   \\
               &\multicolumn{3}{c}{302$\times$265(85.9)}  & 0.31   & 9.8(0.1)  & 3.8  &32.0   &35.2   &18.4   &   \\
               &\multicolumn{3}{c}{339$\times$192(7.5)}   & 0.28   & 9.7(0.1)  & 4.5  &27.7   &33.6   &18.8   &   \\
G192.28-11.33  &\multicolumn{3}{c}{336$\times$255(-77.8)} & 0.32   & 10.1(0.2) & 4.9  &44.9   &40.7   &18.9   &   \\
               &\multicolumn{3}{c}{476$\times$275(63.7)}  & 0.39   & 10.4(0.1) & 4.1  &71.1   &47.7   &19.5   &   \\
               &\multicolumn{3}{c}{707$\times$452(-70.7)} & 0.62   & 10.3(0.1) & 2.6  &175.0  &94.3   &34.7   &   \\
G192.54-11.56  &\multicolumn{3}{c}{605$\times$302(11.3)}  & 0.47   & 10.5(0.1) & 3.9  &112.2  &53.2   &18.4   &   \\
G194.69-16.84  &\multicolumn{3}{c}{205$\times$139(66.2)}  & 0.18   & -2.1(0.1) & 3.0  &5.3    &12.1   &8.5    &   \\
               &\multicolumn{3}{c}{361$\times$237(49.1)}  & 0.32   & -2.1(0.2) & 1.7  &16.0   &22.5   &14.0   &   \\
               &\multicolumn{3}{c}{188$\times$99(70.7)}   & 0.15   & -2.1(0.1) & 2.1  &1.9    &6.4    &5.4    &   \\
G195.09-16.41  &\multicolumn{3}{c}{534$\times$309(59.3)}  & 0.44   & -0.4(0.4) & 3.5  &86.0   &105.3  &58.8   &   \\
               &\multicolumn{3}{c}{679$\times$341(-62.3)} & 0.52   & -0.2(0.4) & 2.8  &115.9  &112.7  &56.5   &   \\
G195.00-16.95  &\multicolumn{3}{c}{611$\times$430(10.5)}  & 0.56   & -2.2(0.3) & 1.4  &69.8   &55.0   &24.5   &   \\
               &\multicolumn{3}{c}{613$\times$449(-82.7)} & 0.57   & -2.2(0.2) & 1.5  &79.1   &56.3   &23.8   &   \\
\enddata
\end{deluxetable}

\clearpage

\section{Appendix: On line images and tables}

\setcounter{figure}{1}

\begin{figure}
\includegraphics[angle=0,scale=0.7]{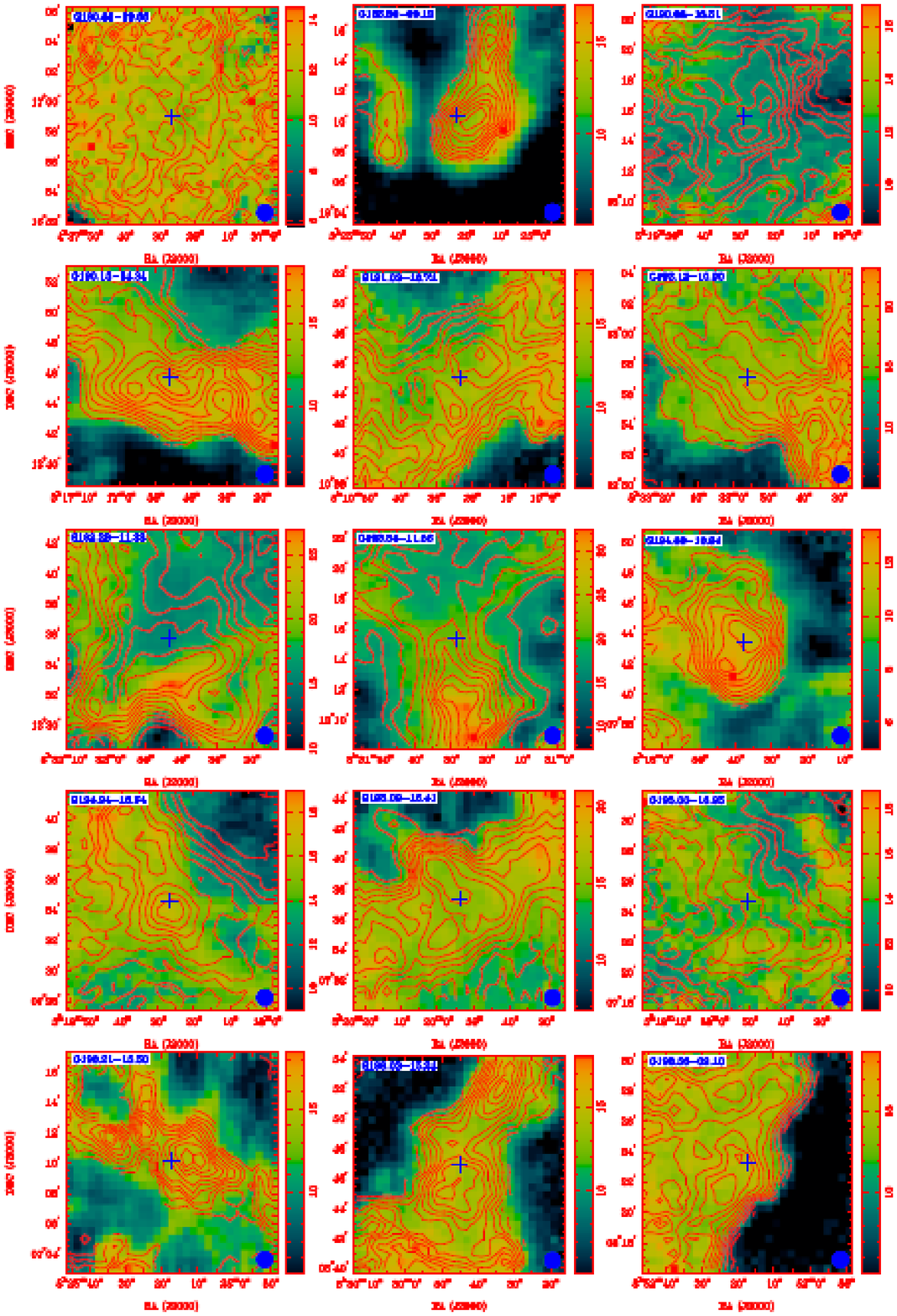}
\caption{The contours represent the column density distribution. The contour levels are from 10\% to 90\% in steps of 10\% of the peak value. The image in color scale shows the distribution of excitation temperature. The cloud names are labeled in the upper-left corner in each panel.}
\end{figure}

\clearpage
\setcounter{figure}{1}

\begin{figure}
\includegraphics[angle=0,scale=0.6]{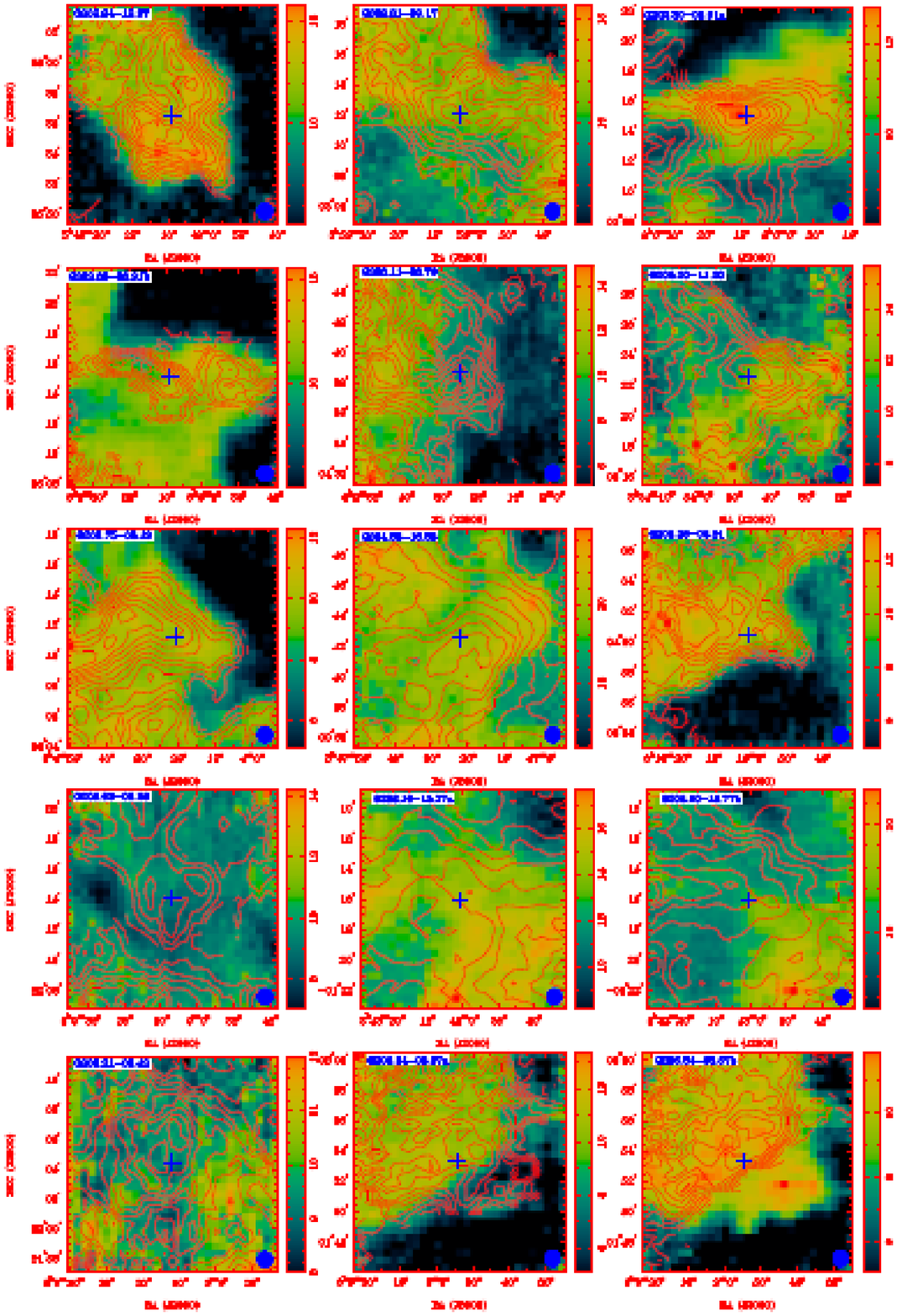}
\caption{continued}
\end{figure}

\clearpage
\setcounter{figure}{1}

\begin{figure}
\includegraphics[angle=0,scale=0.6]{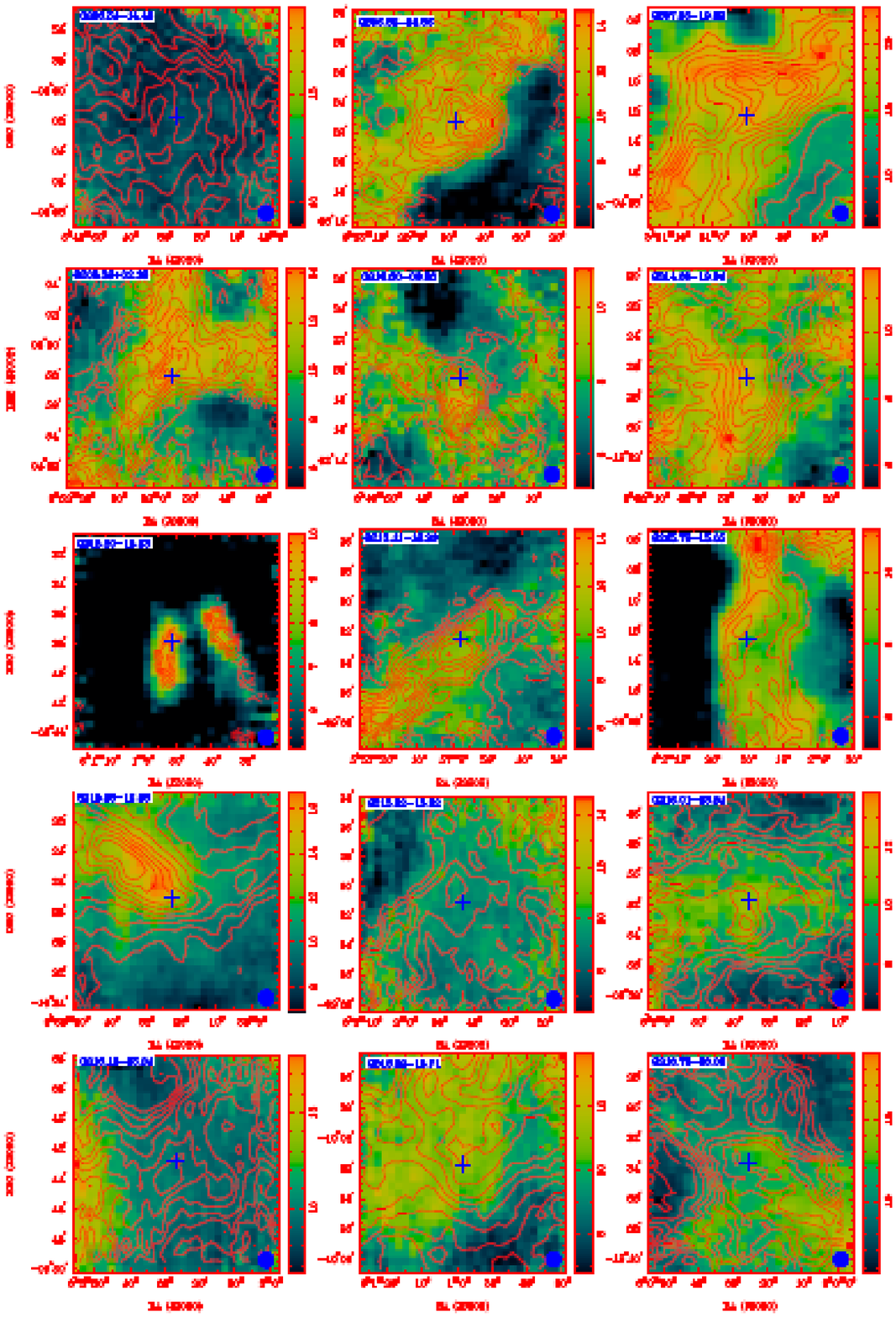}
\caption{continued}
\end{figure}

\clearpage
\setcounter{figure}{1}

\begin{figure}
\includegraphics[angle=0,scale=0.6]{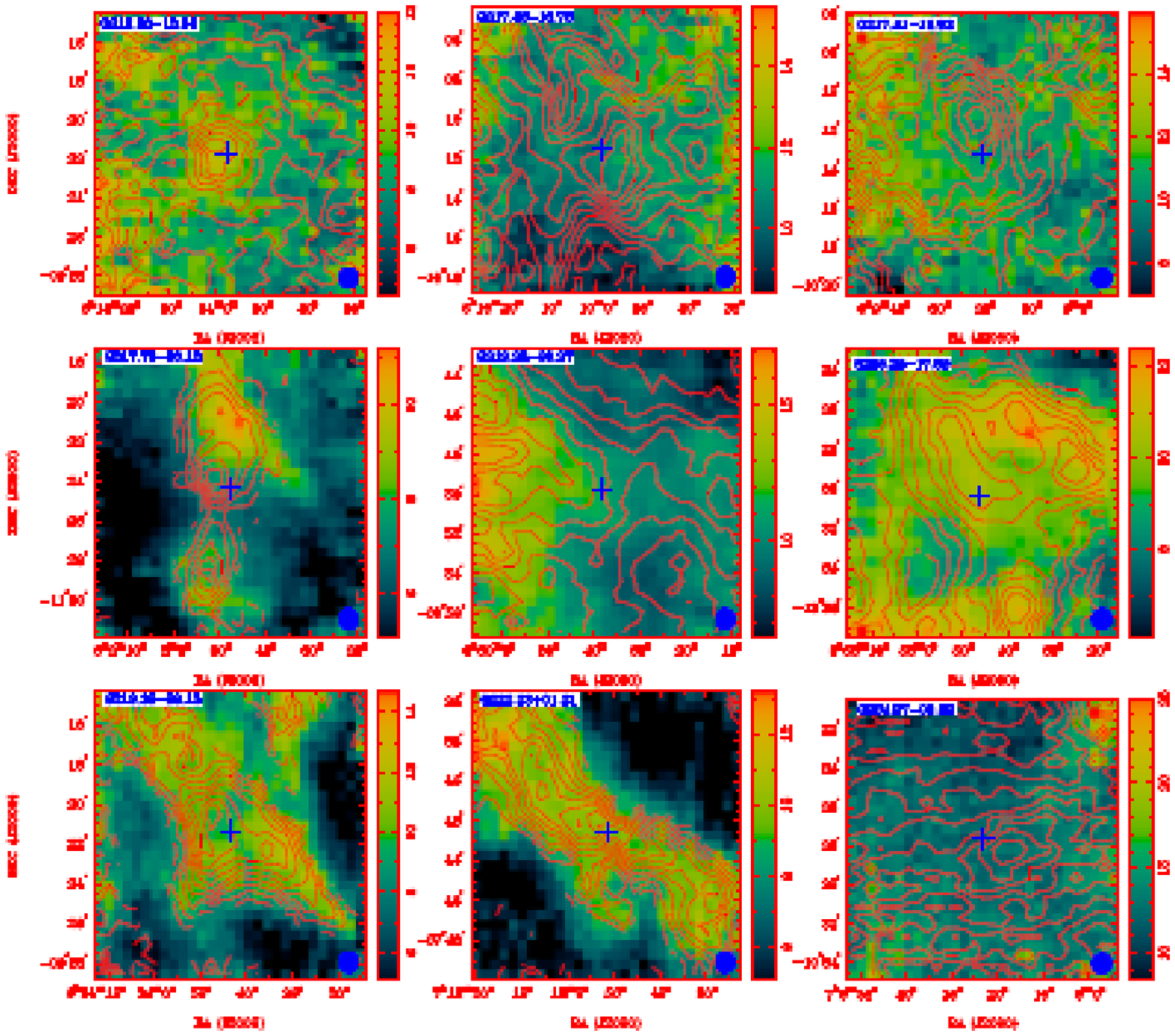}
\caption{continued}
\end{figure}

\clearpage
\begin{figure}
\includegraphics[angle=0,scale=0.6]{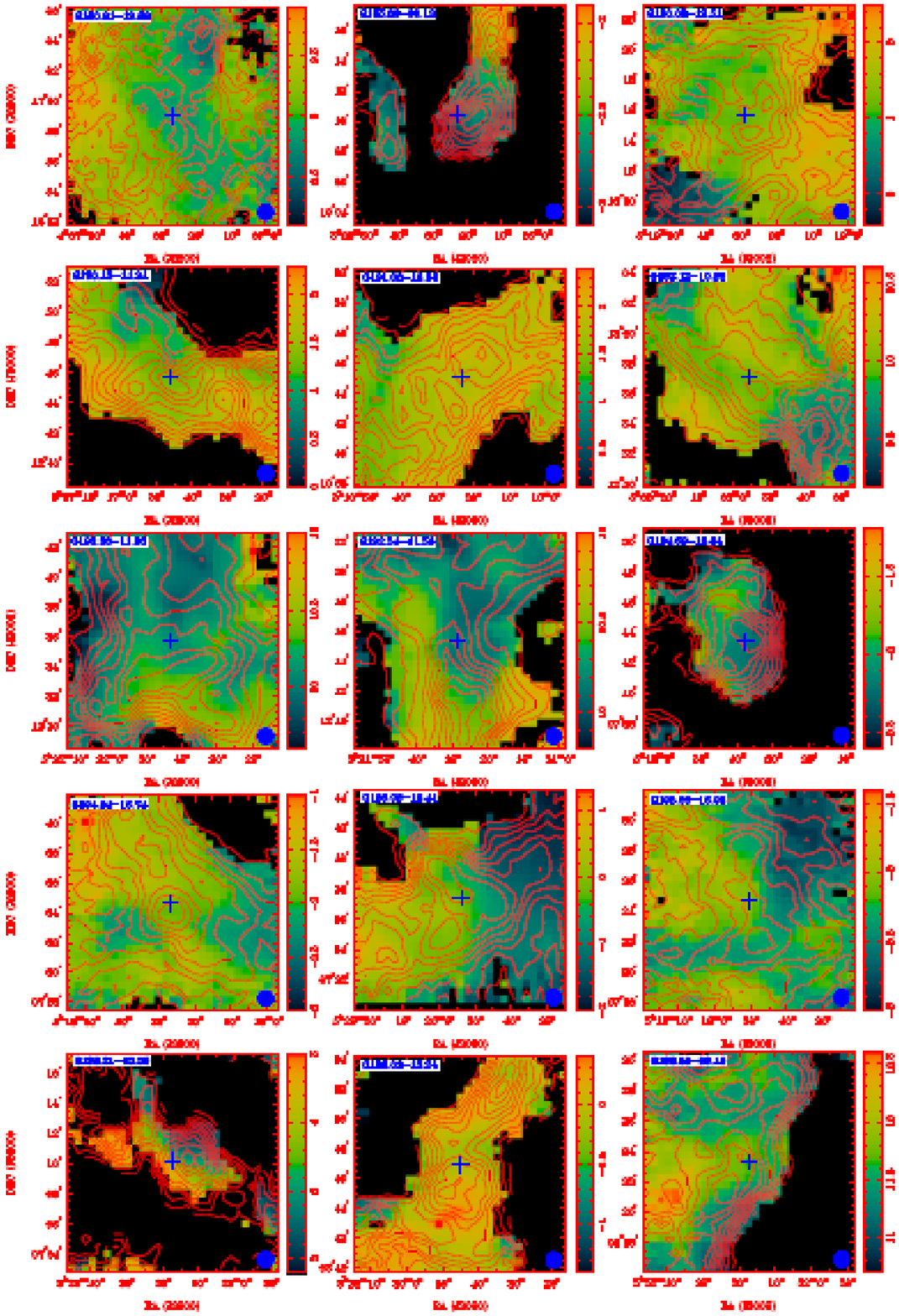}
\caption{The contours represent the column density distribution. The contour levels are from 10\% to 90\% in steps of 10\% of the peak value. The first momentum maps of $^{13}$CO (1-0) emission are shown in color scale. The cloud names are labeled in the upper-left corner in each panel. }
\end{figure}

\clearpage
\setcounter{figure}{2}

\begin{figure}
\includegraphics[angle=0,scale=0.6]{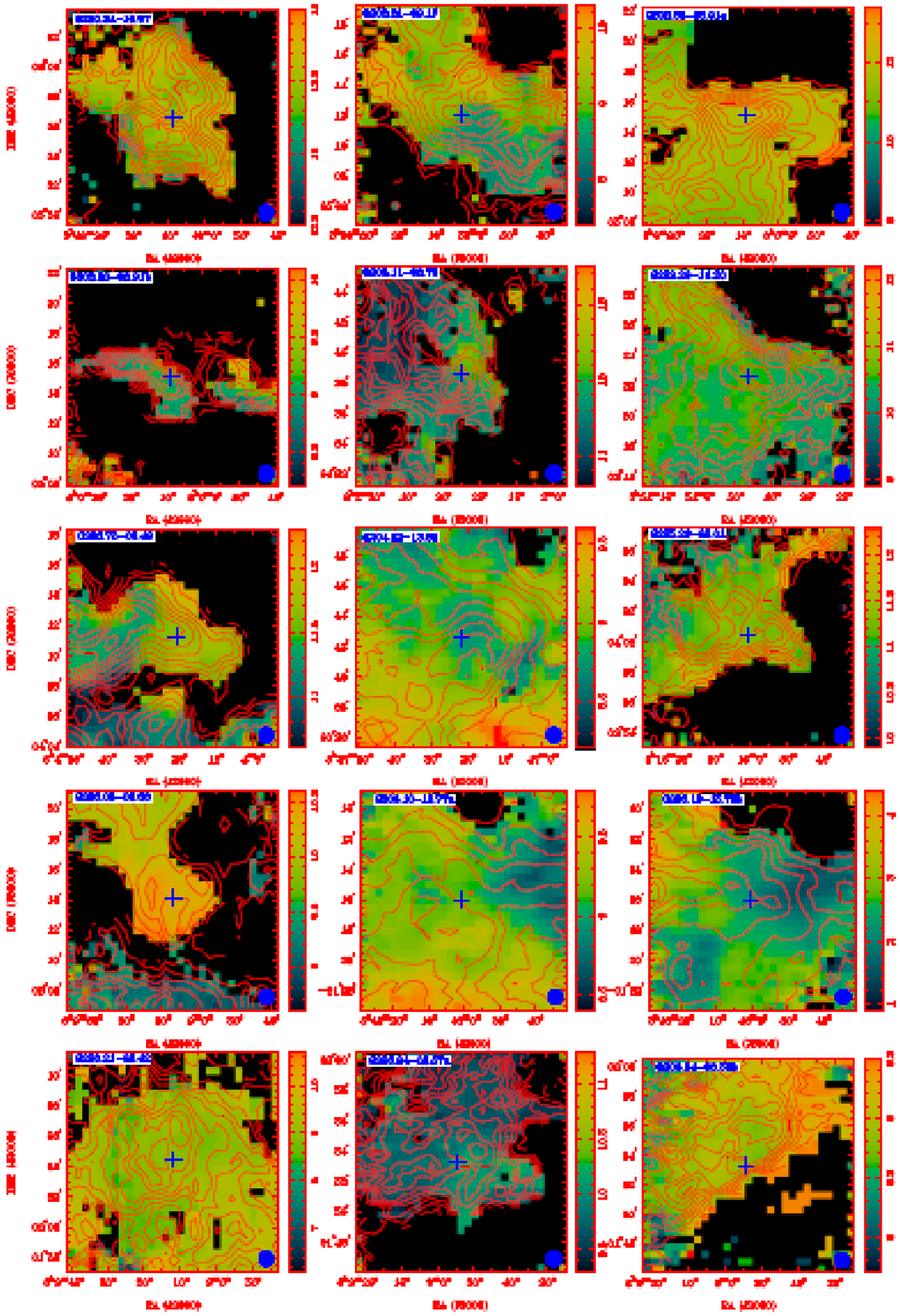}
\caption{continued}
\end{figure}

\clearpage
\setcounter{figure}{2}

\begin{figure}
\includegraphics[angle=0,scale=0.6]{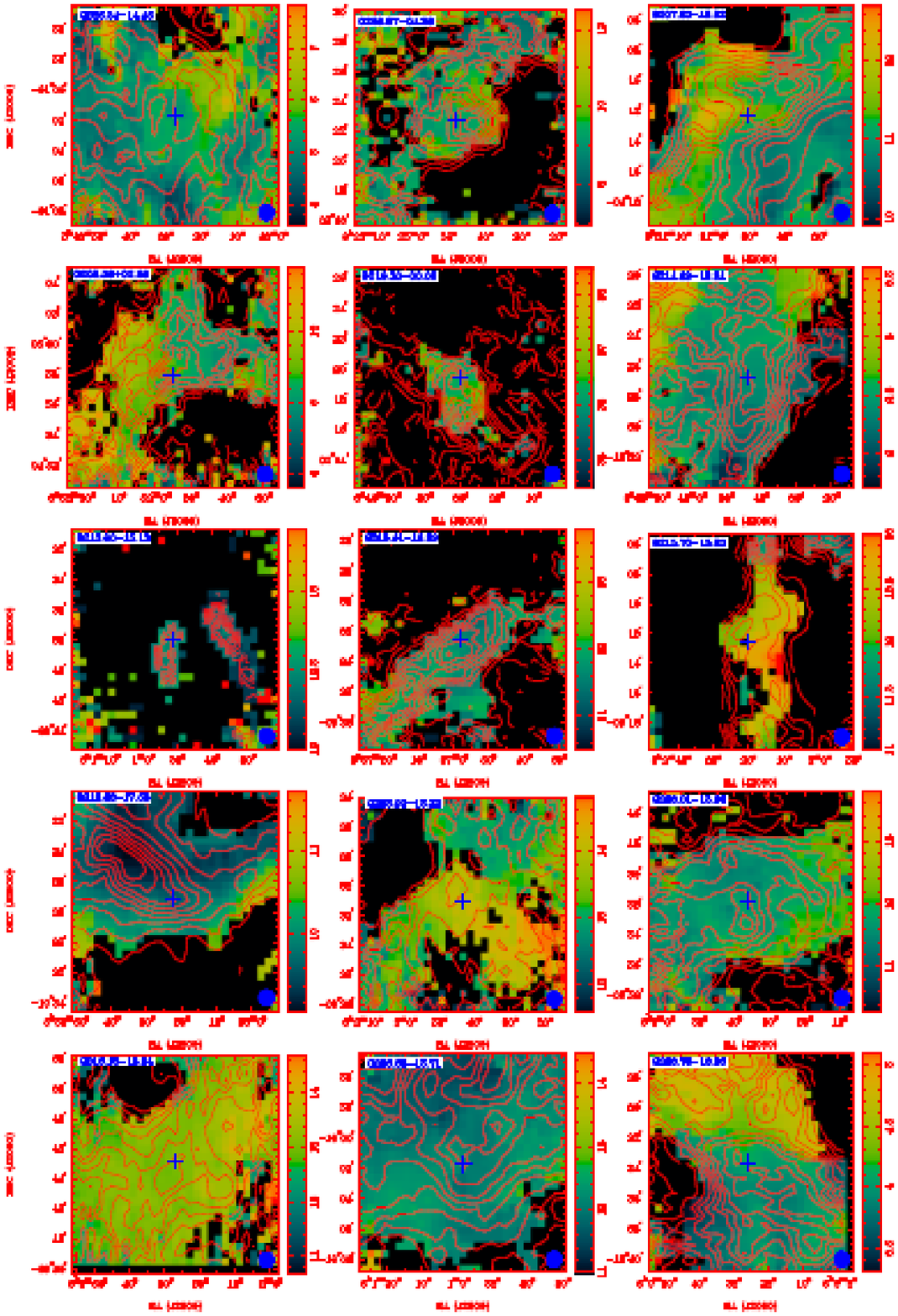}
\caption{continued}
\end{figure}

\clearpage
\setcounter{figure}{2}

\begin{figure}
\includegraphics[angle=0,scale=0.6]{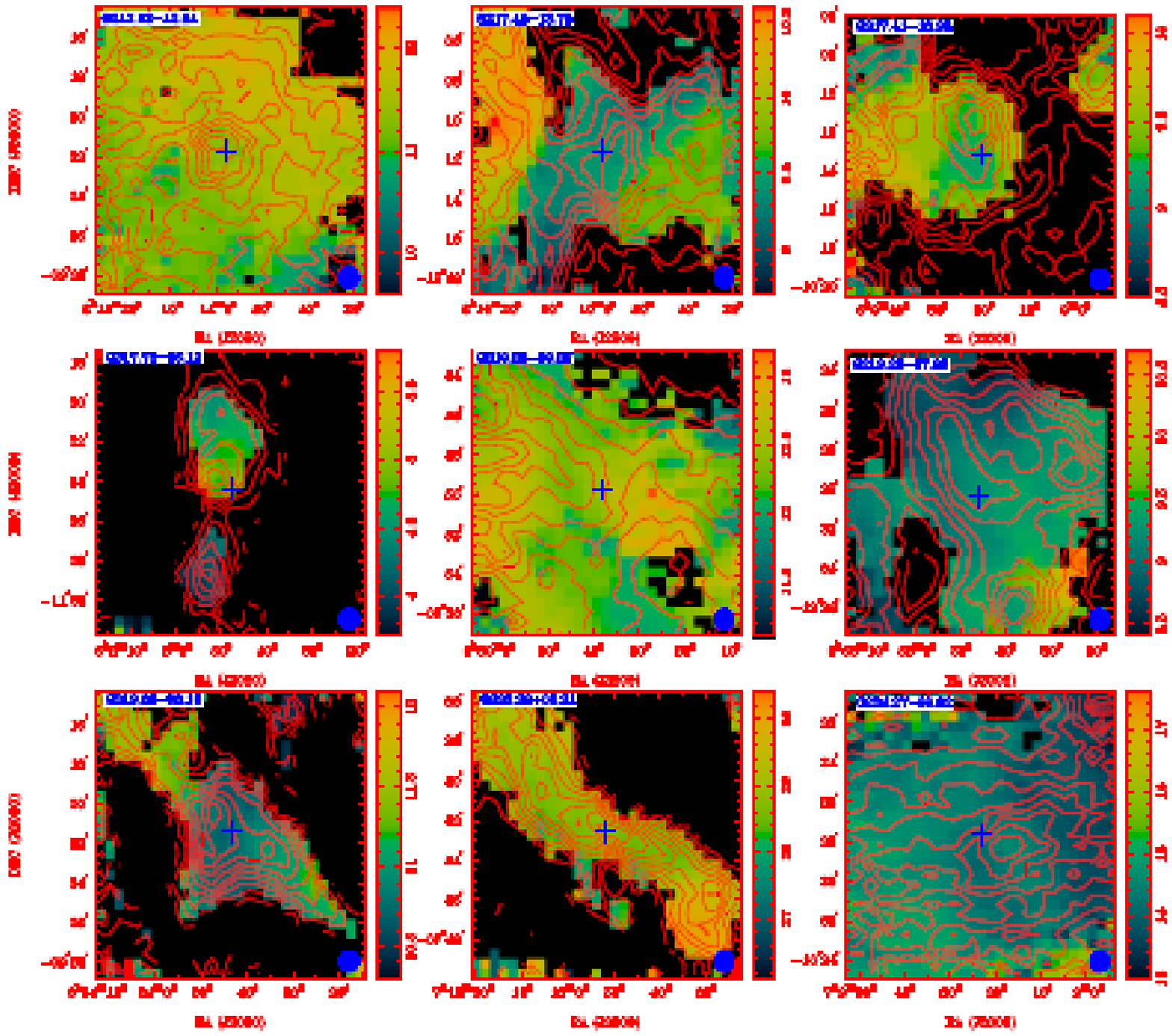}
\caption{continued}
\end{figure}

\clearpage
\begin{figure}
\includegraphics[angle=0,scale=0.6]{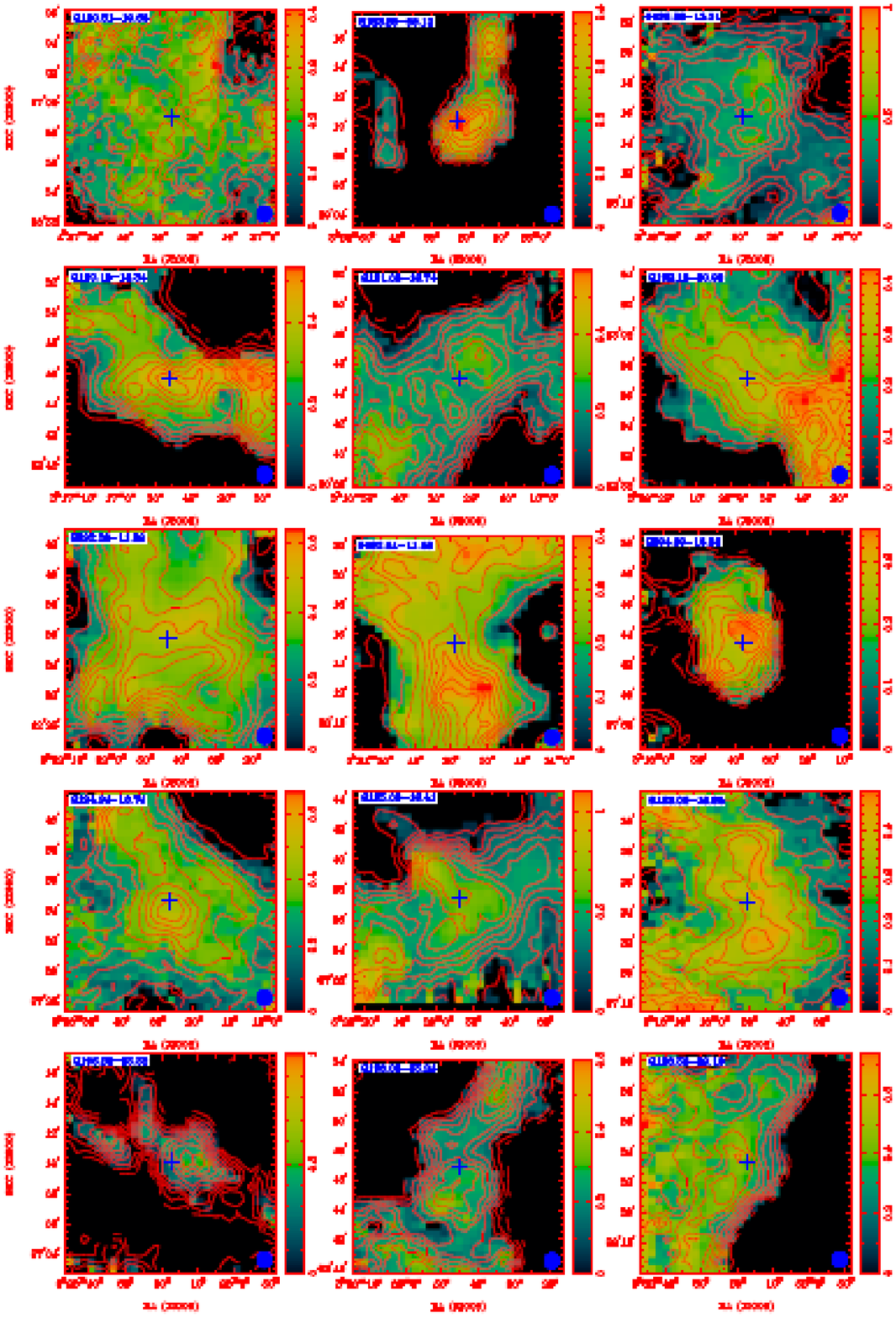}
\caption{The contours represent the column density distribution. The contour levels are from 10\% to 90\% in steps of 10\% of the peak value. The second momentum (velocity dispersion) maps of $^{13}$CO (1-0) emission are shown in color scale. The cloud names are labeled in the upper-left corner in each panel.}
\end{figure}

\clearpage
\setcounter{figure}{3}

\begin{figure}
\includegraphics[angle=0,scale=0.6]{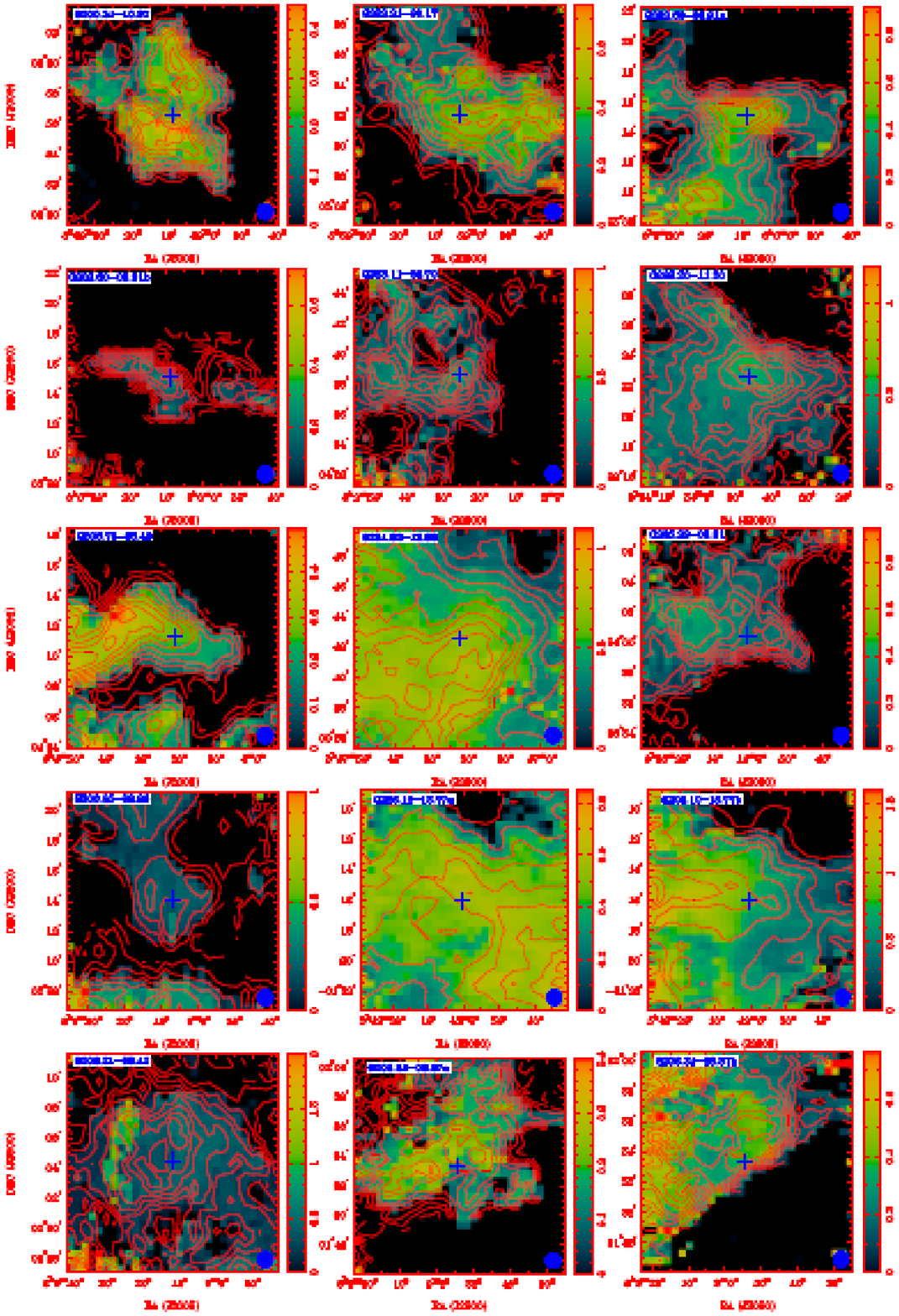}
\caption{continued}
\end{figure}

\clearpage
\setcounter{figure}{3}

\begin{figure}
\includegraphics[angle=0,scale=0.6]{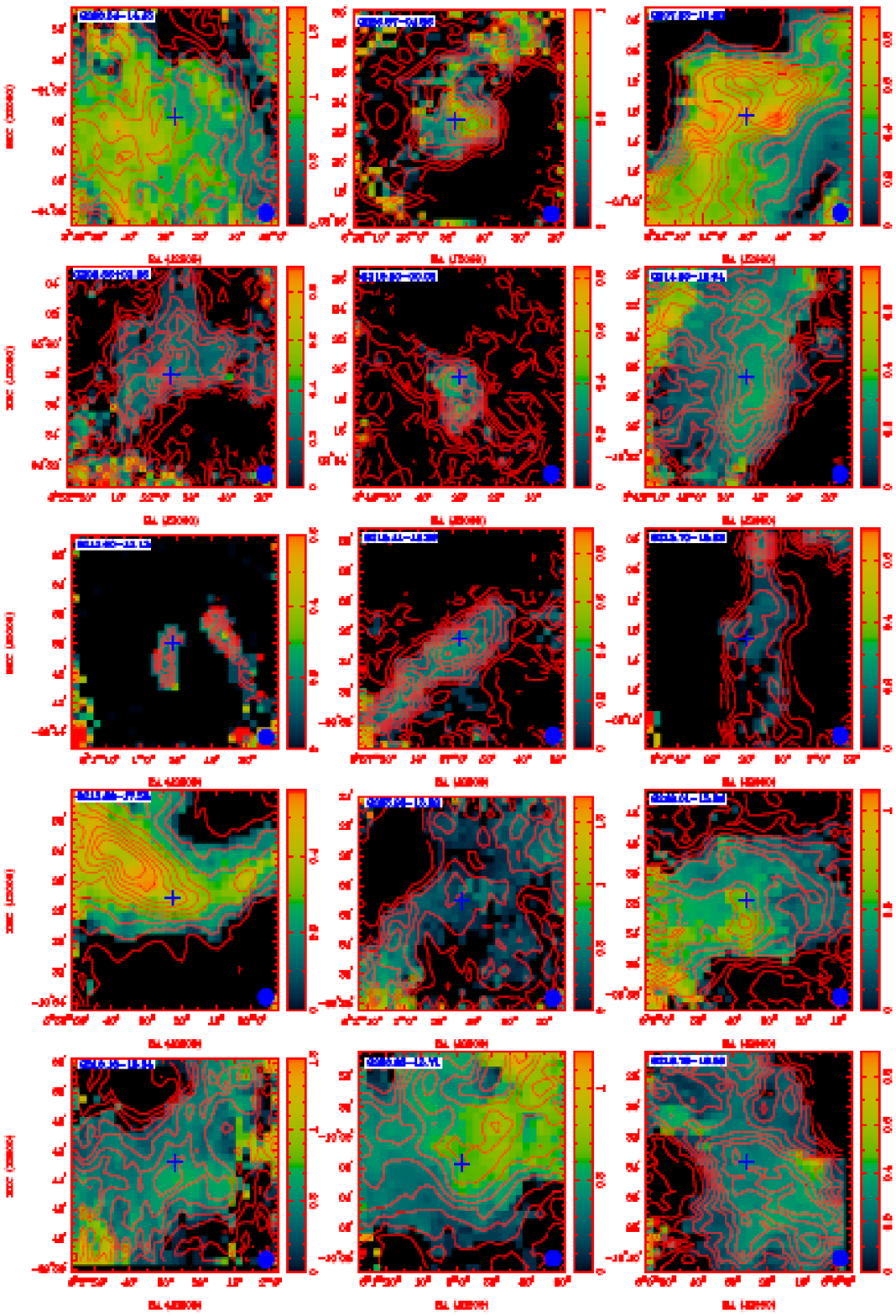}
\caption{continued}
\end{figure}

\clearpage
\setcounter{figure}{3}

\begin{figure}
\includegraphics[angle=0,scale=0.6]{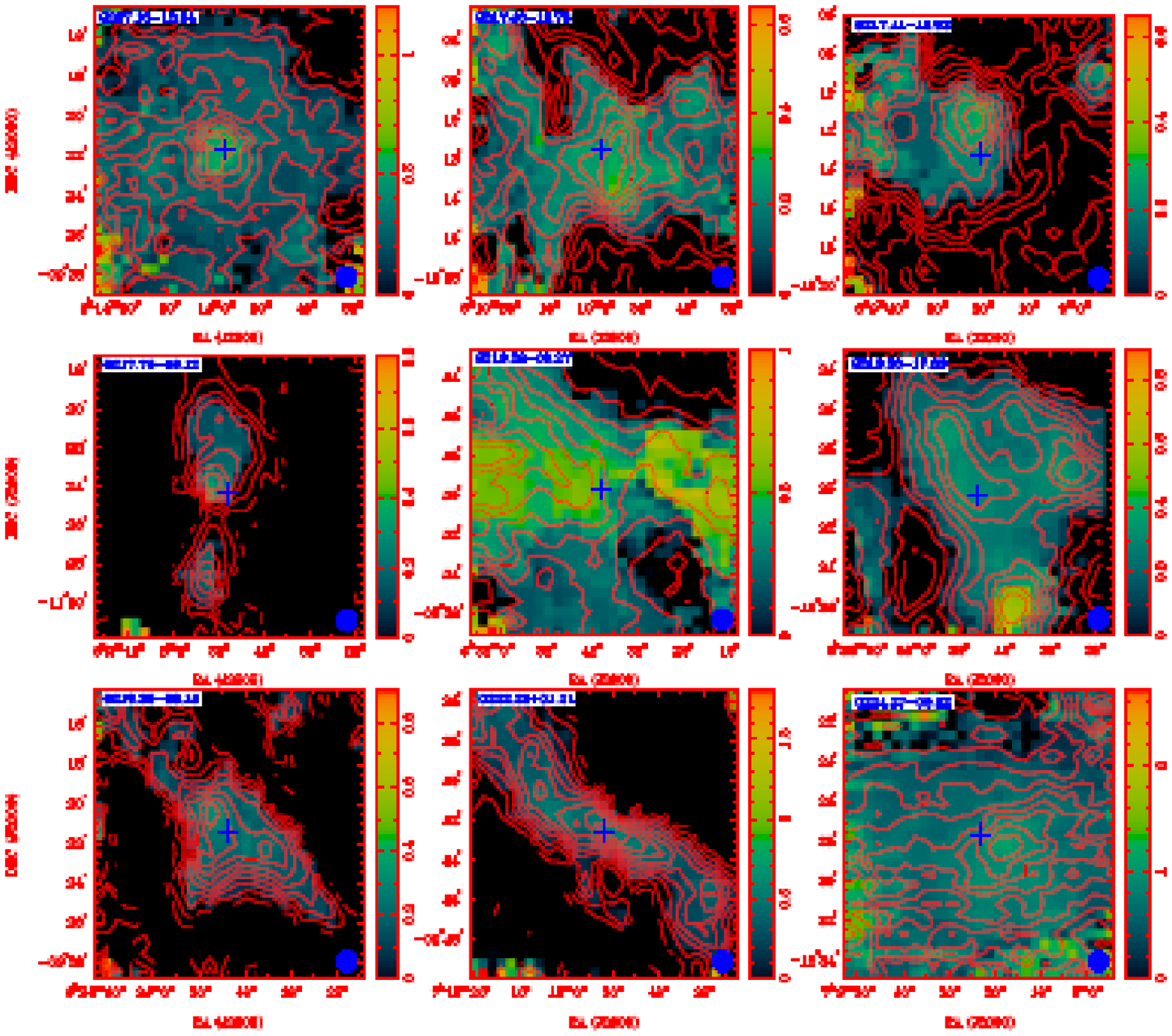}
\caption{continued}
\end{figure}

\clearpage
\setcounter{table}{0}
\begin{deluxetable}{ccrrrrrrrrrrrrrrrrrrcrl}
\setlength{\tabcolsep}{0.05in} \rotate
\tabletypesize{\scriptsize}  \tablecaption{The parameters of the Planck cold clumps surveyed}
 \tablewidth{0pt} \tablehead{
  Name& Glon & Glat &Ra(J2000)  &Dec(J2000) &Ra(B1950)  &Dec(B1950) &V$_{lsr}$ & apflux857 &Remarks\tablenotemark{a}\\
      &($\arcdeg$) &($\arcdeg$) &(h m s) &(d m s) &(h m s) &(d m s) &(km~s$^{-1}$) & (Jy) }
\startdata
G180.81-19.66     & 180.81297    & -19.669159       & 04 37 26.49  & +16 59 01.22  & 04 34 33.78  &  +16 53 02.05   & 8.91  & 102.8 &DNe \\
G185.80-09.12     & 185.80077    & -9.1214981       & 05 25 22.61  & +19 10 27.12  & 05 22 25.97  &  +19 07 49.92   & -2.75 & 74.9  &smm \\
G190.08-13.51     & 190.08543    & -13.516411       & 05 19 28.71  & +13 15 43.32  & 05 16 39.53  &  +13 12 40.97   & 1.28  & 72.9  &DNe,Cld \\
G190.15-14.34     & 190.15135    & -14.342709       & 05 16 45.83  & +12 45 44.57  & 05 13 57.29  &  +12 42 30.58   & 1.32  & 108.8 & \\
G191.03-16.74     & 191.03026    & -16.743393       & 05 10 23.30  & +10 45 03.71  & 05 07 37.24  &  +10 41 22.53   & 1.75  & 46.0  &PoC \\
G192.12-10.90     & 192.12889    & -10.901871       & 05 32 55.83  & +12 57 08.60  & 05 30 06.88  &  +12 55 04.38   & 10.02 & 200.7 &IRAS 05300+1253, smm\\
G192.28-11.33     & 192.2827     & -11.339053       & 05 31 43.36  & +12 35 44.79  & 05 28 54.85  &  +12 33 35.35   & 10.13 & 278.1 &Cld,DNe,PoC \\
G192.54-11.56     & 192.54637    & -11.567412       & 05 31 28.55  & +12 15 22.39  & 05 28 40.45  &  +12 13 11.90   & 10.21 & 208.2 &DNe \\
\enddata
\tablenotetext{a}{Simbad identifiers within 5$\arcmin$. DNe: Dark Cloud (nebula); smm: sub-millimetric source; Cld: Cloud; PoC: Part of Cloud; Rad: Radio-source; RNe: Reflection Nebula; MoC: Molecular Cloud; IRAS: IRAS point source.}
\end{deluxetable}

\clearpage

\setcounter{table}{1}
\begin{deluxetable}{ccccccccccccccccccccccc}
\tabletypesize{\scriptsize} \tablecolumns{16}
\tablewidth{0pc} \rotate \setlength{\tabcolsep}{0.04in}
\tablecaption{Derived parameters of gas emission over the whole clouds} \tablehead{
 Name & \multicolumn{3}{c}{N$_{H_{2}}$}  &\multicolumn{3}{c}{T$_{ex}$} &\multicolumn{3}{c}{$\sigma_{Therm}$} &\multicolumn{3}{c}{$\sigma_{NT}$} &\multicolumn{3}{c}{$\sigma_{3D}$}\\
\cline{2-16}
& mean & \multicolumn{2}{c}{P-value} & mean & \multicolumn{2}{c}{P-value} & mean & \multicolumn{2}{c}{P-value} & mean & \multicolumn{2}{c}{P-value}& mean & \multicolumn{2}{c}{P-value}\\
\cline{2-16} \\
 & (10$^{21}$ cm$^{-2}$) &normal &lognormal & (K) &normal &lognormal & (km~s$^{-1}$) &normal &lognormal &(km~s$^{-1}$) &normal &lognormal &(km~s$^{-1}$) &normal &lognormal
}\startdata
G180.81-19.66  & 2.10$_{-0.32}^{+0.39}$   & 0.16 &0.00  &  12.23$_{-0.55}^{+0.53}$ & 0.77 &0.76  & 0.19$_{-0.00}^{+0.00}$  &0.79 &0.76   & 0.19$_{-0.03}^{+0.02}$   & 0.00  & 0.00  & 0.47$_{-0.04}^{+0.02}$  &0.00  &0.00      \\
G185.80-09.12  & 2.11$_{-0.79}^{+0.66}$   & 0.51 &0.28  &  12.67$_{-1.06}^{+0.93}$ & 0.09 &0.33  & 0.20$_{-0.01}^{+0.01}$  &0.18 &0.33   & 0.22$_{-0.05}^{+0.07}$   & 0.47  & 0.01  & 0.52$_{-0.08}^{+0.09}$  &0.07  &0.31      \\
G190.08-13.51  & 2.45$_{-0.64}^{+0.68}$   & 0.00 &0.10  &  11.88$_{-0.54}^{+0.44}$ & 0.00 &0.01  & 0.19$_{-0.00}^{+0.00}$  &0.00 &0.01   & 0.30$_{-0.09}^{+0.09}$   & 0.03  & 0.00  & 0.63$_{-0.14}^{+0.12}$  &0.00  &0.09      \\
G190.15-14.34  & 2.85$_{-0.79}^{+0.82}$   & 0.15 &0.01  &  14.34$_{-1.09}^{+1.13}$ & 0.20 &0.07  & 0.21$_{-0.01}^{+0.01}$  &0.14 &0.07   & 0.29$_{-0.07}^{+0.05}$   & 0.16  & 0.01  & 0.63$_{-0.10}^{+0.08}$  &0.01  &0.47        \\
G191.03-16.74  & 2.25$_{-0.57}^{+0.52}$   & 0.12 &0.00  &  13.74$_{-1.11}^{+0.92}$ & 0.36 &0.04  & 0.20$_{-0.01}^{+0.01}$  &0.13 &0.04   & 0.21$_{-0.04}^{+0.04}$   & 0.01  & 0.00  & 0.52$_{-0.05}^{+0.05}$  &0.56  &0.12        \\
G192.12-10.90  & 4.14$_{-1.23}^{+1.13}$   & 0.02 &0.02  &  17.23$_{-1.38}^{+1.23}$ & 0.04 &0.29  & 0.23$_{-0.01}^{+0.01}$  &0.11 &0.28   & 0.28$_{-0.05}^{+0.05}$   & 0.00  & 0.11  & 0.62$_{-0.08}^{+0.08}$  &0.00  &0.01        \\
G192.28-11.33  & 5.66$_{-1.41}^{+1.35}$   & 0.06 &0.00  &  18.90$_{-2.04}^{+1.37}$ & 0.02 &0.01  & 0.24$_{-0.01}^{+0.01}$  &0.02 &0.01   & 0.35$_{-0.05}^{+0.06}$   & 0.05  & 0.00  & 0.74$_{-0.07}^{+0.08}$  &0.04  &0.00        \\
G192.54-11.56  & 5.33$_{-1.34}^{+0.58}$   & 0.00 &0.01  &  21.12$_{-2.25}^{+0.79}$ & 0.00 &0.00  & 0.25$_{-0.01}^{+0.01}$  &0.00 &0.00   & 0.28$_{-0.03}^{+0.03}$   & 0.59  & 0.10  & 0.65$_{-0.04}^{+0.04}$  &0.73  &0.79        \\
\enddata
\tablecomments{The errors throughout all the tables are calculated from the first and third quartiles. }
\end{deluxetable}

\clearpage

\setcounter{table}{2}
\begin{deluxetable}{ccccccccccccccccccccccccccccccccccccccccc}
\tabletypesize{\scriptsize} \tablecolumns{35} \rotate
\tablewidth{0pc} \setlength{\tabcolsep}{0.02in}
\tablecaption{Derived parameters of the dense cores} \tablehead{
 Name& \multicolumn{2}{c}{Offset}  & \multicolumn{4}{c}{N$_{H_{2}}$}  &\multicolumn{3}{c}{T$_{ex}$} &\multicolumn{3}{c}{$\sigma_{Therm}$} &\multicolumn{3}{c}{$\sigma_{NT}$} &\multicolumn{3}{c}{$\sigma_{3D}$} \\
\cline{3-19}
& \multicolumn{2}{c}{}&max & mean & \multicolumn{2}{c}{P-value} & mean & \multicolumn{2}{c}{P-value} & mean & \multicolumn{2}{c}{P-value} & mean & \multicolumn{2}{c}{P-value}& mean & \multicolumn{2}{c}{P-value}\\
\cline{3-19} \\
 & \multicolumn{2}{c}{($\arcsec,\arcsec$)} & (10$^{21}$ cm$^{-2}$)& (10$^{21}$ cm$^{-2}$) &normal &lognormal & (K) &normal &lognormal & (km~s$^{-1}$) &normal &lognormal &(km~s$^{-1}$) &normal &lognormal &(km~s$^{-1}$) &normal &lognormal
}\startdata
G185.80-09.12  & \multicolumn{2}{c}{(-69,-19)}   &4.37  & 2.95$_{-0.31}^{+0.72}$   &0.47  &0.02  & 13.10$_{-0.92}^{+0.91}$  &0.21  &0.32  & 0.20$_{-0.01}^{+0.01}$  & 0.26 &0.32  & 0.28$_{-0.04}^{+0.05}$    & 0.22 &0.03  & 0.60$_{-0.05}^{+0.06}$   & 0.30 & 0.11   \\
               & \multicolumn{2}{c}{(-134,290)}  &3.36  & 2.25$_{-0.61}^{+0.62}$   &0.90  &0.90  & 11.89$_{-0.35}^{+0.51}$  &0.48  &0.42  & 0.19$_{-0.00}^{+0.00}$  & 0.45 &0.42  & 0.23$_{-0.04}^{+0.06}$    & 0.86 &0.61  & 0.52$_{-0.06}^{+0.08}$   & 0.87 & 0.91    \\
G190.08-13.51  & \multicolumn{2}{c}{(110,106)}   &4.86  & 3.11$_{-0.67}^{+0.64}$   &0.12  &0.00  & 11.91$_{-0.40}^{+0.35}$  &0.22  &0.32  & 0.19$_{-0.00}^{+0.00}$  & 0.27 &0.33  & 0.36$_{-0.06}^{+0.07}$    & 0.38 &0.00  & 0.71$_{-0.09}^{+0.10}$   & 0.56 & 0.03    \\
               & \multicolumn{2}{c}{(25,-62)}    &4.86  & 3.49$_{-0.40}^{+0.48}$   &0.72  &0.25  & 11.69$_{-0.21}^{+0.20}$  &0.62  &0.73  & 0.19$_{-0.00}^{+0.00}$  & 0.68 &0.74  & 0.39$_{-0.04}^{+0.05}$    & 0.29 &0.02  & 0.75$_{-0.06}^{+0.08}$   & 0.36 & 0.06    \\
               & \multicolumn{2}{c}{(-17,-89)}   &4.86  & 4.03$_{-0.26}^{+0.40}$   &0.99  &0.89  & 11.66$_{-0.18}^{+0.14}$  &0.96  &0.96  & 0.19$_{-0.00}^{+0.00}$  & 0.96 &0.96  & 0.44$_{-0.03}^{+0.04}$    & 0.97 &0.98  & 0.82$_{-0.05}^{+0.06}$   & 0.98 & 0.98    \\
               & \multicolumn{2}{c}{(-22,62)}    &4.86  & 3.04$_{-0.60}^{+0.60}$   &0.14  &0.00  & 11.59$_{-0.31}^{+0.38}$  &0.25  &0.12  & 0.19$_{-0.00}^{+0.00}$  & 0.18 &0.12  & 0.35$_{-0.09}^{+0.08}$    & 0.11 &0.00  & 0.69$_{-0.13}^{+0.12}$   & 0.17 & 0.00    \\
G190.15-14.34  & \multicolumn{2}{c}{(122,-27)}   &5.27  & 3.32$_{-0.48}^{+0.66}$   &0.74  &0.05  & 15.03$_{-0.83}^{+0.85}$  &0.37  &0.50  & 0.21$_{-0.01}^{+0.01}$  & 0.43 &0.50  & 0.31$_{-0.04}^{+0.04}$    & 0.97 &1.00  & 0.66$_{-0.06}^{+0.05}$   & 0.74 & 0.98    \\
\enddata
\end{deluxetable}

\clearpage
\setcounter{table}{3}
\begin{deluxetable}{ccccccccccccccccccccccccccccccccccc}
\tabletypesize{\scriptsize} \tablecolumns{9}
\tablewidth{0pc} \setlength{\tabcolsep}{0.05in}
\tablecaption{Derived parameters of the dense cores} \tablehead{
 Name & \multicolumn{3}{c}{Deconvolved Size} & R & V$_{lsr}$ & n  &\colhead{M$_{LTE}$} &\colhead{M$_{vir}$} &\colhead{M$_{J}$} &Remarks\\
 & \multicolumn{3}{c}{($\arcsec\times\arcsec$($\arcdeg$))}& (pc)  &(km~s$^{-1}$) &(10$^{3}$ cm$^{-3}$) &\colhead{(M$_{\sun}$)} &\colhead{(M$_{\sun}$)} &\colhead{(M$_{\sun}$)}
}\startdata
G185.80-09.12  &\multicolumn{3}{c}{291$\times$176(-42.8)} & 0.22   & -2.8(0.1) & 3.2  &9.6    &18.4   &12.6   &   \\
               &\multicolumn{3}{c}{313$\times$105(-7.6)}  & 0.18   & -2.3(0.1) & 3.1  &4.8    &11.1   &8.4    &   \\
G190.08-13.51  &\multicolumn{3}{c}{577$\times$429(75.8)}  & 0.54   & 1.2(0.2)  & 1.4  &65.6   &63.8   &31.1   &   \\
               &\multicolumn{3}{c}{829$\times$322(-25.5)} & 0.56   & 1.3(0.2)  & 1.4  &70.8   &73.9   &38.3   &   \\
               &\multicolumn{3}{c}{153$\times$121(-39.4)} & 0.15   & 1.3(0.1)  & 5.3  &4.9    &23.3   &26.5   &   \\
               &\multicolumn{3}{c}{627$\times$504(4.1)}   & 0.61   & 1.4(0.3)  & 1.3  &83.8   &68.0   &30.9   &   \\
G190.15-14.34  &\multicolumn{3}{c}{332$\times$223(85.3)}  & 0.30   & 1.5(0.2)  & 2.9  &21.3   &30.1   &17.2   &   \\
               &\multicolumn{3}{c}{518$\times$293(80.0)}  & 0.42   & 1.4(0.1)  & 2.0  &43.8   &45.8   &23.4   &   \\
\enddata
\end{deluxetable}


\begin{thebibliography}{}
{\small
\bibitem[Bally et al.(1987)]{ba87}Bally, J., Stark, A. A., Wilson, R. W., \& Langer, W. D. 1987, \apj, 312, L45

\bibitem[Bally et al.(2005)]{ba05}Bally, John., Cunningham, Nathaniel., Moeckel, Nickolas., Smith, Nathan., 2005, IAUS, 227, 12

\bibitem[Ballesteros-Paredes et al.(2011)]{ba11}Ballesteros-Paredes, Javier., V\'{a}zquez-Semadeni, Enrique., Gazol, Adriana., Hartmann, Lee W., Heitsch, Fabian., Col\'{\i}n, Pedro., 2011, \mnras, 416, 1436

\bibitem[Bertoldi \& McKee(1992)]{be92}Bertoldi, F., \& McKee, C. F. 1992, \apj, 395, 140

\bibitem[Buckle et al.(2009)]{bu09}Buckle, J. V., et al., 2009, \mnras, 399, 1026

\bibitem[Castets et al.(1990)]{cas90}Castets, A., Duvert, G., Dutrey, A., et al. 1990, \aap, 234, 469

\bibitem[Cowie, Songaila \& York.(1979)]{cow79}Cowie, L. L., Songaila, A., \& York, D. G. 1979, \apj, 230, 469

\bibitem[Dame et al.(1987)]{da87}Dame, T. M., Ungerechts, H., Cohen, R. S., de Geus, E. J., Grenier, I. A., May, J., Murphy, D. C., Nyman, L.-A. ., \& Thaddeus, P. 1987, \apj, 322, 706

\bibitem[Dame, Hartmann, \& Thaddeus(2001)]{dame01}Dame, T. M., Hartmann, Dap., Thaddeus, P., 2001, \apj, 547, 792

\bibitem[Du \& Yang(2008)]{du08}Du, Fujun., \& Yang, Ji., 2008, \apj, 686, 384

\bibitem[Elmegreen \& Lada(1977)]{el77}Elmegreen, B. G., \& Lada, C. J. 1977, \apj, 214, 725

\bibitem[Finkbeiner (2003)]{fin03}Finkbeiner, D. P., 2003, \apjs, 146, 407

\bibitem[Froebrich et al.(2007)]{fro07}Froebrich D., Murphy G. C., Smith M. D., Walsh J., Del Burgo C., 2007. \mnras, 378, 1447

\bibitem[Froebrich \& Rowles(2010)]{fro10}Froebrich D., Rowles J., 2010, \mnras, 406, 1350

\bibitem[Guilloteau \& Lucas(2000)]{gui00}Guilloteau, S. \& Lucas, R., 2000, in Astronomical Society of the Pacific Conference Series, Vol. 217, Imaging at Radio through Submillimeter
Wavelengths, ed. J. G. Mangum \& S. J. E. Radford, 299

\bibitem[Garden et al.(1991)]{gar91}Garden, R. P., Hayashi, M., Hasegawa, T., Gatley, I., Kaifu,
N., 1991, \apj, 374, 540

\bibitem[Goodman, Pineda \& Schnee(2009)]{goo09}Goodman A. A., Pineda J. E., Schnee S. L., 2009, \apj, 692, 91

\bibitem[Hennebelle \& Chabrier(2008)]{hen08}Hennebelle, P., \& Chabrier, G., 2008, \apj, 684, 395

\bibitem[Hillenbrand (1997)]{Hil97}Hillenbrand, L. A. 1997, \aj, 113, 1733

\bibitem[Li et al.(2007)]{Li07}Li, D., Velusamy, T., Goldsmith, P. F., Langer, William D., 2007, \apj, 655, 351

\bibitem[Ikeda et al.(2009)]{Ike09}Ikeda N., Kitamura Y., Sunada K., 2009, \apj, 691,1560


\bibitem[Kainulainen et al.(2009)]{ka09}Kainulainen J., Beuther H., Henning T., Plume R., 2009, \aap, 508, L35

\bibitem[Kritsuk, Norman \& Wagner(2011)]{kr11}Kritsuk A. G., Norman M. L., Wagner R., 2011, \apj, 727, L20

\bibitem[Kramer et al.(1996)]{kra96}Kramer C., Stutzki J., Winnewisser G., 1996, \aap, 307, 915

\bibitem[Larson (1981)]{lar81}Larson, R. B., 1981, \mnras, 194, 809

\bibitem[Maddalena et al.(1986)]{ma86}Maddalena, R. J., Moscowitz, J., Thaddeus, P., \& Morris, M. 1986, \apj, 303, 375

\bibitem[MacLaren et al.(1988)]{ma88}MacLaren I., Richardson K.M.,Wolfendale A.W., 1988, \apj, 333, 821

\bibitem[Nagahama et al.(1998)]{na98}Nagahama, T., Mizuno, A., Ogawa, H., \& Fukui, Y. 1998, \aj, 116, 336

\bibitem[Onishi et al.(2002)]{on02}Onishi T., Mizuno A., Kawamura A., Tachihara K., Fukui Y.,
2002, \apj, 575, 950

\bibitem[Peretto \& Fuller(2010)]{pe10}Peretto, N., Fuller, G. A., 2010, \apj, 723, 555

\bibitem[Planck Collaboration et al.(2011a)]{ade11a}Planck Collaboration., Ade, P. A. R., Aghanim, N., Arnaud, M., Ashdown, M., Aumont, J., Baccigalupi, C., Balbi, A., Banday, A. J., Barreiro, R. B., et al., 2011, \aap, 536, 23

\bibitem[Planck Collaboration et al.(2011b)]{ade11b}Planck Collaboration., Ade, P. A. R., Aghanim, N., Arnaud, M., Ashdown, M., Aumont, J., Baccigalupi, C., Balbi, A., Banday, A. J., Barreiro, R. B., et al., 2011, \aap, 536, 7

\bibitem[Ridge et al.(2006)]{rid06}Ridge N. A. et al., 2006, \aj, 131, 2921


\bibitem[Salpeter(1955)]{sal55}Salpeter E.E., 1955, \apj, 121, 161

\bibitem[Sakamoto et al.(1996)]{sak96}Sakamoto, S., Hayashi, M., Hasegawa, T., Handa, T., \& Oka, T. 1994, \apj, 425, 641

\bibitem[Sadavoy et al.(2010)]{Sad10}Sadavoy, Sarah I., Di Francesco, James., Bontemps, Sylvain., Megeath, S. Thomas., Rebull, Luisa M.., 2010, \apj, 710, 1247

\bibitem[Sault et al.(1995)]{sau95}Sault, R. J., Teuben, P. J., \& Wright, M. C. H. 1995, in ASP Conf. Ser. 77, Astronomical Data Analysis Software and Systems IV, ed. R. A. Shaw, H. E. Payne, \& J. J. E. Hayes (San Francisco, CA: ASP), 433


\bibitem[Shimajiri et al.(2011)]{sh11}Shimajiri, Yoshito., Kawabe, Ryohei., Takakuwa, Shigehisa., Saito, Masao., Tsukagoshi, Takashi., 2011, \pasj, 63, 105

\bibitem[Shu, Adams \& Lizano(1987)]{shu87}Shu, Frank H., Adams, Fred C., \& Lizano, Susana., 1987, \araa, 25, 23

\bibitem[V\'{a}zquez-Semadeni (1994)]{va94}V\'{a}zquez-Semadeni E., 1994, \apj, 423, 681

\bibitem[Veltchev., Klessen., \& Clark(2011)]{ve11}Veltchev, Todor V., Klessen, Ralf S., \& Clark, Paul C., 2011, \mnras, 411, 301

\bibitem[Wang et al.(2009)]{wang09}Wang, K., Wu, Y. F., Ran, L., Yu, W. T., Miller, M., 2009, \aap, 507, 369

\bibitem[Williams et al.(1994)]{wil94}Williams J.P., de Geus E.J., Blitz L., 1994, \apj, 428, 693

\bibitem[Wilson et al.(2001)]{wil01}Wilson, T. L., Muders, D., Kramer, C., Henkel, C., 2001, \apj, 557, 240

\bibitem[Wilson et al.(2005)]{wil05}Wilson, B. A., Dame, T. M., Masheder, M. R. W., Thaddeus, P., 2005, \aap, 430, 523

\bibitem[Winnewisser, Churchwell, \& Walmsley.(1979)]{win79}Winnewisser, G., Churchwell, E., Walmsley, C. M., 1979, \aap, 72, 215

\bibitem[Wu et al.(2012)]{wu12}Wu, Y., Liu, T., Meng, F., Li, D., Qin, S.-L., Ju, B.-G., 2012, submitted to ApJ (ApJ86692)

\bibitem[Zinnecker \& Yorke(2007)]{zin07}Zinnecker, H., \& Yorke, H. W., 2007, \araa, 45, 481
}
\end{thebibliography}
\end{document}